# ZettaLith: An Architectural Exploration of Extreme-Scale AI Inference Acceleration

**Kia Silverbrook**[1]

## Abstract

The high computational cost and power consumption of current and anticipated AI systems present a major challenge for widespread deployment and further scaling. Current hardware approaches face fundamental efficiency limits. This paper introduces ZettaLith, a scalable computing architecture designed to reduce the cost and power of AI inference by over 1,000× compared to current GPU-based systems. Based on architectural analysis and technology projections, a single ZettaLith rack could potentially achieve 1.507 zettaFLOPS in 2027 - representing a theoretical 1,047× improvement in inference performance, 1,490× better power efficiency, and could be 2,325× more cost-effective than current leading GPU racks for FP4 transformer inference. The ZettaLith architecture achieves these gains by abandoning general purpose GPU applications, and via the multiplicative effect of numerous co-designed architectural innovations using established digital electronic technologies, as detailed in this paper. ZettaLith's core architectural principles scale down efficiently to exaFLOPS desktop systems and petaFLOPS mobile chips, maintaining their roughly 1,000× advantage. ZettaLith presents a simpler system architecture compared to the complex hierarchy of current GPU clusters. ZettaLith is optimized exclusively for AI inference and is not applicable for AI training.

**Note:** This paper presents a design study and architectural proposal without implementation or simulation validation. All performance projections are based on theoretical analysis and should be interpreted as near the upper bounds, pending experimental verification.

---

## 1 Introduction

Recent developments in AI workloads, particularly agentic AI and reasoning systems, have dramatically increased computational requirements. This paper explores architectural approaches to address these scaling challenges through specialized inference acceleration

ZettaLith is designed to provide this very large increase in inference compute requirements at high efficiency by specializing in efficient inference at the expense of training and HPC and graphics applications.

ZettaLith is designed to achieve 1.507 zettaFLOPS (peak, sparse, FP4) for transformer inference. This approach would enable inference of transformers with up to 20 trillion parameters within a single 84 kW rack, projected to achieve approximately 1,047× higher throughput, 1,490× better energy efficiency, and 2,325× greater cost-effectiveness than leading 2025 GPU racks.

The ZettaLith system is highly specialized for FP4 transformer inference, deliberately sacrificing general-purpose computation capabilities to achieve high efficiency for this increasingly prevalent workload.

The performance benefit stems from eight key innovations working in concert:

1. **WSSCB** (Wafer-Scale Silicon Circuit Board) - A passive silicon substrate that creates an "all-silicon domain" where computation occurs without data leaving the integrated silicon environment.
2. **CASCADE** (Column-Array Systolic Computation with Accumulation During Execution) - A novel matrix multiplication architecture that eliminates inter-chip partial sum transfers, normally responsible for most data movement.
3. **TRIMERA** (TRIchip Module for Exascale Reasoning Applications) - A specialized 3D chip stack optimizing the separation of high-speed computation from memory and control logic.
4. **SHAPE** (Simple Hybrid Array of Processing Elements) – A method of utilizing new CMOS process nodes for production chips 12-18 months before production availability of that node.
5. **HILT** (Hierarchical Integrated Latch Tree) – A special purpose replacement for SRAM using a hierarchy of latches, providing extremely high data bandwidth at very low power consumption, and slightly less chip area than SRAM (note: this is *not* a general purpose alternative to SRAM).
6. **CREST** (Cyclic REdundant Spare Testing) - A fault-tolerance system that continuously monitors and dynamically replaces faulty computation units without performance degradation.
7. **JETSTREAM** (JET Surface Thermal Regulation via Evaporative Array Manifold) - A two phase immersion cooling (2-PIC) system that incorporates a 3D printed titanium manifold to direct a precise submerged array of liquid 2-PIC coolant jets to microchannels etched in the back surface of each TRIMERA or CPU chip stack.
8. **Silicon Springs** - Integrated microstructures that isolate thermal and mechanical stress by orders of magnitude, preventing stress propagation across the WSSCB silicon substrate.

ZettaLith requires ***no fundamental scientific breakthroughs***, instead leveraging innovative architectures and integration methods to extract significantly higher performance using existing and near-term semiconductor processes. All the necessary silicon processes are already available or slated for production availability by 2026.

[1] ZettaLith.com. Disclosure: The author has filed 38 patent applications related to various aspects of the systems described in this paper.

Instead of distributing inference over many racks consuming hundreds of megawatts, ZettaLith collocates the entire workload within one rack with 1,490× less power consumption.

ZettaLith also scales to desktop systems, allowing exaFLOPS AI inference in a 600 W PCIe card. Some of the principles are portable to edge devices, where they would deliver petaFLOPS of AI inference while consuming only a few Watts and around a square millimeter of an advanced smartphone chip.

### 1.1 Development methodology

The development of this architecture followed a systematic methodology of identifying and addressing fundamental bottlenecks in AI inference scaling. Beginning with a target of 1,000× improvement over current systems, each potential barrier was systematically analyzed and addressed, while iteratively adjusting all other aspects to maintain the target improvement.

### 1.2 Development status

This paper presents a theoretical architectural exploration developed through systematic analysis of scaling barriers rather than experimental iteration. The focus was on identifying and resolving potential 'showstoppers' across all system levels—from computational efficiency to thermal management—before proceeding to implementation. While this approach lacks experimental validation, it ensures internal consistency across the highly interdependent system components.

### 1.3 Architecture status

The architectural innovations presented address not only computational challenges but also numerous engineering constraints that would ordinarily prevent scaling to zettaFLOPS levels. These include:

- Power delivery at 114,000 Amps
- Heat dissipation at 321 W/cm²
- Signal integrity across wafer-scale distances
- Manufacturing yield with 31 billion processing elements

Each solution, while sometimes appearing peripheral to the core computational architecture, represents a critical enabling technology without which the system would fail. This holistic approach distinguishes academic architectural proposals from implementable system designs.

### 1.4 Numerical accuracy

The numerical specifications presented throughout this paper maintain mathematical self-consistency across interdependent engineering calculations. The large number of significant digits in various vales reflects this self-consistency and is not intended to imply comparable physical accuracy. As this work represents a pre-implementation design study, readers should assume a typical uncertainty range of ±20% for critical values. Final implementation specifications would naturally evolve during detailed design phases.

## 2 ZettaLith Transformer Inference Engine

### 2.1 ZettaLith

ZettaLith utilizes a distributed array of 31,407 million processing elements (PEs) organized to keep the entire transformer inference resident in a single all-silicon domain without traversing PCBs, backplanes, cables, racks, or optic fibers.

At ZettaLith's core are the CASCADE (Column-Array Systolic Computation with Accumulation During Execution) architecture, TRIMERA (TRIchip Module for Exascale Reasoning Applications) chip stack and WSSCB (Wafer-Scale Silicon Circuit Board), implementing 156 arrays × 24,576 rows × 8,192 columns matrix multiplications simultaneously. This design fundamentally restructures large-scale matrix multiplications by eliminating inter-chip partial sum transfers.

### 2.2 Optimized PEs

ZettaLith achieves its high performance through highly optimized PEs calculating FP4 weights × activations with FP8 accumulation. Each PE has only 505 transistors designed for TSMC's A16 process node (16 Ångstrom = 1.6 nm). Each CASCADE array of 525,312 PEs operates within its own synchronous 12 GHz clock domain spanning just 0.367 mm², isolated from the surrounding 1.5 GHz system environment.

In conventional distributed transformer inference systems, partial sum transfers dominate interconnect bandwidth consumption, accounting for approximately 50% of all data movement between matrix multiply arrays. This occurs because partial sum transfers scale quadratically with model hidden dimension size, while activation transfers and full sums scale only linearly. In ZettaLith, partial sums are normally completed on the TRIMERA chip stacks and consume no inter-chip data fabric bandwidth.

Accumulation of partial sums within a column is FP8. Biases are also FP8 and are added in the output sums HILT recirculation system at the bottom of each column. Non-matrix operations (SoftMax, swiGLU, etc.) and layer sequencing are microcode state machines, creating a flexible hybrid architecture that maximizes acceleration of the most computation-intensive components while maintaining adaptability.

### 2.3 WSSCB interconnect

ZettaLith's passive wafer-scale silicon circuit board (WSSCB) inverts traditional packaging hierarchy, maintaining all inferencing data and computation within a single all-silicon domain at native silicon speeds. The WSSCB serves as an all-silicon substrate that integrates multiple chiplets into a unified computational domain while eliminating conventional PCBs, interposers, and packages. The WSSCB is completely passive and integrates no active logic.

Integrated silicon spring microstructures (Silverbrook, 2000) reduce thermal and mechanical stress propagation in the WSSCB by orders of magnitude, limiting thermal and stress propagation regions to chip-scale islands less than 2 $cm^2$ .

### 2.4 System reliability

System reliability is enhanced through multiple fault-tolerance mechanisms, including CREST (Cyclic Redundant Spare Testing), which continuously monitors and dynamically replaces faulty CASCADE array columns without service interruption.

### 2.5 Chip-to-chip fabric

ZettaLith's 156 TRIMERA chip stacks and 16 CPU chip stacks communicate via in-silicon 39 TB/s vertical and 11 TB/s horizontal chip-to-chip links using UCIe 2.0 (Universal Chiplet Interconnect Express) pathways. This could provide the 7,800 TB/s inter-chip bandwidth used by 156 TRIMERA stacks, each with 201,326,592 active processing elements, to function cohesively as a transformer inferencing system in a single all-silicon domain.

### 2.6 External connectivity

External connectivity includes 16× PCIe 6.0 channels providing 2 TB/s bandwidth (primarily for SSD access) and 32× 800 GbE connections delivering 25.6 Tb/s for optional multi-rack expansion. A single ZettaLith independently handles transformers up to 20 trillion parameters (for maximum memory version. Standard version handles 5 trillion parameters).

### 2.7 High-precision current regulation

Power is distributed through 86 precision power supply PCBs connected to the WSSCB, featuring 1,032 TLVR (Trans-Inductor Voltage Regulator) modules positioned within 24 mm of their respective silicon loads, with current primarily conducted through solid copper busbars to minimize power loss.

### 2.8 Advanced thermal management

Thermal management is achieved through JET Surface Thermal Regulation via Evaporative Array Manifold (JETSTREAM). The system employs an additively manufactured titanium manifold that directs 172 precision-tuned two-phase immersion coolant jets at silicon heatsink fins deeply etched as microchannels in the back surface of the TRIMERA and CPU chip stacks.

ZettaLith draws 84 kW for computation (97 kW including power conversion overhead from 48V DC to chip-level voltages), delivering more than 10 petaFLOPS/Watt for trillion-parameter model inference.

## 3 ZettaLith Differentiation

A SOTA GPU rack of 2025 is used for comparison of FP4 sparse PFLOPS performance. These performance advantages apply only to FP4 transformer inferencing tasks. Unlike GPUs ZettaLith cannot perform AI training, general HPC workloads, or other compute tasks outside its specialized domain. This highly specialized design is what enables its 1,490× power efficiency for transformer inference.

### 3.1 Eight key differences

Table 1 illustrates how eight key technological differences combine multiplicatively to deliver ZettaLith's performance advantage compared to state-of-the-art GPU-based transformer inference.

The comparison with A SOTA GPU rack spans three critical metrics: raw performance (PFLOPS), power efficiency (PFLOPS/W), and cost effectiveness (PFLOPS/$).

The cost model that forms the basis of cost effectiveness comparisons will be supplied as supplemental data to this paper.

These factors interact non-orthogonally, with improvements in one area often enabling or amplifying improvements in others. The individual contribution values are approximations, adjusted so that their product matches the total system performance differentials, which are calculated directly. The totals are objective calculations, but the subjectivity and interactions of the individual sources of the difference makes those numbers only useful as a "sanity check".

**Table 1: ZettaLith Compared to SOTA GPU rack**

| Source of difference | Performance | PFLOPS/W | PFLOPS/$ |
|---|---|---|---|
| WSSCB | 3.86 × | 3.86 × | 3.86 × |
| JETSTREAM | 2.08 × | 1.28 × | 2.08 × |
| TRIMERA, HILT & SHAPE | 2.94 × | 2.94 × | 2.94 × |
| Optimized PE | 1.92 × | 1.92 × | 2.97 × |
| Specialized for FP4 Inference | 3.12 × | 4.37 × | 4.37 × |
| CASCADE | 3.91 × | 4.75 × | 3.91 × |
| Rightsized HBM | 1.00 × | 1.00 × | 1.51 × |
| TSMC A16 Process | 1.89 × | 2.56 × | 1.29 × |
| **Total comparison** | **1,047 ×** | **1,490 ×** | **2,325 ×** |

### 3.2 ZettaLith eliminates most data center infrastructure for transformer inference

A key aspect of ZettaLith is that all inferencing for transformers up to 20 trillion parameters can be calculated on one integrated silicon structure – a passive WSSCB with 172× modules of HBM and/or HBF and logic chiplet stacks.

This is in comparison to current GPUs, where trillion parameter class transformers must be inferenced using data communications on a hierarchy of levels – from on-silicon buses on GPU chips, to USR connections on silicon interposers, to PCB connections for GPUs on a board, to backplanes connecting modules in a server, to copper cable connections between servers in a rack, to optic fiber connections between racks in a datacenter. Each level of the hierarchy increases system complexity, substantially increases power consumption, and reduces transformer inference performance.

The complex infrastructure of high-speed data switches, CPUs, racks, pods, and complex scheduling software can be largely eliminated if the entire transformer can be inferenced at speed on a single WSSCB. This is the essence of the ZettaLith approach described herein.

### 3.3 ZettaLith integration

ZettaLith achieves its scale, and much of its efficiency, from calculating the entire transformer inference in a single silicon domain, operating at native silicon speeds, power, and component density. This is achieved by 344 advanced chip stacks (172 logic and 172 HBM) attached ZettaLith's wafer-scale silicon circuit board (WSSCB) - a passive silicon substrate analogous to a printed circuit board but fabricated using semiconductor processes. Containing no transistors - only interconnects - the WSSCB supports attachment of chiplets and chip stacks with standard microbumps, replacing conventional PCB, package substrate and silicon interposer functions in a single integrated structure. The passive WSSCB essentially functions as an extremely high performance PCB.

Unlike Cerebras wafer-scale chips, ZettaLith's WSSCB has no active transistors across its area, avoiding the yield and fault-tolerance challenges of wafer-scale active devices. Instead, it leverages TSMC's CoWoS-S process (or equivalent) adapted to integrate the equivalent of silicon interposers, chip packages, PCBs, backplanes, racks, and pods into a single structure. Thermal and mechanical stresses are reduced by orders of magnitude by rows of flexible silicon springs etched through the silicon substrate.

### 3.4 Error rates and CREST fault tolerance

HBMs and UCIe 2.0 have quite extensive error detection and correction. This is essential, as they are used for HPC and financial computing. However, transformer inference is different. Occasional bit-flipping caused by radioactivity or cosmic rays are very unlikely to have noticeable effect on the output of a transformer inference. Cosmic ray-induced bit flips ($\sim 10^{-15}$ errors/bit/day at sea level) are negligible for inference accuracy. Transformers tolerate small activation/weight deviations well (e.g., <0.1% accuracy drop per $10^{-6}$ error rate in GPT-3 simulations). Conventional EDAC (Error Detection and Correction) is not required, but correcting for defective processing elements is essential, as there are 31,407 million fused multiply-accumulate (FMA) processing elements in a ZettaLith.

The ZettaLith TRIMERA stack uses novel cyclic redundant spare testing (CREST) for dynamic array redundancy and runtime error detection. The system leverages spare CASCADE array columns - which are required to improve manufacturing yield - to detect and correct operational faults during run-time inference. CREST can detect and correct many faulty CASCADE columns of PEs at run-time with no degradation in performance. CREST is described in detail below.

While conventional EDAC is not used in the transformer matrix multiplies, the CPU stacks have full EDAC capability. This is necessary, as they run software where a single bit-error can change an instruction and cause a program to crash.

### 3.5 WSSCB fault tolerance

The WSSCB incorporates 100% wire-level redundancy to accommodate as many as tens of thousands of defects across the wafer. To result in an open circuit, wires in the WSSCB RDL require at least two open circuit manufacturing defects on different layers affecting the same wire. Those defects must also be within a few µm of each other to cause an actual open circuit or significantly increased resistance of the connection.

Even in the presence of the unlikely wire defects, the WSSCB can still be used, as almost all the wires have redundancy at the interface level. All the major interfaces of the TRIMERA - the HBM4, UCIe-2.0, include wire level fault tolerance. Even if wire faults exceed this level of fault tolerance, the ZettaLith will only lose one of the 172 SCB module locations, less than 1% of the ZettaLith's performance. With this level of fault tolerance, it is possible that the WSSCB will effectively yield at 100% before chip attachment.

Using this fault tolerance, in combination with silicon springs, ZettaLith is insulated from the common pitfalls of large area active designs, dramatically increasing yield and reducing performance risk.

### 3.6 TSMC A16 or A14 nodes

The main matrix multiply die, the TRIMERA SLD, is intended to be manufactured on the most advanced CMOS node available, to maximize performance within area and power constraints. For the 2027, that is TSMC's A16 node, or, if fully utilizing the SHAPE advantage, TSMC A14 node. If TSMC's A16 or A14 node are not available, the design can be adapted to an older node (e.g., N2, N3, N4) or Samsung or Intel's foundry service with an appropriate performance adjustment. This does not seriously undermine overall commercial viability.

In either scenario, the TRIMERA stacks are intensively tested after bonding, using standard test protocols. This ensures that defective chips stacks are intercepted prior to final integration of KGD on the WSSCB.

As a result, the only silicon device with an area larger than chiplet size is the passive WSSCB substrate, which has a large minimum CD of around 0.5 µm and is highly fault tolerant.

.

## 4 Transformer FP4 Compatibility

ZettaLith trades off flexibility for increased performance. It is optimized for transformer inference in FP4 format and can't run any other format. Transformers must therefore either be converted to FP4 or effectively trained in FP4 using quantization-aware training (QAT), where a model is trained end-to-end under simulated low-precision conditions (Jacob et al., 2017). Various systems have been derived to quantize transformer models after training, including GPTQ (Frantar et al., 2022), ZeroQuant (Ren et al., 2022), and SmoothQuant (Xiao et al., 2022). Transformers are proving to be remarkably resilient to extreme quantization, with good performance being achieved even with ternary weights, where weights can have one of only three values (-1, 0, +1) known as 1.58 bit precision. With FP4 precision, weights can have any of 16 different values.

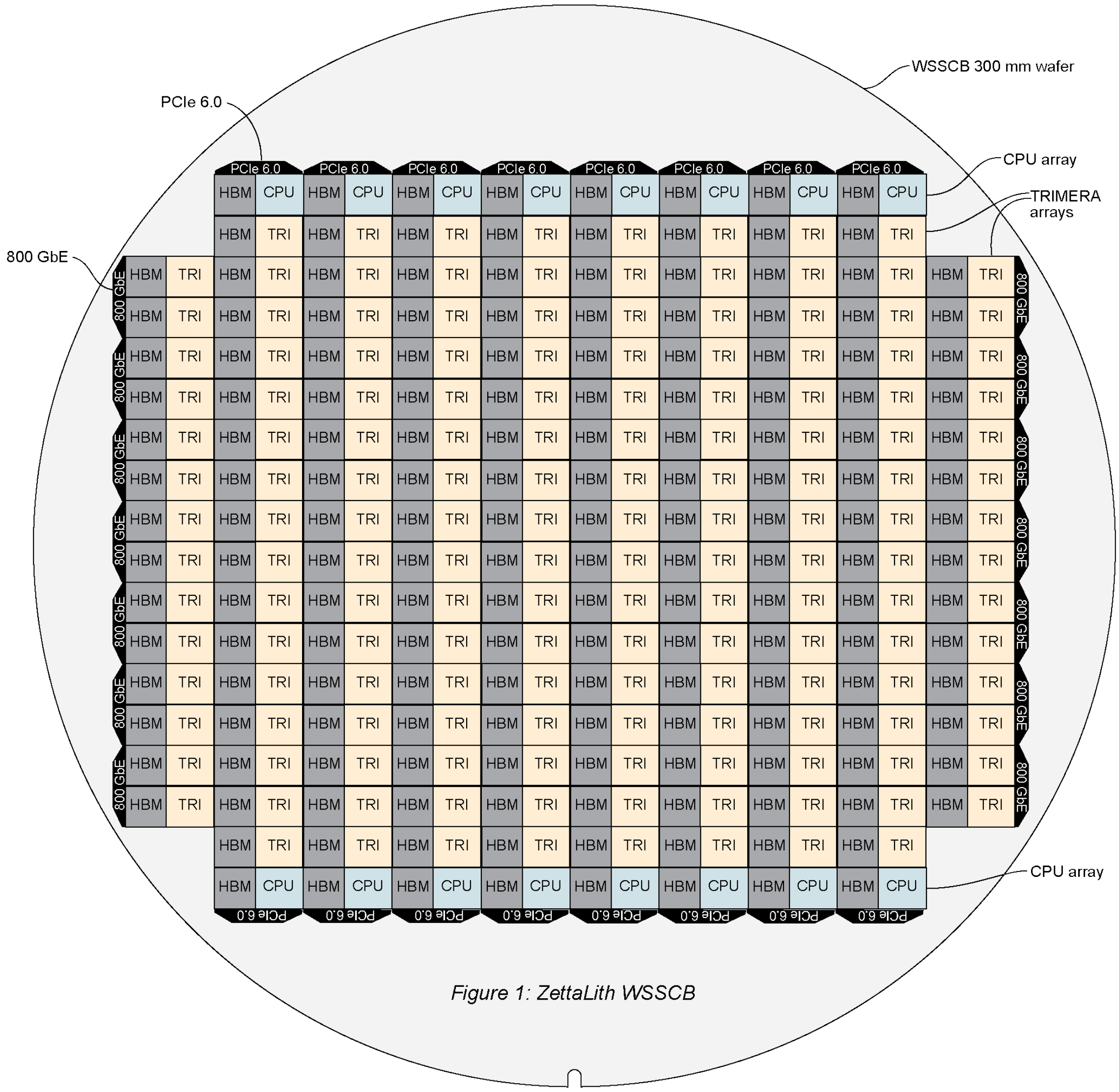

*Figure 1: ZettaLith WSSCB*

## 5 The Wafer-Scale Silicon Circuit Board (WSSCB)

Figure 1 illustrates ZettaLith implementation on a 300 mm silicon wafer-scale silicon circuit board (WSSCB), accommodating an array of SCB modules. The central portion has 156 systolic array compute modules, with 8×1 arrays of CPU modules above and below. TSV connections lead to 800 GbE and PCIe 6.0 PCBs, facilitating high-speed external communication.

This heterogeneous architecture enables a complete large-scale computing system on a single WSSCB, with data fabric connections providing cohesive operation.

WSSCB solves the yield, thermal stress, physical stress, breakage and testing problems with large silicon interposers, and solves the high current power supply problem by integrating many PSU PCBs using column grid array attachment.

The WSSCB has µm-scale routing pitches, mechanical and thermal stress-relief structures, and integrated redundancy for each wire. Consequently, high defect densities can be tolerated with no loss of function. The result is a high-yield, passive and robust large silicon substrate providing the interconnections, power distribution, and mechanical support for a large array of active chiplet stacks.

The WSSCB uses near-full-thickness silicon. This is viable because the TSVs are not used for high speed signals within the array – only for power supply and relatively low speed signals. This, in turn, is because the WSSCB takes the role of silicon-performance PCB, not silicon interposer.

Multiple PCBs are connected to the one silicon substrate, as opposed to multiple chips being attached to one PCB. This makes the silicon thickness irrelevant to high speed signal propagation, keeping all high speed signals contained to the front surface RDL of the WSSCB, the TRIMERA stacks, and the HBM stacks.

### 5.1 WSSCB testing

A WSSCB is a passive silicon device with literally tens of millions of short wire segments connecting pairs of microbumps. It is untestable by conventional semiconductor ATE. This paper describes a simple MEMS probe with tens of thousands of integrated MEMS elastic spring probes that can test an entire WSSCB with 100% coverage in a few minutes. As there are no active components on the WSSCB, the test system is very simple – only testing for wire opens and shorts. No test vectors or complex ATE equipment are required.

### 5.2 Silicon Circuit Boards

Silicon circuit boards (SCBs) enable high system integration through direct silicon-based interconnection. While sharing some characteristics with silicon interposers, SCBs represent a fundamental shift in electronic system architecture, replacing traditional PCBs as the primary integration platform.

Conventional electronic systems employ a hierarchical structure where silicon chips mount to silicon interposers, which mount to package substrates, which in turn mount to PCBs. Silicon interposers provide high-density interconnects between chips but remain limited in size due to manufacturing constraints. The SCB architecture inverts this hierarchy - instead of mounting silicon components to PCBs, the PCBs (primarily for power delivery) mount to a large silicon substrate.

### 5.3 Mechanical stress

Mechanical stress and warpage present challenges in larger silicon structures. The CTE and temperature mismatch between silicon, attached dies, and substrate materials creates stress that scales with distance from the neutral point. This stress can impact both manufacturing yield and long-term reliability of connections.

### 5.4 Thermal expansion

Thermal expansion effects become particularly significant as silicon substrate size increases. The absolute movement from center to edge grows linearly with distance, potentially exceeding the strain limits of conventional interconnect structures. This movement can stress bump interfaces and affect signal integrity across temperature variations. In existing systems, repeated temperature swings can cause elasto-plastic strain in solder joints. The ZettaLith WSSCB is designed to eliminate this problem

### 5.5 Yield

Manufacturing yield has been another key constraint. The probability of defects increases dramatically with substrate area, affecting both RDL processing and TSV formation. This exponential relationship between size and yield has made larger silicon substrates economically impractical using conventional approaches.

### 5.6 SCB solutions

The SCB architecture and manufacturing methods address these fundamental challenges through several innovations, with stress relief structures playing a particularly crucial role. These stress relief structures are MEMS silicon springs fabricated directly in the SCB substrate. The springs include Fermat-Archimedean spiral springs for regions requiring maximum compliance with minimal signal routing, V-beam springs for areas requiring high-density signal routing such as HBM interfaces, and folded beam springs for regions with intermediate signal routing. These, and other spring structures can be used in a single design and can readily be automated as libraries in EDA software. Silicon springs enable ZettaLith's large passive silicon substrates to tolerate thermal gradients and mechanical stresses without cracking, warping, or causing excess elasto-plastic strain of microbump and CGA solder connections.

Other innovations include redundant interconnect schemes, specialized handling techniques, and thermal management approaches that enable practical implementation of large-scale silicon substrates.

## 6 Silicon Springs

Figures 2a to 2k illustrate various stress relief structures integrated into the SCB architecture, which are essential for managing thermal expansion and mechanical stress across large silicon areas while maintaining electrical connectivity.

- Figure 2a presents a 1×4 SCB module array, showing the placement of stress relief structures throughout the SCB. These structures are critical for maintaining mechanical stability and electrical continuity across the multi-module array. The springs limit the thermal expansion and warpage stress zones to one HBM or chiplet stack (e.g. TRIMERA stack) footprint, 1.21 cm$^2$ and 1.43 cm$^2$ respectively. This is considerably smaller than the stress zones encountered by current silicon interposers.

The SCB module includes silicon springs – mechanical structures etched completely through the silicon wafer that provide thermal and mechanical stress relief. This stress relief can isolate sources of thermal and mechanical stress by orders of magnitude, effectively limiting propagated stress to chiplet scale regions of approximately 1 cm$^2$.

### 6.1 V-beam silicon springs for areas with dense wiring

- Figure 2b shows an array of V beam silicon springs

specifically designed for regions requiring high interconnect density in the RDL. The V beam configuration incorporates bent channels etched through silicon while accommodating multiple local interconnects. What appears as black bars in this diagram is 16 parallel wires for each of six RDL layers, so 96 wires for each black bar.

### 6.2 Fermat-Archimedean silicon springs for 3 axis compliance

- Figure 2c provides an enlarged view of a line of Fermat-Archimedean (FA) spiral silicon springs decoupling stress from one portion of an SCB to another. The combination of Fermat spiral geometry with Archimedean spiral arm spacing creates a structure that could provide optimal stress relief while maintaining consistent spacing between adjacent arms. The stiffness of the FA spiral can be tuned over a very large range by changing the width of the spiral arms and the number of turns of the spiral. The thickness of the spiral is the wafer thickness and cannot be altered without adding manufacturing complexity.

The FA spiral combines the properties of Fermat and Archimedean spirals to create a structure that can effectively absorb stress in X, Y, and Z directions simultaneously while maintaining a compact footprint. The Fermat-Archimedean spiral springs are ideal for areas of low or zero density of wiring in the RDL. This FA spiral might be referred to as a double spiral, except the term double spiral is indefinite, as it is commonly used to refer to at least three different structures, of which only the FA spiral version is suitable for this application.

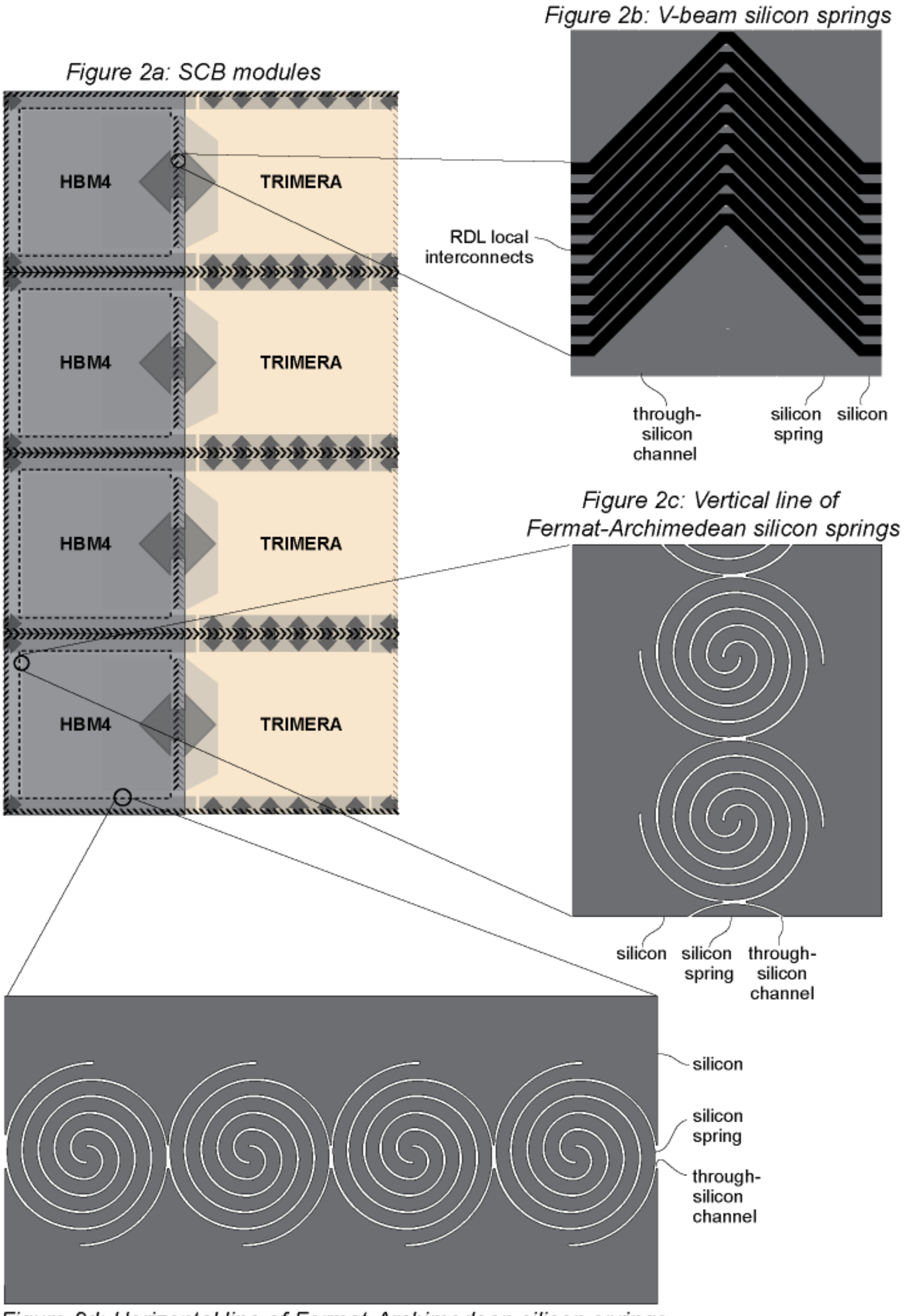

Figure 2d: Horizontal line of Fermat-Archimedean silicon springs

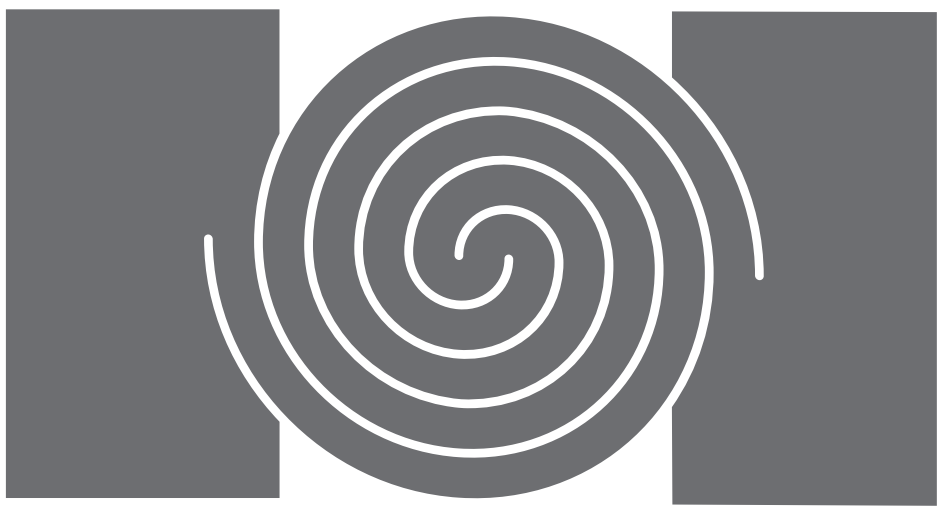

*Figure 2e: unstressed FA spiral*

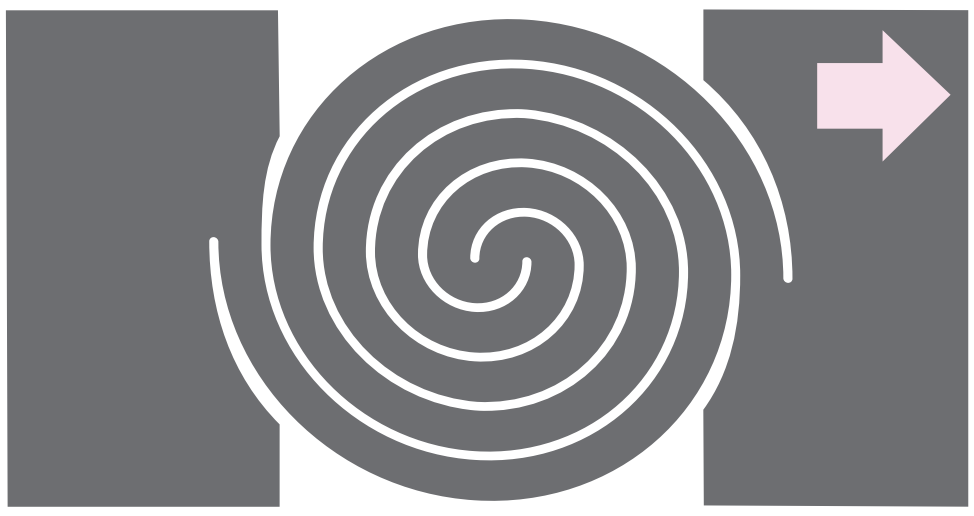

*Figure 2f: tensile stress*

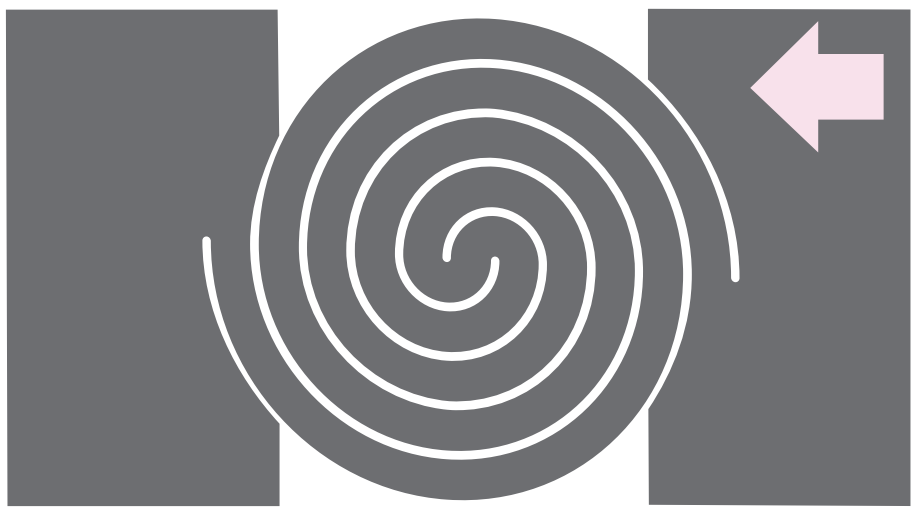

*Figure 2g: compressive stress*

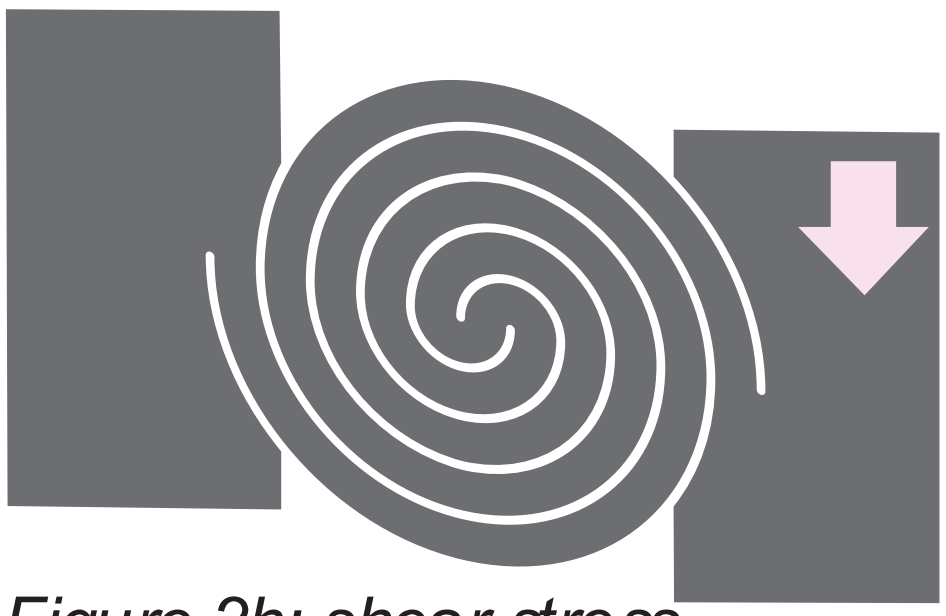

*Figure 2h: shear stress*

- Figure 2d depicts the same FA spiral structure aligned in the X direction, demonstrating how the design can be oriented to surround locations of chips attached to the SCB.

### 6.3 Critical importance of silicon springs to WSSCB viability

These stress relief structures represent a critical innovation in enabling large-scale silicon integration, allowing the SCB to maintain reliable operation despite the significant thermal and mechanical stresses inherent in large scale silicon substrates. FA spirals can reduce mechanical and thermal stress propagation by orders of magnitude compared to solid silicon. The stress propagation may be made arbitrarily low by tuning the FA spiral – the more turns the spiral has, and the thinner the spiral arms, the more compliant the silicon spring becomes.

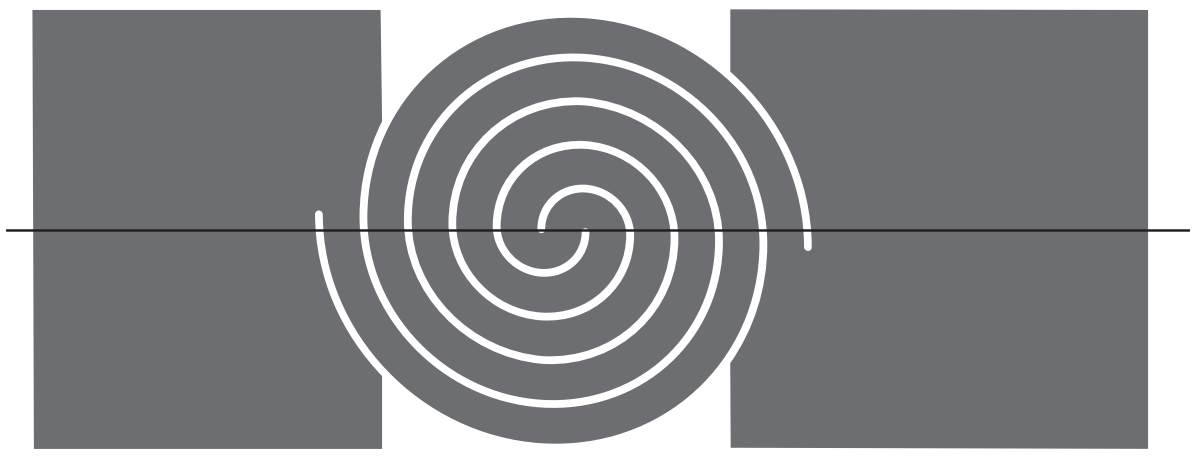

*Figure 2i: line of cross section*

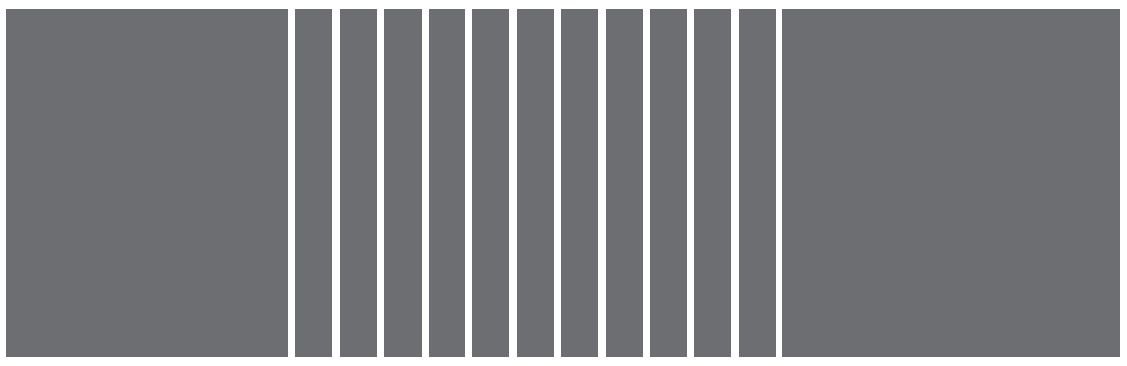

*Figure 2j: cross section of FA spiral*

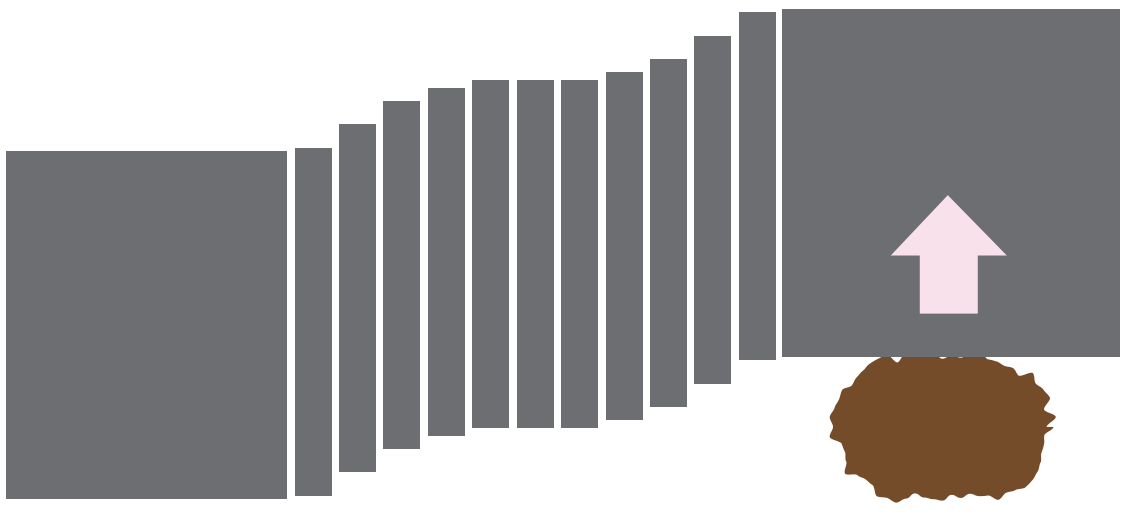

*Figure 2k: Z deflection by 250 µm particle*

- Figure 2e shows the unstressed shape of a FA spiral silicon spring.
- Figure 2f shows strain resulting from tensile stress in the +X direction. Without the FA spirals, this amount of strain would shatter the silicon. With the FA spirals, this strain is well within the elastic limit of the silicon.
- Figure 2g shows strain resulting from compressive stress in the -X direction.
- Figure 2h shows strain resulting from shear stress in the Y direction.
- Figure 2i shows the line where the cross section shown in Figure 2j is taken from. The thickness of the silicon is nominally 710 µm, a relatively arbitrary value less (due to wafer processing of the WSSCB) than the typical 775 µm full silicon wafer thickness.
- Figure 2k shows the WSSCB in the presence of a large particle of around 250 µm between a handling chuck and the WSSCB. The Z deflection enabled by FA spirals is significant, as this can allow elastic deflections of hundreds of µm with minimal stress. This allows the WSSCB to be robust against significant deflections caused by the chip attach or PSU attach processes, as well as particulate contamination in handling when the WSSCB is handled outside a cleanroom. Such a situation would destroy the WSSCB were it not for the presence of the silicon springs.

## 7 Fault tolerance in WSSCB wiring

Figure 3a and 3b illustrate a method for achieving fault tolerance in RDL wiring without increasing the total number of metal layers or significantly impacting electrical characteristics.

Figure 3a shows a conventional four-layer RDL stack with, for example, $n$ µm wide signal lines at $2n$ µm pitch. Metal layer M1 contains wires A, B, C, and D running in one direction, while metal layer M2 contains wire I running orthogonally. Metal layer M3 contains wires E, F, G, and H, with metal layer M4 containing wire J running orthogonally

Figure 3b demonstrates the fault-tolerant configuration using the same four metal layers. Each signal is implemented as a pair of parallel wires of $0.5n$ µm width and $n$ µm pitch on adjacent metal layers, connected periodically by vias. Wires A through H are now arranged in metal layer M1 and M2 at half the original width and pitch, each wire in M1 connected to its counterpart in M2 by a via. Wire I is shown on both metal layers M3 and M4, connected by vias. Wire J has the same configuration as wire I but is not visible as it is located directly behind wire I in the diagram view.

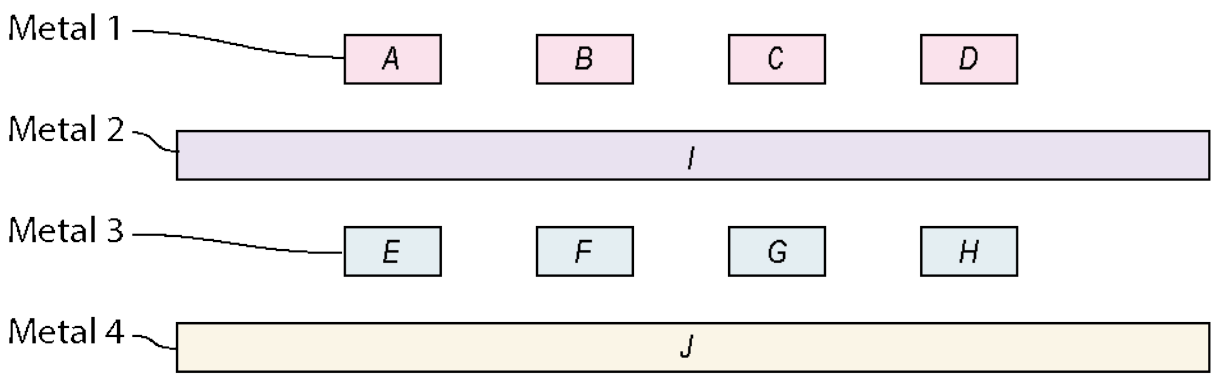

*Figure 3a: Cross section of four layers of wiring*

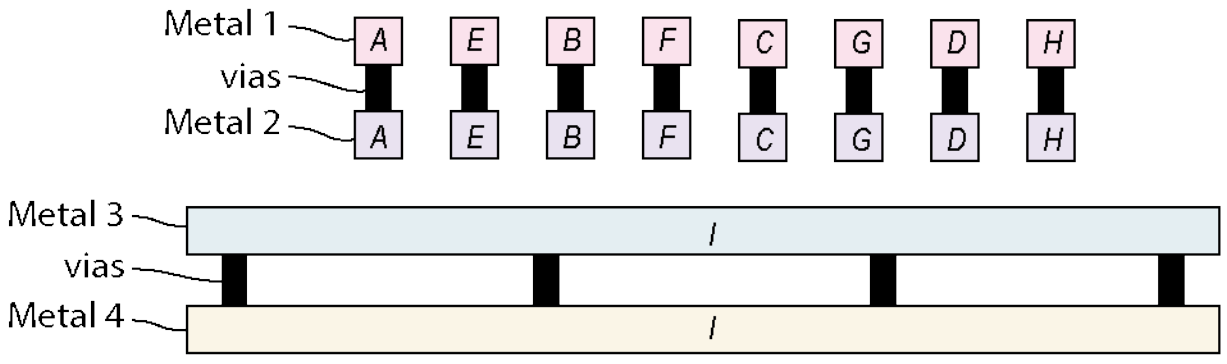

*Figure 3b: Four layers of wiring with fault tolerance*

This redundant configuration could provide:

1. Protection against open-circuit defects with minimum change in resistance, as current can route around defects through the connecting vias;
2. Maintained signal resistance equivalent to the n µm single traces, as the 2 parallel $0.5n$ µm lines provide largely the same total cross-sectional area;
3. No increase in total RDL thickness or layer count; and
4. Compatibility with existing 65 nm CMOS fab equipment.

This system achieves very high fault tolerance allowing high yields even of WSSCBs with millions of wires between microbump landing pads. Assuming short circuits are detected during optical inspection of each layer, and automatically laser ablated, the system is highly tolerant of open circuits. For an open circuit in a layer to cause an actual open circuit in the wire, there must be another open circuit on the matching layer affecting the same wire between the same set of vias. For random defects the chance of this happening is vanishingly remote. The two masks for adjacent layers will be similar, but typically not identical. Even if the masks are identical, the same mask should not be used for the two layers, as a mask defect can provide a correlated open circuit on both layers, causing an actual defect in the SCB.

This approach achieves fault tolerance through geometric reconfiguration rather than through additional process steps or materials. Parasitic capacitance is increased between wires running parallel to each other (potentially 4 times higher due to the combination of halved spacing and doubled layer interaction) but reduced between orthogonal wires (potentially halved). The increase in parasitic capacitance between parallel wires must be considered for high-speed signals.

## 8 WSSCB structure and manufacturing

Figure 4a shows a cross section of a small portion of a WSSCB attached to a TRIMERA stack.

The WSSCB cross section shows an almost full thickness 300 mm silicon wafer of approximately 710 µm thick silicon. The WSSCB wafer contains integrated decoupling capacitors and power/ground TSVs. High speed signals between HBM4 stacks (not shown) and TRIMERA stacks, and between adjacent TRIMERA stacks attached to the WSSCB travel in the ultra-short range (USR) RDL signal wires.

The WSSCB contains silicon springs etched through the wafer at the spring gaps. These silicon springs may be FA spiral silicon springs, V beam silicon springs, folded beam silicon springs, or any other configuration of silicon spring appropriate to the design.

An RDL-silicon indent prevents stress concentrators formed from overhang of the RDL layer into the spring gap, which could potentially cause delamination or crack propagation.

An elastomeric underfill prevents ingress of the liquid coolant into the WSSCB and its attached chips, without interfering with the elastic deformation of the silicon springs. The underfill is not required for thermal or mechanical reasons and is just a precaution against contaminants. It should be eliminated if the manufacturing process and 2-PIC coolant are sufficiently clean.

TRIMERA stack is connected to the WSSCB through microbump copper pillars joined by solder to microbump landing pads of the redistribution layer (RDL), which also contains signal wires and edge seals.

The WSSCB has UBM pads for connecting the CGA pillars of the PSU PCBs, the 800 GbE PCBS, and the PCIe 6.0 PCBs.

### 8.1 Detailed process flow

A detailed manufacturing process flow for a WSSCB has been developed. This is out of scope this this paper but is available in supplementary material.

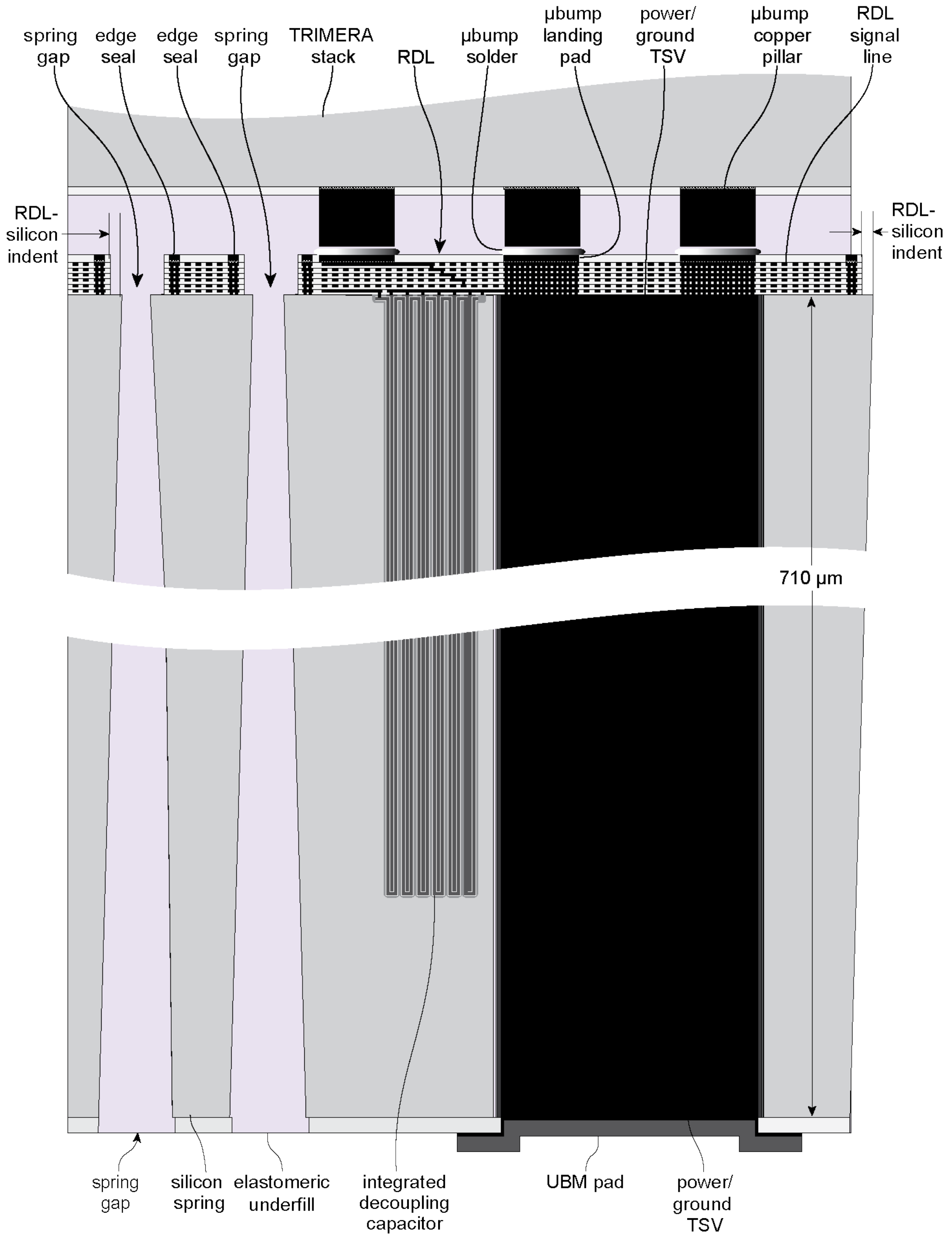

*Figure 4: WSSCB cross section*

# 9 TRIMERA

The ZettaLith TRIMERA stacks are CASCADE arrays of FP4 processing elements. Other systems using ZettaLith construction can use different TRIMERA stacks, such as BitNet b1.58 CASCADE arrays, FP8 CASCADE arrays, HPC stacks or DSP stacks for various applications.

The FP4 TRIMERAs are designed as a Simple Hybrid Array of Processing Elements (SHAPE). They contain edge-to-edge CASCADE arrays of FP4 PEs. This achieves maximum performance and simplicity. The SLD contains 203 million FP4 PEs, each being 505 transistors. There are no bond pads, no TSVs, no SRAM, no analog, and nothing that requires synthesis or standard cells.

All connections to any other circuitry is via hybrid bonding to the HILT die. The SLD can be designed for a new process without waiting for standard cells, SRAM, or analog/mixed-signal qualification, or IP blocks for complex designs such as processors or high speed interfaces. All such circuits are in the mainstream process BID or the HILT dies, which can potentially remain unchanged over multiple generations of SOTA process nodes.

While back-side power is scheduled to be available for the A16 node, this is not used. Power is delivered via hybrid bonding to the front side of the wafer.

The SLD is intentionally very simple and highly repetitive. This is to make it extremely fast to design, and to port to new processes. It also substantially reduces mask calculation time and cost, which is significant at SOTA logic nodes.

## 9.1 Main signal interconnects of TRIMERA

Figure 5 illustrates the fundamental signal interconnect architecture within an SCB module of a WSSCB, showing how high-bandwidth memory (HBM) interfaces, logic processing, and I/O functions are integrated through interconnection paths.

The HBM stack connects to the BID through HBM channels in the WSSCB. The SLD is integrated with the HILT via very high density face-to-face hybrid bonds providing millions of high-density, low-latency vertical connections between the SLD and the HILT. The HILT is integrated with the BID through back-to-back TSV-to-TSV hybrid bonding.

The module maintains connectivity with adjacent SCB modules through UCIe 2.0 connections in the WSSCB between BIDs in all four orthogonal directions: leftward, rightward, topward, and bottomward.

This interconnect architecture would enable the creation of a scalable computing platform where multiple modules can work together cohesively. The combination of high-bandwidth memory interfaces, advanced SLD logic processing, and mainstream BID functions, all connected through high-density on-silicon connections, could provide a balanced architecture that can be replicated across the WSSCB.

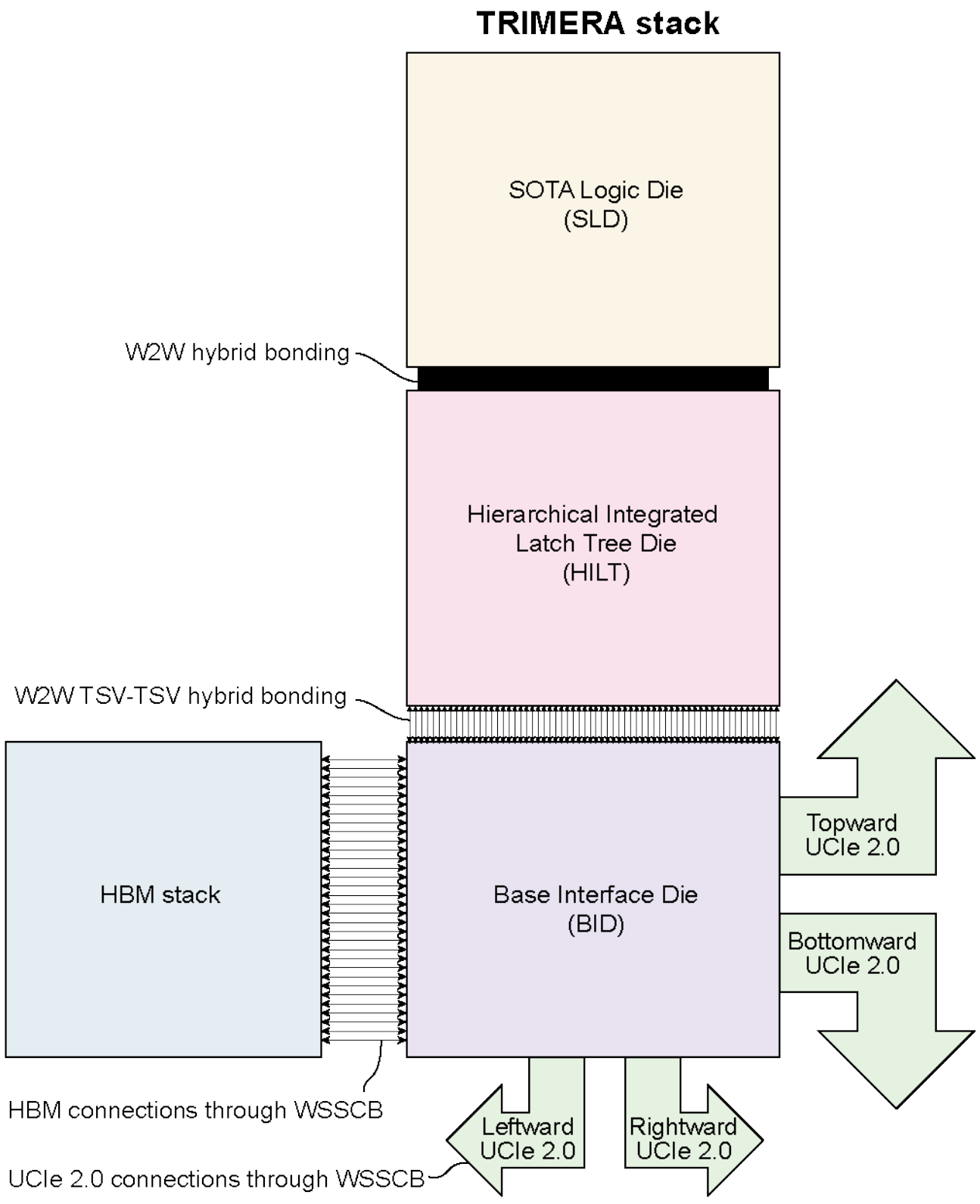

*Figure 5: TRIMERA SCB module*

## 9.2 No New Reticle Stitching

Reticle stitching is a significant annoyance. The WSSCB substrate is fabricated using mature 65 nm DUV lithography. TSMC's established multi-reticle stitching techniques, already proven for large silicon interposers (e.g., in CoWoS-S packaging), resolve any wafer-scale patterning challenges.

The small size of the chiplets in the TRIMERA and CPU stacks do not require reticle stitching. No novel stitching processes are required for ZettaLith.

## 9.3 CASCADE Array Columns

Figure 7a shows an SLD as a grid of CASCADE arrays right to the edge of the SLD die, minus allowance for saw streets and seal rings. Both the SLD and the HILT dies are highly fault tolerant. The BID is in a mainstream process. Wafer level process checks are done using test circuits at the wafer edge and in the center of the wafer.

Once TRIMERA stacks are hybrid bonded, the SLD and HILT can be tested by probing the microbumps on the frontside of the BID, connected by back-to back hybrid bonding of TSVs to the HILT, and from there by front-to-front hybrid bonding to the SLD. The SLD and HILT dies are more difficult to thoroughly test, due to the lack of extensive BIST and JTAG circuitry, which is in the BID.

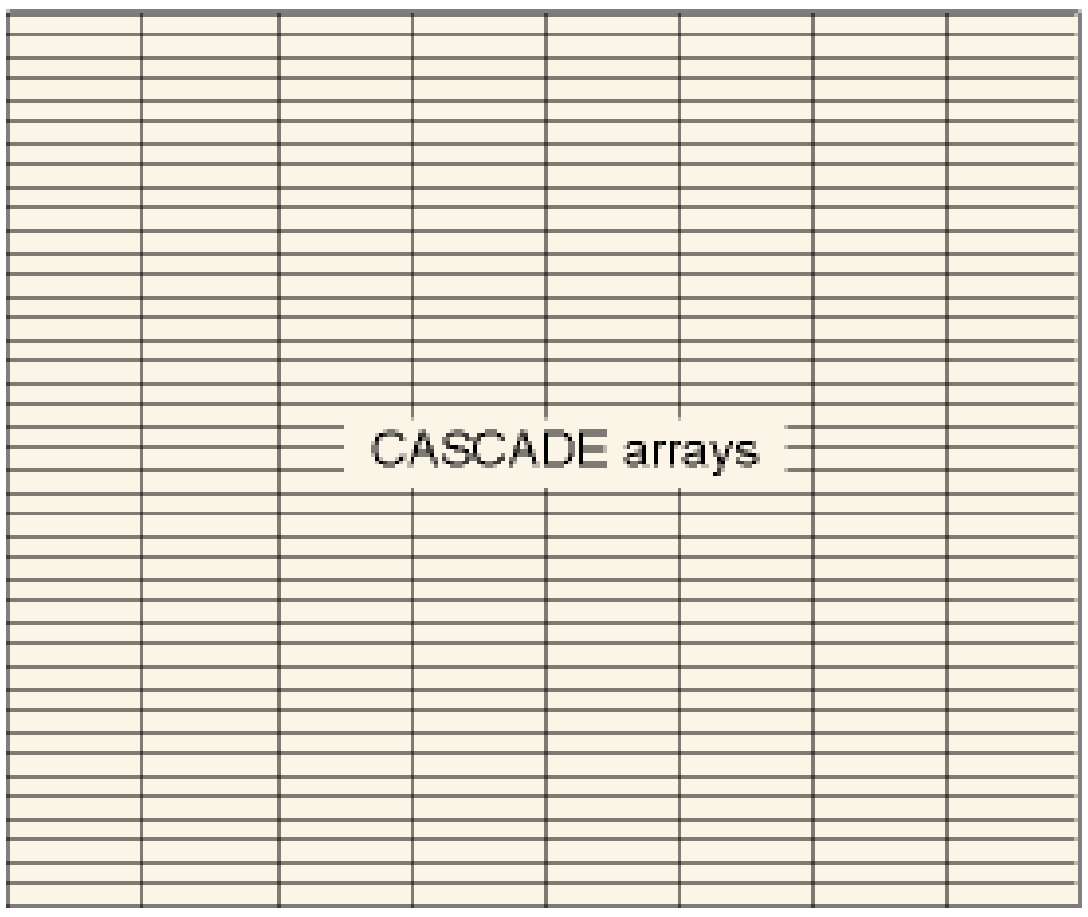

*Figure 6a: SLD die*

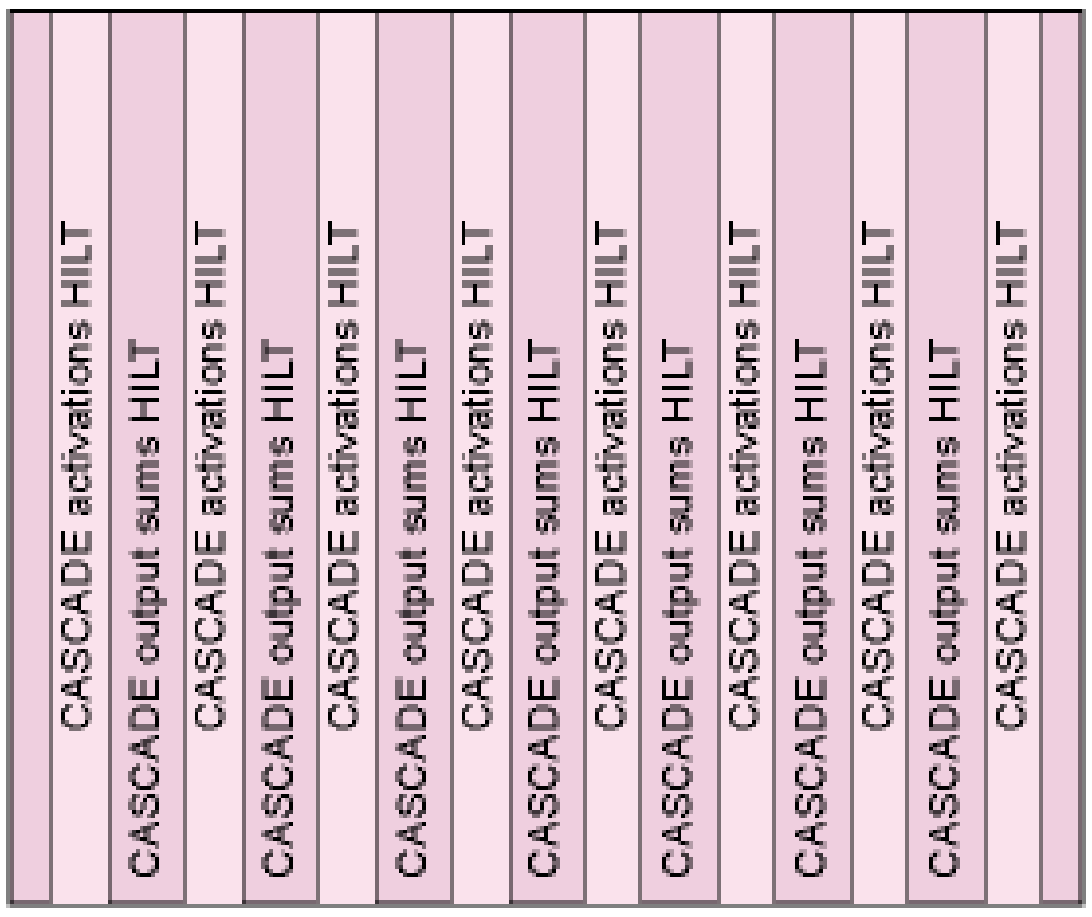

*Figure 6b: HILT die*

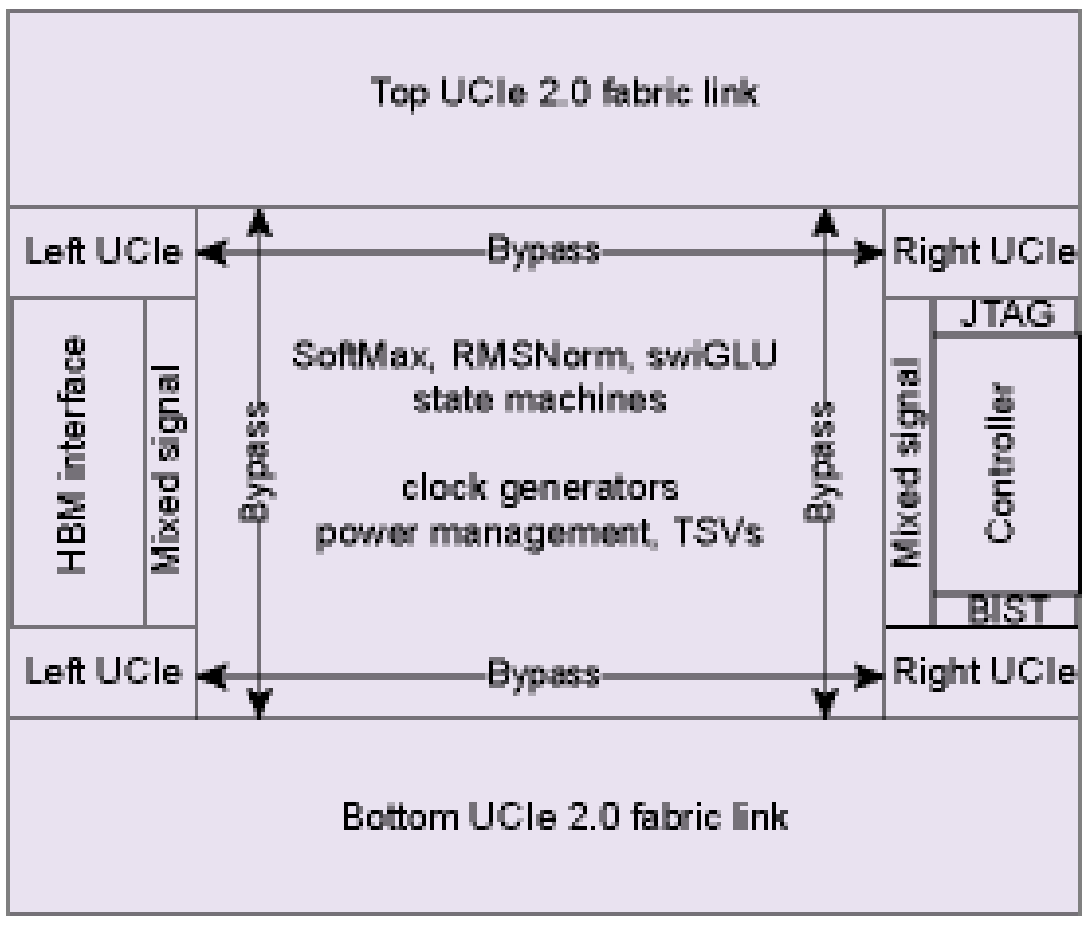

*Figure 6c: BID die*

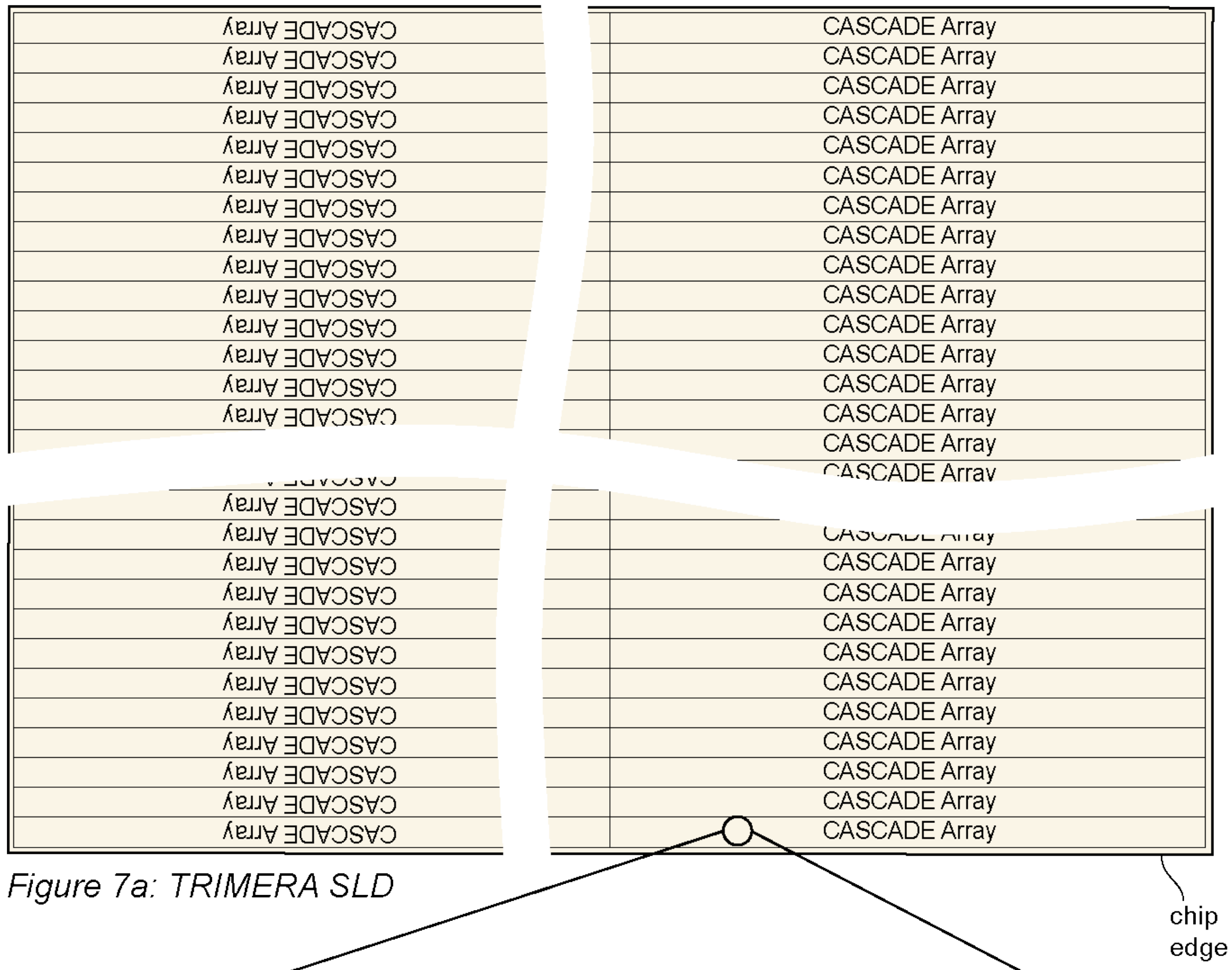

*Figure 7a: TRIMERA SLD*

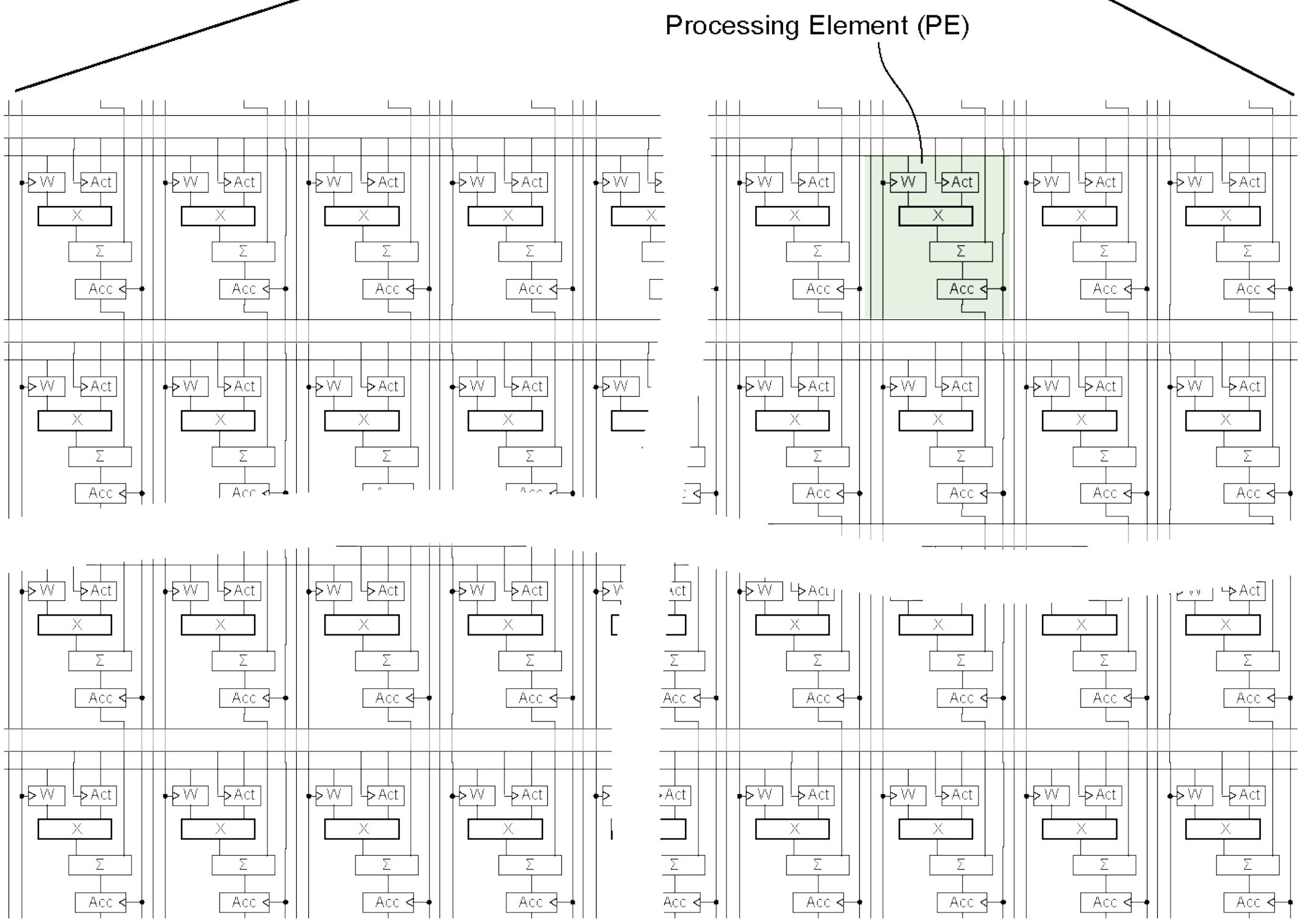

*Figure 7b: CASCADE PEs*

Figure 7b shows part of the CASCADE arrays showing the FP4 PEs.

## 10 The core FP4 processing element

A survey of quantization methods for efficient neural network inference can be found in (Gholami et al., 2021).

The FP4 PE forms the core computational unit of the CASCADE array, replicated 201 million times in the TRIMERA SLD, and 31,407 million times in a WSSCB ZettaLith. Having 32.4 billion active processing elements simultaneously calculating the transformer at 12 GHz is the reason why ZettaLith performance is so high.

The processing element is extremely simple compared to GPU cores or DSP cores, with only 505 transistors per PE. There are no instructions, no branching operations, no cache, and intra-PE wires and inter-PE wires are sub-micron in length.

The TRIMERA SLD contains these FP4 PEs and little else. Even the memory required to feed activations to the CASCADE arrays and collect sums is not in the SLD – it is in the HILT die which is face-to-face hybrid bonded to the SLD.

The SLD is deliberately designed to be as simple as possible, using the SHAPE (Simple Hybrid Array of Processing Elements) system. This dramatically reduces design time and cost, mask-making time and cost, and facilitates early transition to the latest SOTA process. Most of the system complexity is in the BID and HILT, not the SLD.

| Table 2: ZettaLith FP4 PE | | |
|---|---|---|
| **Aspect** | **Value** | **Unit** |
| Total ZettaLith Modules | 172 | modules |
| TRIMERA modules | 156 | TRIMERAs |
| CPU modules | 16 | CPU stacks |
| TRIMERA die area for each of SLD-SRAM-BID | 143 | $mm^2$ |
| Power-limited operational clock frequency | 12.0 | GHz |
| 2-PIC limited max silicon power | 85 | kW |
| Max power available for CASCADE Arrays | 72 | kW |
| PE area | 0.70 | $\mu m^2$ |
| PE power at chosen clock frequency | 2.3 | μW |
| Max PEs that fit in SLD die area | 205 | million PEs |
| Max active PEs within power or area limit | 203 | million PEs |
| Active CASCADE array columns | 8,192 | columns |
| CASCADE rows (PEs in a CASCADE column ) | 64 | rows |
| Active PEs in a CASCADE array | 524,288 | PEs |
| Active CASCADE arrays in TRIMERA | 384 | arrays |
| Active CASCADE matrix rows in TRIMERA | 24,576 | rows |
| Active CASCADE PEs in TRIMERA | 201 | million PEs |
| Percentage spare space on SLD | 0.8% | cushion |
| Performance of 1 PE (1 MAC = 2 Ops) | 24 | GFLOPS |
| SLD performance (sparse) | 9,664 | PFLOPS |
| SLD performance (dense) | 4,832 | PFLOPS |
| SLD CASCADE array power | 458 | W |
| SLD power density | 321 | $W/cm^2$ |
| CPU-TRIMERA data fabric bandwidth | 624 | TB/s |
| WSSCB ZettaLith active PEs | 31,407 | million PEs |
| WSSCB ZettaLith performance (sparse) | 1,507 | exaFLOPS |
| WSSCB ZettaLith performance (dense) | 753 | exaFLOPS |
| WSSCB ZettaLith PE power | 72 | kW |
| WSSCB ZettaLith power | 84 | kW |
| WSSCB ZettaLith current at 1.1V (I/O, CPU, SRAM) | 12 | kA |
| WSSCB ZettaLith current at 0.7V (PEs) | 102 | kA |
| WSSCB total ZettaLith current | 114 | kA |

### 10.1 Structure of CASCADE arrays

Figure 8 shows a block diagram of parts of two adjacent CASCADE arrays of FP4 PEs, each PE having:

- an FP4 weight latch;
- an FP4 activation latch, which is the final stage of the activation latch tree;
- an FP4 × FP4 multiplier, with FP5 approximated result;
- an FP8 plus FP5 saturating adder; and
- an FP8 accumulator register.

### 10.2 Activation HILTs

There is one activations HILT memory for each of the 24,576 rows of the CASCADE arrays on the TRIMERA stacks. The HILT memory takes the place of SRAM, but has far higher bandwidth, smaller bit-cell size, and far lower power. However, the HILT is not a random access memory, but more akin to a large FIFO, but with a tiny fraction of the latches toggling as opposed to a FIFO, where all the latches toggle. The activations HILT memory uses:

- activations HILT stage 1 with 131,072 tri-state latches, each storing one bit of the B × L 4-bit activations. The tri-state latches have 8 transistors each and are approximately comparable to an SRAM bit cell. The tri-state outputs are transmission gates implementing a 16:1 multiplexer;
- activations HILT stage 2 with 8,192 latches with tri-state outputs forming 16:1 multiplexers;
- activations HILT stage 3 with 512 latches with tri-state outputs forming 16:1 multiplexers;
- activations HILT stage 4 with 32 latches with tri-state outputs forming 8:1 multiplexers; and
- activations HILT stage 5 with 4 latches interfacing with the activations broadcast latch tree on the SLD.

The activation broadcast latch tree takes the FP4 output of the activations HILT stage 5 latches and replicates the one activation to be provided simultaneously to all 8,208 columns (including spare/CREST columns) of the cascade array. In the array, this activation is multiplied by 8,192 specific weights and accumulated into 8,192 partial sums. The partial sums flow down the CASCADE arrays until each of 24,576 activations from successive activation HILTs and activation broadcast latch trees has been multiplied by its appropriate weight and accumulated as 8,192 output sums and stored in their appropriate HILTs. The stages of the activations HILT and broadcast latch tree are shown in Table 7.

## 10.3 CASCADE inter-array mechanism with CREST

The CASCADE inter-array mechanism is shown in Figure 8 between the first and second CASCADE array of the TRIMERA stack. Such a mechanism occurs between each of sequential pairs of the 384 CASCADE arrays in the chip stack. The CASCADE inter-array mechanism uses 8,208 copies of each of:

- a previous-array column segment latch;
- a CREST multiplexer. Under CREST software control, this selects either the previous column to the left, the previous direct column, or the previous column to the right to be added to the output of the current direct column. The operation of the CREST mechanism is shown in Figure 10a to Figure 10g;
- a CASCADE array adder, which adds the previous array (after CREST selection) to the current array; and
- a current-array column segment latch. This directly feeds the previous-array column segment latch of the next array, resulting in only the wire delay the length of 64 PE's (the number of rows in a column) between the previous latch and the current latch, which should enable timing closure at 12 GHz. If not, the rows in a CASCADE array can be reduced with a consequent increase in number of CASCADE arrays with little consequence.

## 10.4 Partial Sum Accumulation

The CASCADE array takes FP4 weights and activations and accumulates sums in FP8. Accumulating sums in INT8 is an alternative, but INT8 provides a smaller dynamic range, so it makes it more difficult for the quantized transformer to maintain accuracy.

The ZettaLith FP8 arithmetic is not IEEE 754 compliant, as this is not required for transformer inference, and ZettaLith is not a general purpose GPU.

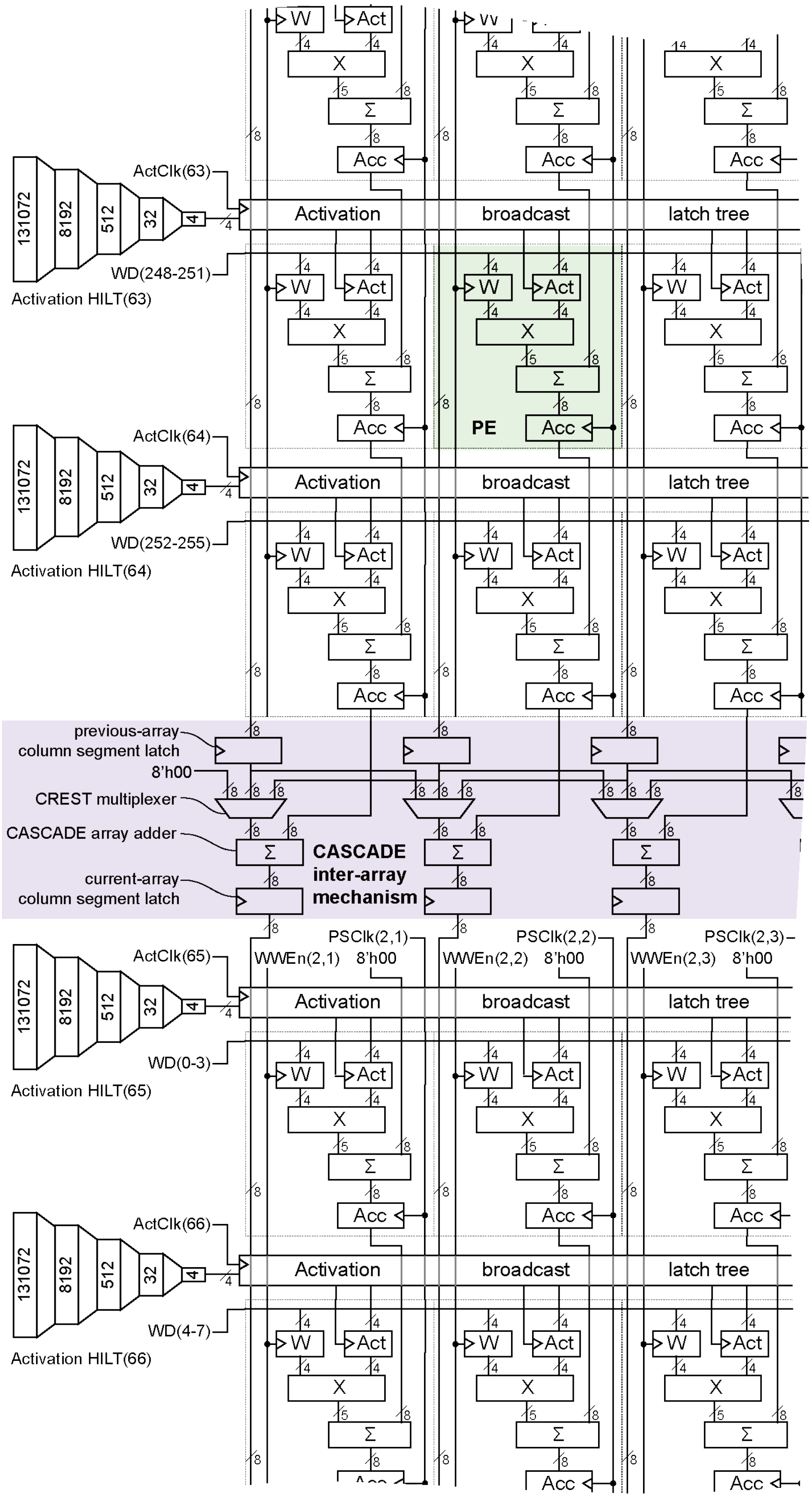

*Figure 8: the bottom of CASCADE array 1 and top of CASCADE array 2*

## 10.5 FP4 PE transistor count estimate

Table 3 shows the transistor counts of two versions of the FP4 PE. The column Full CMOS shows the number of transistors if the PE were to be automatically synthesized as standard cells. The hybrid pass transistor column shows the number of transistors if the PE is extensively and intensively optimized for full-custom implementation.

This version is to be optimized for architecture, GAAFETs, power, performance, area, and simulation efficiency, with the objective of obtaining the best possible PPA. The performance of ZettaLith depends upon whether the final TRIMERA stacks are power or area limited, and whether the PE meets 12 GHz timing closure. As there are 31,407 million copies of the PE in a ZettaLith, this optimization is commercially very valuable.

The transistor count of a single PE in the CASCADE array is estimated in Table 3. The minimum transistor count for functionality is 470 transistors. A cushion of 5% transistors is added to account for extra transistors which may be required to achieve timing closure or reduce power consumption, to make the total of 494 transistors. This is the actual transistors of each PE. However, a further 11 transistors is added as a share of the inter-array CASCADE and CREST mechanism to obtain the effective average of 505 transistors per PE used for the calculations in this document. While not strictly the number of transistors in a PE, using 505 makes the calculations consistent and avoids a long-winded explanation every time it is used, or the introduction of yet-another acronym.

**Table 3: PE transistor count**

| **CMOS style** | **Full CMOS** | **Hybrid Pass** |
|---|---|---|
| **Registers** | **Transistors** | **Transistors** |
| 4-bit weight latch | 48 | 24 |
| 4-bit activation latch | 48 | 24 |
| Proportion of activation broadcast latch tree | 24 | 12 |
| 8-bit partial sum accumulator | 96 | 48 |
| **FP4 x FP4 with FP5 result** | | |
| XOR gate (for sign) | 6 | 4 |
| Exponent adder (simplified 2-bit) | 24 | 14 |
| Mantissa processing (simplified) | 12 | 8 |
| Zero / special case detection | 8 | 6 |
| Result selection MUX | 20 | 12 |
| **FP8 + FP5 adder with FP8 result** | | |
| Sign Extraction and Comparison | 12 | 8 |
| Optimized Exponent Handling | 42 | 32 |
| Mantissa Alignment Shifter | 48 | 30 |
| Mantissa Addition/Subtraction | 80 | 52 |
| Simplified Normalization | 70 | 46 |
| Truncation Logic | 24 | 16 |
| Exponent Adjustment and Overflow | 70 | 48 |
| Saturation Circuit | 70 | 46 |
| Final Result Encoding | 50 | 34 |
| **Clock buffers** | | |
| Weight clock inverter-buffer | 2 | 2 |
| Activation clock inverter-buffer | 2 | 2 |
| Accumulator clock inverter-buffer | 2 | 2 |
| Minimum transistors | **758** | **470** |
| Cushion | 5% | 5% |
| Total transistors in a PE | **796** | **494** |
| CASCADE and CREST mechanism | 1216 | 668 |
| Shared by 64 rows in a CASCADE array | 19 | 11 |
| **Total transistors apportioned to a PE** | **815** | **505** |

## 10.6 PE optimizations

The architectural optimizations and trade-offs include:

- The output of the multiplier is FP5, in E3M1 format, where ideally it should be FP7. The 1 bit mantissa of an FP4 E2M1 number (with implied leading 1) has two states: (1.0, 1.1). The product of two such numbers has four states: (1.0, 1.1, 1.1, 10.01), requiring a 3 bit mantissa. 10.01 is truncated to 10 and normalized to 1.0 (with an increase in exponent), allowing a 1 bit mantissa. The difference is minor and should be inconsequential in transformer calculation.
- The adder is not a full FP8 adder, but is FP8 + FP5, as the only source of the number to be added to the FP8 partial sum is the FP5 output of the multiplier. This saves many transistors.
- The adder is truncated, not rounded. Again, this saves many transistors.
- The 1 bit full adders used are CLRCL, used for its 10T design, high speed and suitability for GAAFET process nodes. CLRCL directly uses pass-transistor structures to convey signals, often resulting in fewer intermediate nodes storing charge. Hence, CLRCL can achieve higher speed, provided the pass-transistor network is optimally sized, and threshold drops are mitigated. This requires careful transistor scaling to ensure clean output levels. Alternative well-known 10T alternative designs include 13A and SERF. Newer full adder circuits specifically designed for GAAFET may emerge, and these should also be considered.
- There is no direct reset of the accumulator. Reset should not be required for normal operation, but if it is required for testing a zero condition can be flowed down the CASCADE column.
- The Activation clock and Accumulation clock are separate, allowing them to be carefully phased to present the multiply result and the partial sum input to the adder simultaneously, almost doubling the effective cycle time.
- The accumulator is a D latch. Timing closure would likely be easier if it were an edge triggered flip flop. However, this would add another 48 transistors to the PE, and therefore significantly reduce performance of ZettaLith, so it should be avoided by extensive optimization of the PE.
- The circuit is specifically designed for 12 GHz operation, instead of "as fast as possible". If timing closure can't be achieved at 12 GHz, the operating frequency can be reduced, or a pipeline register can be added to the PE. These decisions should be made after optimization, layout and SPICE simulation of the PE, using the PDK appropriate to the node chosen.

- Power and ground are directly and independently provided to each CASCADE column of 64 PEs (32,284 transistors) via a hybrid bond pair and metal stack from the power and ground metal planes of the chip, which have on-chip decoupling capacitance. This is to reduce pattern-sensitive ground-bounce. It is also to make the simulation of a single PE highly representative of every PE. This actually improves the simulated timing of the SPICE simulation, as without this high power supply regularity, the SPICE simulation results would need to be derated to accommodate differing power and ground IR droop and inductance variations. With these 3.1 million power stacks and 3.1 million ground stacks, the SPICE simulation of a CASCADE column hard macro can be used without derating it according to its position in the array.
- Connections within PEs are on-chip connections in metal 1 (M1) or metal 2 (M2), typically around 100 nm long. Connections between PEs within an on-chip CASCADE array are also around 100 nm, typically in M1 or M2. Each transistor in the PE should be optimally sized for PPA.
- GAAFET (Gate all around FETs) are assumed. This analysis should be derated if FinFET is used.
- A dataflow architecture with wave pipelining is not used due to simulation complexity and noise sensitivity but can be used to improve clock frequency and power consumption at the expense of more difficult design.

## 10.7 Relevance of a tiny PE

**Table 4: FP4 PE silicon area**

| Aspect | Value | Unit |
|---|---|---|
| TSMC N2 standard cell (SC) density | 313 | $MTr/mm^2$ |
| Projected TSMC A16 standard cell density | 344 | $MTr/mm^2$ |
| Transistors in a PE | 505 | Tr |
| Minimum area | 1.47 | $\mu m^2$ |
| Full custom density improvement over SC | 2.1 | x |
| Optimised full custom area | 0.70 | $\mu m^2$ |
| Total number of PEs in a CASCADE array | 525,312 | PEs |
| Area of a 12 GHz clock domain | 0.367 | $mm^2$ |

This PE is very simple and small and is replicated 201 million times on the TRIMERA SLD chip. It is worthwhile to extensively optimize this small PE for the latest SOTA process for each technology the CASCADE arrays may be ported to.

As the PE and inter-array CASCADE and CREST mechanism are practical to implement as a hand-tuned full-custom designs, the SLD can be implemented very early in the availability of a new SOTA process. It can predate the availability of standard cells, I/O, SRAM, mixed signal SIP as well as through-silicon vias (TSV).

As explained elsewhere, all the hard-to port and complex elements reside in the HILT and BID die. The SLD is therefore simple, being millions of PEs and little else. Hybrid bonding could provide the large number of connections that connect the data storage circuits of the HILT with the calculation arrays of the SLD.

## 10.8 Clock frequency

To run a clock at 12 GHz across an entire wafer is impractical. But this is not what ZettaLith does. The maximum size of a synchronous clock domain in the SLD is 0.367 $mm^2$ , the size of a single CASCADE array. Data transferred between columns of CASCADE arrays is resynchronized using inter-array CASCADE circuits, and the HILT and activation broadcast latch tree circuits. The remainder of the CASCADE array support system runs at 1.5 GHz (one-eighth the CASCADE clock) but can readily be adapted for lower or higher clock rates.

Synchronization between SLD chips is via UCIe 2.0, where each UCIe link has its own clock domain and is also synchronized using FIFOs. Therefore, 0.367 $mm^2$ is the maximum area that the 12 GHz clock skew and jitter is relevant to. This should be readily achieved in the 16 Å node or 14 Å nodes, but this must be determined by post-layout simulation achieving acceptable jitter and skew using the TSMC A16 PDK or A14 PDK.

The phase of the 12 GHz clock can be minutely different for each CASCADE column, to average out the 12 GHz current consumption and essentially eliminate ripple at the clock frequency. With as many as 8,192 independent phases per chip (one per active column) the ripple can be dramatically reduced both locally at the mm scale, and globally across the whole chip. Conveniently, the clock phases can be simply produced by differential gate delays in the activation broadcast latch trees.

The high clock frequency of the ZettaLith CASCADE arrays is made possible because the FP4 PE is very small, has no branching logic, is not programmable, is in a CASCADE array, is heavily optimized, and has a tiny synchronous clock domain. Around 201 million PEs can be incorporated into the CASCADE arrays of the 143 $mm^2$ SLD at 12 GHz.

In a best-case PVT corner (high-speed transistors, nominal or slightly higher supply, low temperature, minimal IR drop), the 2 bit adders, register latching, logic and sub-micron local routing might close timing in ~30–40 ps. This corresponds to a clock frequency of ~25–33 GHz in an ideal scenario.

However, margin above 12 GHz is required to accommodate:

- Process variations (slow corner vs. typical/fast corner)
- Temperature fluctuations
- Power supply variations (e.g., dips below nominal 0.7 V)
- Clock skew
- Clock jitter

These uncertainties can add 50%–100% overhead in real production, meaning that a design which can theoretically run at ~25–33 GHz in best-case conditions is conservatively operated at 12 GHz to ensure yield across all manufactured parts and environmental conditions.

## 10.9 12 GHz clock feasibility

While 12 GHz may appear ambitious compared to conventional CPUs or GPUs that operate at 3-5 GHz, it's important to note fundamental differences in circuit complexity. Modern CPU cores typically contain 100 million to 500 million transistors with complex control paths and branch prediction. In contrast,

CASCADE PEs are tiny, with around a millionth the number of transistors, and execute a fixed multiply-accumulate operation with no branching.

Precedents for operating PEs at or above 12 GHz include:

- 32-bit adders operating at 12 GHz (Agah et al, 2007), and carry-lookahead adders reaching 15 GHz, both in 65 nm CMOS technology. The 15 GHz carry-lookahead adder utilized low-voltage-swing pass-transistor logic, a specialized circuit technique aimed at minimizing delay that is potentially applicable to this PE.
- Baud-rate SerDes transceivers, such as a 12.5 Gb/s design in 65nm CMOS (Harwood et al, 2007) employ digital FFE and DFE blocks whose arithmetic units (including adders) operate at the line rate (12.5 GHz). Cadence's 224G SerDes PHY IP, which involves extremely high-speed DSP, is designed for TSMC's 3nm process node.
- Analog Devices AD9986 RF DAC/ADC explicitly features a 48-bit Coarse Digital Up Converter (CDUC) NCO (phase accumulator/adder) with a maximum clock rate of 12 GHz.
- DDFS MMICs with 9-bit pipelined accumulators operating at clock frequencies around 11.9 GHz to 12.3 GHz. (Yu et al, 2008).

In research environments, examples of 12 GHz PEs extend as far back as 2007, and in CMOS nodes as large as 0.18 µm. As tiny PEs are insignificant fractions of ASICs in advanced CMOS nodes, they are now rarely mentioned in the literature. It is only because ZettaLith has so many of them and relies upon fast tiny PEs as the primary source of high performance, that they are significant in the ZettaLith architecture.

### 10.10 Design and timing closure

Designing digital circuits for operation at 12 GHz requires a holistic approach that extends far beyond standard logic gate implementations. It involves the judicious selection of appropriate logic structures, the strategic application of architectural parallelism and pipelining, highly customized physical layout to mitigate parasitic effects, and the design of robust, high-performance clocking networks. These specialized techniques are essential to harness the speed potential of advanced semiconductor transistors and to overcome the numerous physical challenges encountered at such high frequencies.

To be commercially viable, the complexity of the circuit required must be low enough so that the effort is worth the return. The effort to design a CPU core with 200 million+ transistors is not commercially viable, while a ZettaLith PE with 505 transistors is.

To establish timing closure at such a high clock frequency, it is necessary to design, lay-out, optimize, and simulate the PE using the PDK for the target process (TSMC A16 or TSMC A14 in this case, but any process can be targeted with appropriate change in PE PPA). Several iterations and refinements will be required. As the PE is the fundamental determiner of ZettaLith performance it is worth assigning a crack team of full-custom engineers a year or so to complete this task.

The isolated 0.367 mm$^2$ synchronous domains ensure that clock skew minimization and jitter control remain manageable engineering challenges rather than fundamental physical limitations.

If, despite this, a 12 GHz clock cannot be achieved, one fallback is to simply reduce the clock frequency. This has the disadvantage of proportionally reducing ZettaLith performance but the advantage of also reducing power consumption. A 6 GHz ZettaLith with 0.75 zettaFLOPS of sparse FP4 inference at 48 kW compute power and a maximum power density of 160 W/cm$^2$ has far lower execution risk.

Another fallback is to use dataflow and wave pipelining. A CASCADE column of FP4 PEs is highly suited to a dataflow architecture using wave pipelining. However, dataflow architectures and wave pipelining are more complex, and simulation tools are not well adapted to them. The entire CASCADE column would need to be simulated at the SPICE level, instead of just a single PE. As a dataflow architecture is unlikely to be required it is not currently recommended or further addressed in this document. However, a small parallel investigation could be warranted, as it is possible that a substantial improvement in performance is available with this approach.

**Table 5: FP4 PE Power Consumption**

| Aspect | Value | Unit |
|---|---|---|
| Transistors in a PE | 505 | Tr |
| Gate capacitance per transistor | 0.06 | fF |
| Total gate capacitance | 30 | fF |
| Parasitic capacitance of 100 nm M1 | 0.02 | fF |
| Total local interconnect | 10 | fF |
| Clock driving overhead | 6 | fF |
| Total capacitance of a PE - standard cell | 46 | fF |
| Full custom optimization factor | 2.2 | x |
| Total capacitance of a PE - full custom | 21 | fF |
| Operating voltage | 0.7 | V |
| Operating frequency | 12.0 | GHz |
| Baseline activity factor | 0.10 | α |
| Sparsity after Top-K sparsification | 90% | |
| Zero weight activity factor | 0.04 | α |
| Average activity factor | 0.046 | α |
| Peak matrix multiply use | 75.3% | |
| Power of a PE in TSMC N3E | 4.3 | µW |
| Scaling to TSMC A16 | 53% | |
| Power of a PE in TSMC A16 | 2.3 | µW |

### 10.11 Power dissipation limited clock frequency

The power dissipation of the SLD chip is 458 Watts, with a power density of 321 W/cm$^2$, requiring JETSTREAM cooling.

Power supply IR variations across the chip are minimized by direct metal stacks to each CASCADE column from the power and ground planes of the chip. All chips are supplied with optimized 2-PIC cooling jets irrespective of where they are on the WSSCB, due to the JETSTREAM manifold.

Chips which don't meet 12 GHz can be binned for use in ZettaLiths that operate at lower clock speeds.

While the PE is initially intended for 12 GHz operation, the system is power dissipation limited and can potentially operate at higher clock speeds as faster transistors become available without increasing power dissipation in subsequent CMOS generations.

The silicon area of single PE and an entire CASCADE array is estimated in Table 4. The transistor density that TSMC gives for a process is for high density standard cell. Optimized full custom of a small repeating cell can achieve substantially higher transistor densities.

### 10.12 FP4 PE power consumption estimate

The power consumption of a single PE in the CASCADE array is estimated in Table 5. In digital CMOS circuits, power consumption is dominated by dynamic switching power. This is governed by the equation $P = \alpha CV^2 f$, where $\alpha$ represents the switching activity factor, $C$ is the node capacitance, $V$ is the supply voltage, and $f$ is the operating frequency.

### 10.13 Sparsity

Sparsity in AI transformers refers to the strategic design of network architectures that selectively activates a limited subset of parameters or connections during processing, thereby reducing computational and memory demands while maintaining or improving overall model performance. (Fuad et al., 2023) provides a survey on sparsity explorations in transformer-based accelerators.

The percentage zero weights used in Table 5 is the worst case of the typical 90%-95% range of sparsity after Top-K sparsification of quantized transformers. ZettaLith hardware automatically uses the natural arbitrary sparsity of a quantized transformer or Top-K sparsified transformer to reduce power, but not to increase performance. The zero weight calculation takes the same time as any other weight.

Using sparsity (e.g. by re-organizing weights and activations to create blocks of zero weights, by MoE and other higher level means of skipping large parts of a transformer calculation) can also be used to effectively increase inference speed and reduce inference power. These optimizations are implemented at the high level configuration of the transformer inference, not at the PE level, and do not affect PE design.

## 11 SHAPE: Simple Hybrid Array of Processing Elements

SHAPE represents a novel processing architecture wherein an ultra-dense extremely regular array of PEs operating at a high clock frequencies in a logic die is synchronized, managed, and interfaced via a hybrid bonded memory and control die. While the SLD operates at 12 GHz, the HILT operates at 1.5 GHz and the Base Interface Die (BID) operates asynchronously at normal CMOS clock frequencies. The BID is used for all standard circuits including complex logic, I/O, analog, and mixed signal circuits. The BID is intended to be re-usable across designs – e.g. the CPU stacks should be able to use identical BIDs.

The HILT die is produced using a CMOS process optimal for low leakage high density logic, mostly operating at 1.5 GHz (one eighth the SLD clock frequency). Millions of fine-pitch hybrid bonded interconnects directly couple the SLD CASCADE arrays to the HILT die. This would enable low-latency delivery of activation data to the CASCADE arrays, and collection of complete output sums data from the arrays. The HILT die also could provide essential functions such as clock distribution, signal conditioning, power management, and temperature sensing.

The BID hosts all the peripheral logic and complex control circuitry required to drive the TRIMERA stack arrays. The BID also could provide essential functions such as clock distribution, signal conditioning, power management, and high-speed I/O, offloading all complex digital operations from the SLD.

This separation of functions could provide multiple benefits beyond pure area efficiency. The mainstream process node of the BID is inherently better suited for analog and mixed-signal circuits, offering superior power efficiency, better noise characteristics and lower leakage for I/O functions. Similarly, cells in mainstream nodes benefit from years of optimization for density and reliability, while avoiding the increasing complexity of SRAM implementation in advanced nodes. Through-silicon vias (TSV) are also confined to the mainstream process BID and the HILT die, where they don't consume valuable SLD real estate, and don't complicate or delay the SLD manufacturing process.

### 11.1 SHAPE is intended to enable early TTM

The SHAPE system achieves TTM advantages through its TRIMERA architecture using hybrid bonding. While conventional integrated circuits - even those using advanced packaging techniques - require extensive qualification of complex components such as PLLs, SRAM arrays, standard cell libraries, EDA toolchains, and I/O and ESD structures, SHAPE strategically eliminates these dependencies to enable design and production of chips in advanced nodes before these are available for regular production.

### 11.2 Production before standard cell libraries are available

Traditional semiconductor designs follow digital design flows that require mature standard cell libraries and associated synthesis capabilities - components typically unavailable until 9-12 months after a new process node is defined. SHAPE circumvents this constraint by employing a radically simplified SLD design consisting almost exclusively of highly replicated, minimalist processing elements (PEs). These PEs are deliberately architected to be sufficiently simple for manual design by experienced circuit engineers, eliminating dependencies on automated synthesis and standard cell libraries while still leveraging the performance benefits of cutting-edge process technology.

SHAPE's multi-die architecture could provide another critical advantage: the BID and HILT are implemented in in production, well-characterized process nodes with established design tools and IP blocks. This approach allows the BID and HILT development and validation to proceed in parallel with - and be completed ahead of - the SLD's availability. When the advanced process node becomes production-ready, only the SLD requires fabrication using the new technology, while the fully-validated BID and HILT designs can already be production-ready.

### 11.3 Production before SRAM, PLL, I/O, analog, mixed signal, bond pads, TSVs, and IP blocks are available

By strategically partitioning functionality between the dies, SHAPE eliminates the need to implement and qualify complex components in the advanced node: high-precision PLLs, I/O structures, SRAM arrays, analog/mixed-signal circuits, bond pads, and TSVs. These components typically require multiple design iterations and extensive characterization in any new process node, often becoming critical path elements for commercial deployment.

### 11.4 Reduced design and verification cycles

The simplified SLD design dramatically reduces design and verification cycles. Rather than synthesizing and validating millions of unique logic paths across a complex SoC, engineers need only optimize a single PE containing a few hundred transistors, replicate it across the die, and add a small amount of full custom inter-array logic. This focused approach accelerates time-to-silicon compared to conventional flows, with verification complexity reduced by several orders of magnitude.

### 11.5 Reduced mask calculation

Further time savings occur during mask preparation. For leading-edge nodes (such as TSMC A16) employing EUV lithography with double patterning, mask set generation represents one of the most computationally intensive and iterative aspects of tape-out, typically requiring 2-3 months from initial data preparation to production-ready masks. The highly regular, replicated structure of the SLD significantly reduces computational complexity for optical proximity correction (OPC), verification, and hotspot detection compared to conventional designs with diverse structures and varying pattern densities across the die.

### 11.6 Combined TTM advantage

These combined advantages enable SHAPE designs to commence high-volume production immediately when a new process node reaches initial production capability, providing a TTM advantage of 12-18 months compared to conventional design approaches.

The economic model further supports early adoption: since ZettaLith system costs are dominated by HBM memory and packaging, the incremental cost impact of lower initial SLD yields is manageable. ZettaLith's architecture requires only approximately 40% of yielded die area from a full wafer, making it economically viable to commence SLD production during the early stages of process maturity. This approach would enable deployment of leading-edge silicon technology substantially ahead of competitors using conventional design methodologies.

SHAPE can reduce TTM substantially compared to a SoC. For production in 2027, SHAPE makes the use of TSMC A14 viable, even though TSMC A14 is not scheduled for production until 2028. SHAPE can potentially utilize TSMC's A14 node a year or two ahead of its volume production schedule.

### 11.7 Compatibility of a pre-designed BID with a new SLD

The only specific design requirement imposed by SHAPE on the SLD die is the external connections of the CASCADE arrays, and the *exact* (x,y) tiling pitch of the arrays. Provided that the CASCADE array circuit interface and tiling dimensions are maintained, variations in the PEs circuit or layout between the already finalized HILT and a new pre-production SOTA process can be accommodated by the metal wiring within the unit cell of the SLD.

In contrast, even a tiny deviation in tiling pitch will accumulate across the array, leading to cumulative wiring skew between SLD unit cells that would make the wiring of each cell different, thereby invalidating the SPICE simulation of a unit cell, and invalidate a hard-macro repetition of the cell across the chip.

If the new SOTA process is used to reduce power and increase speed at the same area, then the TRIMERA array can take full advantage of a next generation CMOS process extremely early, without redesigning the HILT die or the BID.

# 12 Systolic Arrays

The concept of systolic arrays was introduced by H. T. Kung and C. E. Leiserson in 1978. Their seminal work (Kung et al., 1978) was the first to describe systolic architectures for VLSI – an array of simple processing elements that rhythmically compute and pass data to neighbors. This laid the foundation for using systolic arrays as a cost-effective high-performance design for specialized computations in hardware.

## 12.1 Systolic Arrays in AI and Transformer Inference

Decades later, systolic arrays became vital in AI accelerators. A prime example is Google's Tensor Processing Unit (TPU). The first-generation TPU (Jouppi et al., 2017) was built for neural network inference and featured a 256×256 systolic array of 8-bit multipliers (65,536 MACs) as its heart. This matrix-multiply unit achieved ~92 TeraOps/s and demonstrated the advantage of systolic dataflow for deep learning workloads. The TPU's success – providing better latency and energy-efficiency for DNN inference than general CPUs/GPUs – was a seminal deployment of systolic arrays in AI hardware.

Given the rapid development of Transformer-focused hardware, comprehensive reviews have emerged. (Kachris, 2025) provides a recent survey of hardware accelerators for LLM transformers, with an emphasis on systolic-array-based designs and other specialized architectures.

## 12.2 ZettaLith: Very Large Arrays

ZettaLith extends the performance advantages of systolic arrays through:

1. Specialization for FP4 multiply with FP8 accumulation;
2. SHAPE ultra-dense simple PEs;
3. CASCADE column-oriented architecture;
4. TRIMERA chip stack optimization;
5. CREST fault tolerance; and
6. WSSCB integration.

ZettaLith implements 156 TRIMERA chip-stacks each with 384 CASCADE arrays of 524,288 PEs for a total of 31,406,948,352 simultaneously operating PEs in an all-silicon domain.

## 13 CASCADE

ZettaLith implements CASCADE (Column-Array Systolic Computation with Accumulation During Execution) for matrix multiplication through a large scale column-oriented array architecture. This approach differs significantly from traditional systolic array implementations, optimizing for on-chip computation without inter-chip partial sum transfers, while enabling the CREST real-time redundancy system.

Though organizationally distinct, the design maintains mathematical equivalence to conventional systolic multiplication while eliminating partial sum transfers and activation fill skew and while offering superior fault tolerance for large arrays.

### 13.1 Final summation of CASCADE arrays

Figure 9 shows a block diagram of the end of the CASCADE arrays. The last two rows of the 24,576 rows of the CASCADE arrays are shown for context. The previous array column segment latches, CREST multiplexers, CASCADE array adders, and current array column segment latches of the last CASCADE array are also shown.

There is one output sum HILT memory for each of the 8,208 (8,192 plus 16 spares) columns of the CASCADE arrays on the TRIMERA stacks. Each output sum HILT memory uses:

- output sum HILT stage 1 with 262,144 tri-state latches, each storing one bit of the B × L 8-bit output sum;
- output sum HILT stage 2 with 16,384 latches with tri-state outputs forming 16:1 multiplexers;
- output sum HILT stage 3 with 1,024 latches with tri-state outputs forming 16:1 multiplexers;
- output sum HILT stage 4 with 64 latches with tri-state outputs forming 8:1 multiplexers; and
- output sum HILT stage 5 with 8 latches interfacing with the recirculating sum mechanism on the SLD.

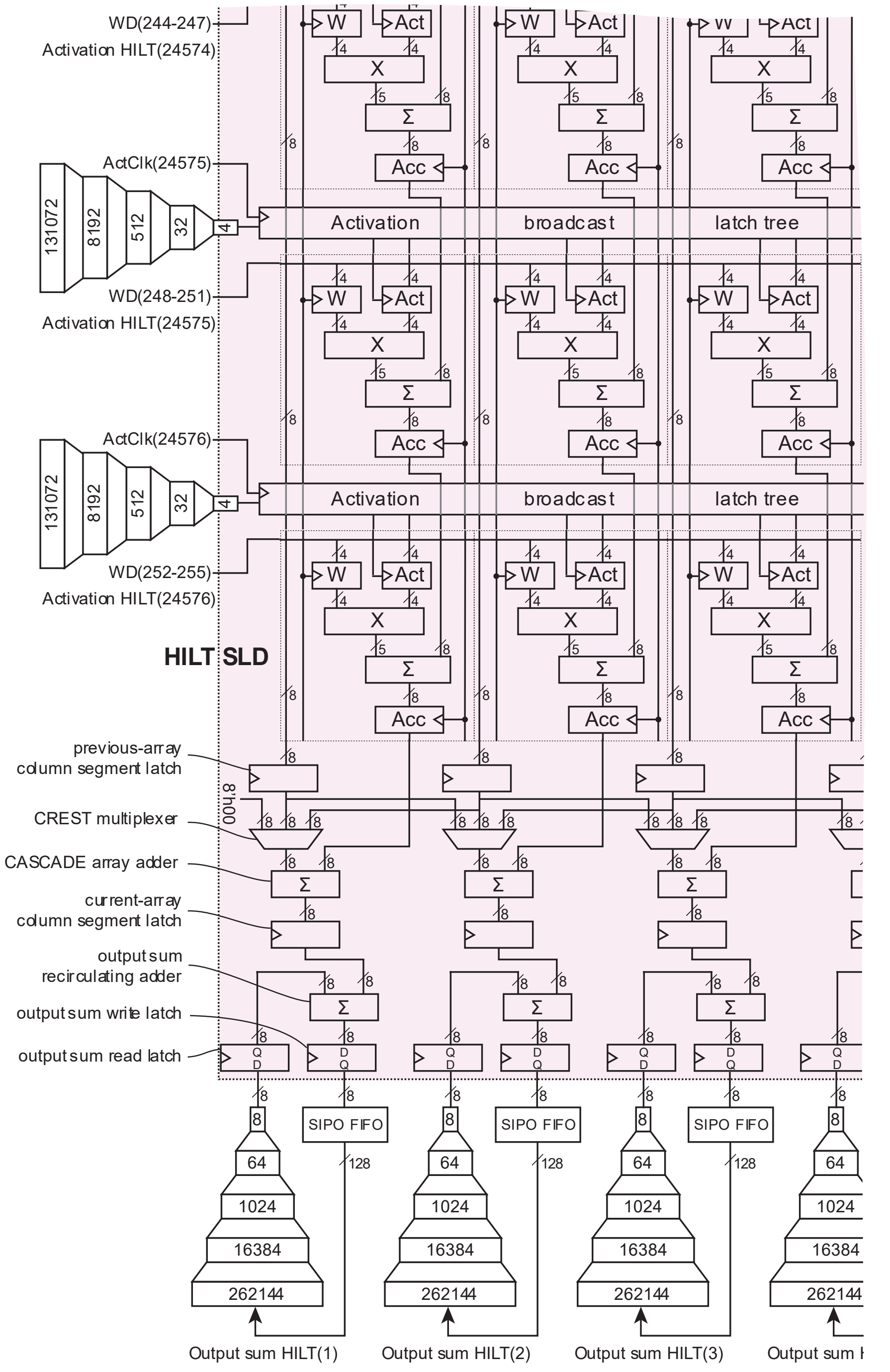

*Figure 9 The bottom of CASCADE array 384 and some of the 8,192 output sum memories*

The final adder stage adds the results of the CASCADE calculations for the columns to the existing contents of the output sum HILT memories. If the CASCADE calculation is the first pass of a transformer matrix multiply involving biases, then the biases for the batches can be loaded into the output sum HILTs and these will be automatically added to the final sum. On subsequent passes, the sums for each batch are accumulated in the output sum HILTs. The output sum accumulation mechanism reads the output sum HILTs as described above, and:

- latching the stored value in the output sum read latch;
- adding the current CASCADE column sum using the output sum recirculating adder;
- latching the result in output sum write latch; and
- converting the calculation frequency from the SLD frequency to the HILT frequency using the output sum write SIPO FIFO.

For consistency with the remainder of the SLDS, the recirculating sum mechanism can be moved from the SLD to the HILT. If this is done, the older process of the HILT should be considered, and the speed of the mechanism should be reduced with a concomitant increase in parallelism. This is straightforward and a reduction of speed and increase of parallelism has the advantage of reducing the final stages of the output sum HILT and the FIFO.

## 13.2 CASCADE step-by-step computational process

Table 6: CASCADE calculations of a single wave on a TRIMERA. TRIMERA runs 32768 waves simultaneously, offset by one 12 GHz clock.

| | CASCADE array 1 | | | CASCADE array 2 | | | • • • | CASCADE array 384 | | |
|---|---|---|---|---|---|---|---|---|---|---|
| Clock | Column 1 | • • • | Column 8192 | Column 1 | • • • | Column 8192 | • • • | Column 1 | • • • | Column 8192 |
| 1-17 | Load batch 1-8 activations 1 from HILT | | | Load batch 1-8 activations 65 from HILT | | | • • • | | • | |
| 18-24 | Broadcast activations | | | Broadcast activations | | | • • • | | • | |
| 25 | A(1)W(1,1) + | • • • | A(1)W(1,8192) + | A(65)W(65,1) + | • • • | A(65)W(65,8192) + | • • • | | • | |
| 26 | A(2)W(2,1) + | • • • | A(2)W(2,8192) + | A(66)W(66,1) + | • • • | A(66)W(66,8192) + | • • • | | • | |
| 27 | A(3)W(3,1) + | • • • | A(3)W(3,8192) + | A(67)W(67,1) + | • • • | A(67)W(67,8192) + | • • • | | • | |
| • | • | | • | • | | • | • • • | | • | |
| • | • | | • | • | | • | • • • | | • | |
| • | • | | • | • | | • | • • • | | • | |
| 88 | A(64)W(64,1) + | • • • | A(64)W(64,8192) + | A(128)W(128,1) | • • • | A(128)W(128,8192) | • • • | | • | |
| 89 | ► | ►►► | ► | Σ(A(1)W(1,1) to | • • • | Σ(A(1)W(1,8192) to | • • • | | • | |
| 90 | | | | ► | ►►► | ► | • • • | | • | |
| • | | | | | | | • • • | | • | |
| • | | | | | | | • • • | | • | |
| • | | | | | | | • • • | | • | |
| 383-400 | | | | | | | • • • | Load batch 1-8 activations 24513 from HILT | | |
| 401-407 | | | | | | | • • • | Broadcast activations | | |
| 408 | | | | | | | • • • | A(24513)W(24513,1) + | • • • | A(24513)W(24513,8192) + |
| 409 | | | | | | | • • • | A(24514)W(24514,1) + | • • • | A(24514)W(24514,8192) + |
| 410 | | | | | | | • • • | A(24515)W(24515,1) + | • • • | A(24515)W(24515,8192) + |
| • | | | | | | | • • • | • | | • |
| • | | | | | | | • • • | • | | • |
| • | | | | | | | • • • | • | | • |
| 471 | | | | | | | • • • | A(24576)W(24576,1) | • • • | A(24576)W(24576,8192) |
| 472 | | | | | | | ► | Σ(A(1)W(1,1) to | • • • | Σ(A(1)W(1,8192) to |
| 473 | | | | | | | ► | | Batch 2 sums | |
| 474 | | | | | | | ► | | Batch 3 sums | |
| • | | | | | | | ► | | • | |
| • | | | | | | | ► | | • | |
| • | | | | | | | ► | | | |
| 33240 | | | | | | | ► | | Batch 32768 sums | |
| 33241-33256 | | | | | | | | | SIPO FIFO | |
| 33257-33260 | | | | | | | | | Write to output sums HILT | |

Table 6 shows the calculation of large matrix multiplications using the ZettaLith implementation of CASCADE system on a single TRIMERA chip stack. In this case, 32,768 batches (and/or input tokens) of an array of 24.576 activations × 8,192 columns is being calculated in 33,260 clock cycles (2.77 µs). This time is used to read 805,306,368 activations from activation HILT, perform 13,194,139,533,312 FLOPs, and write the sums to the output sums HILT. As 33,260 clock cycles would normally be enough for 13,392,244,899,840 FLOPs, this matrix multiplication operates at 98.52% efficiency. Each of the 32,768 batches (and/or input tokens) in a TRIMERA stack is calculated simultaneously, offset by one clock. Also, each of the 156 TRIMERA stacks in a ZettaLith can perform matrix multiplies of this size simultaneously.

A summary of the CASCADE calculation shown in Table 6.

- **Clock 1:** CASCADE array 1 and 2 both start on clock 1, as their sum in the CASCADE inter-array mechanism is aligned. Subsequent CASCADE arrays start on subsequent clocks, i.e. CASCADE array 3 starts on clock 2 through to CASCADE array 384 which starts in clock 383. This is because their sums in the CASCADE inter-array mechanism are sequential.
- **Clocks 1 to 17** are used to load B(1-8) A(1) – activations(1) – from HILT memory. This has a latency of 16 clocks, but a throughput of 16 billion activations per second. That is, B(1)A(1) is available on clock 17, but subsequent batches of A(1) are available on subsequent clocks of the CASCADE array from the activation HILT(1). Simultaneously in overlapping access cycles, B(1)A(2) is available on clock 18 from activation HILT(8), and subsequent batches of A(2) are available on subsequent clocks. Every 8 clocks, the activation HILTs read a new set of 8 batches of activations until all 32,768 batches in HILT are read. (Note: "batches" are actually B × L - a combination of batch size and token length).
- **Clocks 18 to 24** are used to broadcast (A1) to all 8,192 columns of the CASCADE array using the activation broadcast latch tree (Figure 8, clocks 18 to 24 of Table 7). A(2) is broadcast on the next clock to row 2, and subsequent activations are broadcast on subsequent clocks. The activation broadcast latch tree is a pipeline, so new results are available to each of the 8,192 columns every clock. The total rate of activations for a single TRIMERA SLD is 8,192 columns × 24,576 rows × 12 GHz = 2,415,919,104,000,000,000 activations per second.
- **Clock 25** is the first clock of computation. Row 1 of CASCADE columns 1 to 8,192 multiply A(1) by the weights for each column – W(1,1) to W(1,8192).

- **Clock 26** is the second clock of computation. Row 2 of CASCADE columns 1 to 8,192 multiply A(2) by the weights for each column – W(2,1) to W(2,8192) and accumulate the result with the results of A(1)W(1,1) to A(1)W(1,8192).
- **This continues until Clock 88**, the last calculation of the first CASCADE array. Row 64 of CASCADE columns 1 to 8,192 multiply A(64) by the weights for each column – W(64,1) to W(64,8192) and accumulate the result with the ongoing sums for column 1: ΣA(1)W(1,1)…A(63)W(63,1) through to column 8,192: ΣA(1)W(1,8192)…A(63)W(63,8192).
- **Clock 89** adds the accumulation of one CASCADE array with the next CASCADE array which was being calculated simultaneously. Thus, at clock 89, the calculation wave for batch 1 gives the 8,192 column sums ΣA(1)W(1,1)…A(128)W(128,1) through to ΣA(1)W(1,8192)…A(128)W(128,8192). The calculation wave for batch 2 is proceeding is one clock behind.
- **On clock 472** batch 1 is complete, with the 8,192 column sums being: ΣA(1)W(1,1)…A(18432)W(24576,1) through to ΣA(1)W(1,8192)…A(24576)W(24576,8192). The FP8 results from each column are then added to the accumulated sums in the output sums HILTs (or biases if it is a first pass calculation) and written back to the output sums HILT at a 1 GHz rate, after being expanded to 128 bits wide by a SIPO FIFO.
- **On clock 473** batch 2 is complete.
- **On clock 33,240**, all 32,768 batches are complete.
- **By clock 33,260** the last of the 32,768 batches has been written to output sums HILT.

Of course, it is not necessary to calculate all 32,768 batches of 24,576 activations × 8,192 columns each time. Control circuitry should be added to allow appropriate subsets of the maximum calculation.

### 13.3 Parallel adder tree alternative

The partial sums from each CASCADE array are added sequentially. If they were added in parallel using an adder tree, the entire computation would be complete in 32,885 clock cycles, resulting in 99.64% efficiency. However, this would complicate chip layout, with each successive pair of additions being over greater physical distances. Pattern dependent ground bounce would also be exacerbated. At 12 GHz clock frequency, such complications could lead to significant difficulties. Therefore, CASCADE uses sequential additions, at the expense of 1.12% efficiency.

### 13.4 Summary of CASCADE technique

The CASCADE mechanism occurs across two chips in the TRIMERA stack- the SLD for computation and storage of weights, and the HILT die for storage of batches of activations and output sums. Some characteristics include:

- **Column Oriented**: Each column of the output is calculated independently, with no cross-column calculation except for CREST nearest neighbor multiplexing every 64 rows.
- **Weight-Stationary Design**: The entire weight matrix of 201,326,592 FP4 weights is preloaded into the array before computation begins, and remains unchanged during the calculation of a batch.
- **Direct Weight Loading**: Weight loading occurs asynchronously directly from HBM without requiring intermediate cache storage.
- **Parallel Partial Sum Propagation**: After multiplication with stored weights, partial sums propagate vertically down each column independently.
- For arrays up to 24,576 rows (activations), or batches less than 32,768 the partial sums do not need to be transferred from chip to chip, only the completed sums from the 8,192 columns.
- **Broadcast Activation Flow**: Unlike conventional horizontal activation pipeline flow, a single FP4 activation value enters simultaneously at the PEs of all 8,208 (8,192 plus 16 spares) columns. While this is a little more complex in hardware than "systolically pumping" the activations from left to right through the array, it is worth the extra hardware complexity to avoid the delay in activation availability, and the complexity of skewed data. The activation broadcast is accomplished via a 8-level fan-out tree of latches, distributing one activation value across all columns each clock cycle. The 32,768 batches of 24,576 activations are entered into all columns simultaneously at the 12 GHz CASCADE array clock frequency, using 24,576 activation HILTs and 24,576 activation broadcast latch trees. The broadcast latch tree, shown in Table 7, is used instead of a bus, even though the simpler bus structure would be functionally equivalent. A bus would result in significant (and insurmountable, in TSMC A16 or A14) propagation delay, IR drop, fan-out and ground bounce difficulties operating at the SLD's 12 GHz clock frequency.

### 13.5 Advantages of CASCADE

This full-array column-oriented approach offers critical advantages:

1. **Simplified Accumulation**: Final results accumulate automatically without complex sharding of submatrices and stitching accumulation processes.
2. **Minimized Inter-Chip Communication**: In most circumstances, no partial sums need to be transferred between chips during computation. This dramatically reduces chip-to-chip bandwidth requirements compared to traditional architectures.
3. **Reduced Output Bandwidth**: With only complete sums output after 33,260 cycles, the output data rate is vastly lower than systems that must transfer partial sums.
4. **Memory Efficiency**: Weights reside directly within the CASCADE array, eliminating the need for duplicate weight storage in cache SRAMs. Weights are loaded into the CASCADE arrays asynchronously using the HBM4 data paths or transferred between TRIMERA stacks at 39 TB/s.

5. **Superior Fault Tolerance**: With no cross-column communication, the CREST redundancy system can
6. independently validate and substitute spare CASCADE columns for any detected faults, maintaining computational throughput despite silicon defects.

### 13.6 CASCADE Rows, Columns and Arrays Tradeoff

The number of active PEs on a TRIMERA stack is the product of the 64 rows in a CASCADE array, the 8,192 active columns in a CASCADE array, and the 384 CASCADE arrays in a TRIMERA SLD. There is a significant degree of flexibility in choosing these numbers.

1. The number of rows in a CASCADE array stack primarily affects the SLD chip layout and the effectiveness of the CREST mechanism. Increasing the number of rows in a CASCADE array reduces the number of CASCADE inter-array mechanisms but reduces the level of fault-tolerance provided by CREST and makes the SLD physical layout more sensitive to chip dimensions.
2. Increasing the number of active CASCADE columns proportionally reduces the number of CASCADE rows, given a constant number of PEs available on the SLD. It also proportionally increases the number of output sum HILT memories and reduces the number of activation HILT memories.
3. Increasing the number of CASCADE arrays on the chip requires either a decrease in the number of rows or the number of columns in each array, with appropriate changes in the number of activation HILTs and output sum HILTS.
4. There is a broad fitness peak for these three values, so they can be optimized together with relatively little consequence.

### 13.7 ZettaLith Aggregation of TRIMERA Stacks

While a single TRIMERA stack is optimized for 8,192 columns, there are 156 TRIMERA stacks in a ZettaLith, allowing for up to 1,277,952 columns to be calculated simultaneously, without requiring transfer of partial sums. The entire ZettaLith enables batches of 32,768 arrays of 24,576 activations × 8,192 columns x 156 TRIMERAs (1,029,142,883,598,340 FLOPs) to be calculated in 33,260 clock cycles (HILT to HILT) at 98.52% efficiency.

This is the limit of ZettaLith calculation without transfer of activations, weights, or sums between TRIMERA stacks. Larger matrix calculations require transfer of data between TRIMERA stacks using the 7,800 TB/s data fabric bandwidth of the WSSCB.

**Table 7: Activations HILT and broadcast latch tree**

| Clock | Phase | Activations | Spare | Bits | Fanout | Clock gen. |
|---|---|---|---|---|---|---|
| 1-3 | Read MUX | 32,768 | 1 | 131,072 | 0.0625 | 1 |
| 4-7 | Read MUX | 2,048 | 1 | 8,192 | 0.0625 | 1 |
| 8-11 | Read MUX | 128 | 1 | 512 | 0.0625 | 1 |
| 12-15 | Read MUX | 8 | 1 | 32 | 0.1250 | 1 |
| 16 | Read MUX | 1 | 1 | 4 | 1 | 1 |
| 17 | HILT->SLD | 1 | 1 | 4 | 2 | 1 |
| 18 | Broadcast | 2 | 1 | 8 | 2 | 2 |
| 19 | Broadcast | 4 | 1 | 16 | 2.25 | 4 |
| 20 | Broadcast | 8 | 1 | 36 | 4 | 9 |
| 21 | Broadcast | 32 | 1 | 132 | 4 | 33 |
| 22 | Broadcast | 128 | 1 | 520 | 4 | 130 |
| 23 | Broadcast | 512 | 4 | 2,064 | 4 | 516 |
| 24 | Broadcast | 2,048 | 8 | 8,224 | 4 | 2,056 |
| 25 | PE | 8,192 | 16 | 32,832 | Within PEs | 8,208 |

# 14 Transformer Inference

Model sizes for OpenAI and Anthropic LLMs are not publicly available, so this calculation is based on Meta's Llama 3.1 405B, the largest open source LLM currently available.

Table 8 shows the basic model size characteristics of Llama 3.1 405B.

Table 9 shows the order of calculation size, and the number of calculations required to calculate the model at the batch size and context length of Table 8.

| **Table 8: Example Transformer calculation** | | |
|---|---|---|
| **Llama 3.1 405B Core Model Parameters** | **Size** | **Symbol** |
| Model dimension | 16,384 | d |
| Number of attention heads | 128 | h |
| Vocabulary size | 128,000 | V |
| Number of layers | 80 | N |
| Feedforward dimension | 65,536 | 4d |
| Max context length | 128,000 | max L |
| Batch size used in this example | 1,024 | B |
| Context length (average used in example) | 2,000 | L |

## 14.1 DeepSeek R1 MHLA

DeepSeek R1 introduces multi-head latent attention (MHLA) to the transformer equation. As this makes a large reduction to the computation of the Query, Key and Value matrices, MHLA is likely to become a standard part of most, if not all, LLM transformers. This is not included in the tables for Llama 3.1 405B, as this model predates DeepSeek's innovation of MHLA.

The ZettaLith architecture is flexible enough to incorporate innovations such as MHLA.

## 14.2 Weight Reuse

A key aspect of achieving the performance levels described in this specification is the efficient management of memory bandwidth relative to computational throughput. The ZettaLith architecture's 1,047× improvement in inference performance compared to contemporary systems is achieved with a substantially smaller increase in HBM bandwidth, necessitating substantially higher efficiency in weight utilization.

This efficiency, commonly known in the industry as weight reuse, represents the number of times that a weight can be reused from on-chip storage between times that it must be loaded from off-chip memory – in this case HBM4 stacks.

Balancing weight reuse so that the HBM bandwidth is sufficient to feed the extremely high calculation rates of the CASCADE arrays can be achieved in ZettaLith by having larger batch sizes. The usual disadvantage of large batch sizes – increased latency – does not really apply with ZettaLith. Being 1,047× faster allows 1,047× the batch size with the same latency.

Tables 9 and 10 show how a batch size of 1,024 creates a good match between the time to compute a batch of inferences and the time to load the weights of those inferences from HBM, with both operations taking around the same time for this batch size.

| **Table 9: Example Transformer Inference Calculations** | | | **Llama 3.1 405B** | **Table 10: Weights loading** | | |
|---|---|---|---|---|---|---|
| **Inference - FLOPS** | **Calculation Order** | **OPs** | **% of total** | **Weight Order** | **Weights** | **% of total** |
| Token embedding lookup | O(B x L x d) | 3.36E+10 | 0.000005% | O(V x d) | 2.10E+09 | 0.60% |
| Positional encoding (RoPE) | O(B x L x d) | 3.36E+10 | 0.000005% | O(1) | - | - |
| Layer normalization (pre-attention) | O(B x N x L x d) | 2.68E+12 | 0.00038% | O(N x d) | 1.31E+06 | 0.00038% |
| Multi-head attention (Q, K, V projections) | O(B x N x L x 3 x d x d) | 1.32E+17 | 18.61% | O(N x 3 x d x d) | 6.44E+10 | 18.52% |
| Attention score | O(B x N x h x L x L) | 4.19E+13 | 0.006% | O(1) | - | - |
| SoftMax of attention scores | O(B x N x h x L x L) | 4.19E+13 | 0.006% | O(1) | - | - |
| Value weighting and concatenation | O(B x N x h x L x L x (d/h)) | 5.37E+15 | 0.76% | O(1) | - | - |
| Output projection | O(B x N x L x d x d) | 4.40E+16 | 6.20% | O(N x d x d) | 2.15E+10 | 6.17% |
| Residual connection (post attention) | O(B x N x L x d) | 2.68E+12 | 0.00038% | O(1) | - | - |
| Layer normalization (pre-FFN) | O(B x N x L x d) | 2.68E+12 | 0.00038% | O(N x d) | 1.31E+06 | 0.00038% |
| Up projection FFN | O(B x N x L x d x 4d) | 1.76E+17 | 24.81% | O(N x d x 4d) | 8.59E+10 | 24.70% |
| Gate projection FFN | O(B x N x L x d x 4d) | 1.76E+17 | 24.81% | O(N x d x 4d) | 8.59E+10 | 24.70% |
| SwiGLU Nonlinearity | O(B x N x L x 4d) | 1.07E+13 | 0.0015% | O(1) | - | - |
| Down projection FFN | O(B x N x L x 4d x d) | 1.76E+17 | 24.81% | O(N x 4d x d) | 8.59E+10 | 24.70% |
| Residual connection (post FFN) | O(B x N x L x d) | 2.68E+12 | 0.00038% | O(1) | - | - |
| Final layer normalization | O(B x L x d) | 3.36E+10 | 0.000005% | O(d) | 1.64E+04 | 0.000005% |
| Output projection (autoregression) | O(B x 1 x d x V) | 2.15E+12 | 0.00030% | O(V x d) | 2.10E+09 | 0.60% |
| Total for a batch of 1024 inferences | | 7.09E+17 | 100% | Total weights | 3.48E+11 | 100% |
| ZettaLith peak performance | | 1,507,534 | PFLOPS | HBM bandwidth | 2.56E+14 | Bytes/s |
| Target percentage of peak PFLOPS | | 80% | | Weight data | 1.74E+11 | Bytes |
| Time to compute a batch of 1024 inferences | | **0.00059** | Seconds | Weights from HBM | **0.00068** | Seconds |

## 15 The Base Interface Die (BID)

Figure 6c illustrates the basic contents of the BID, which integrates multiple interface blocks and memory elements in a mainstream process node. This is not a floor plan of the chip, but an approximate use of chip area per function, and approximate arrangement of microbonds to the SCB. The die includes:

1. HBM4 interface, also applicable to HBF
2. A central controller managing die operations
3. A configuration NVM for the central controller
4. Mixed signal circuits containing:
    a. Analog components and PLLs
    b. Temperature sensors and thermal management
    c. Clock generation and distribution
    d. Power management
    e. Power-on reset and initialization circuits
5. System monitoring and telemetry
6. JTAG interface for external testing and debugging
7. BIST controller for built-in self-testing
8. Error logging memory
9. ESD protection circuits for the signal TSVs
10. TSVs to convey signal and power connections to the reverse side of the die.
11. Very high bandwidth UCIE 2.0 data fabric links to the next BID above and below.
12. Split high bandwidth UCIE 2.0 data fabric links to the next BID to the left and to the right. These are split to make room for the HBM4 interface, which must be in this location due to the layout of the TRIMERA stack to the HBM stack.
13. AI specific engines may be on the BID, but are preferably on the HILT, depending on available space:
    a. **SoftMax** state machines
    b. **RMSNorm** state machines
    c. **SwiGLU** state machines
    d. A final **image decoder/VAE** for image applications

The BID design includes UCIe-to-UCIe module bypass paths in both horizontal and vertical directions, enabling faulty modules to be mapped out with only a tiny amount of the BID functional. Mapping out the SCB module is the default mode until the BID passes boot-up tests, allowing the modules to be mapped into the array only if they are functional. These bypass circuits, consuming only µW of power, are powered by neighboring modules. In this way, module arrays can be actively mapped out for fault tolerance even if it is the module's power supply that has failed.

| **Table 11: HILTs supporting the CASCADE Arrays** | | |
|---|---|---|
| **Values in Common** | **Value** | **Unit** |
| Batch size x input token length in HILT | 32,768 | B x L |
| Active CASCADE array columns | 8,192 | columns |
| Spare CASCADE columns for CREST | 16 | columns |
| Columns per CASCADE array | 8,208 | columns |
| Rows per CASCADE array | 64 | rows |
| CASCADE array size | 525,312 | PEs |
| CASCADE arrays in a TRIMERA | 384 | arrays |
| Total CASCADE rows in a TRIMERA | 24,576 | rows |
| PEs in a TRIMERA | 201,719,808 | PEs |
| TRIMERA total spare columns for CREST | 6,144 | columns |
| CASCADE array clock in SLD chip | 12 | GHz |
| Clocks to output delay without CASCADE | 24,616 | clocks |
| Clocks to output delay with CASCADE | 488 | clocks |
| HILT and BID chips clock speeds | 2 | GHz |
| HILT unit cell (D latch plus transmission gate) | 8 | Tr |
| Full custom HILT bit cell in TSMC N2 | 0.012 | $\mu m^2$ |
| HILT overhead (decoders, clock buffers) | 16% | |
| **Input Activations HILTs** | | |
| Activation HILT storage tristate latches | 131,072 | bits |
| Activation HILT stage 2 tri-state latches | 8,192 | bits |
| Activation HILT stage 3 tri-state latches | 512 | bits |
| Activation HILT stage 4 tri-state latches | 32 | bits |
| Activation HILT output bit width (1 row) | 4 | bits |
| Activation HILT total tri-state latches | 139,812 | bits |
| CASCADE array activation HILT bits | 8,388,608 | bits |
| CASCADE activation HILT bitcells area | 102,098 | $\mu m^2$ |
| CASCADE activation HILT total area | 121,545 | $\mu m^2$ |
| TRIMERA bit-width of all activation HILTs | 3,221,225,472 | bits |
| Total TRIMERA activation HILT area | 47 | $mm^2$ |
| **Output sum HILTs** | | |
| Output sums HILT storage tristate latches | 262,144 | bits |
| Output sums HILT stage 2 tri-state latches | 16,384 | bits |
| Output sums HILT stage 3 tri-state latches | 1,024 | bits |
| Output sums HILT stage 4 tri-state latches | 64 | bits |
| Output sums HILT output bit width (1 column) | 8 | bits |
| Output sums HILT total tri-state latches | 279,624 | bits |
| CASCADE output sums HILT bits | 2,151,677,952 | bits |
| CASCADE output sums HILT bitcells area | 26,188,078 | $\mu m^2$ |
| CASCADE output sums HILT total area | 31,176,283 | $\mu m^2$ |
| Total TRIMERA output sums HILT area | 31 | $mm^2$ |
| Output sums SIPO FIFO | 8 | :128 |
| **Weights are stored directly in the CASCADE arrays** | | |
| **Total HILT for TRIMERA** | | |
| TRIMERA activation HILT data | 384 | MBytes |
| TRIMERA output sums HILT data | 257 | MBytes |
| TRIMERA total HILT data | 641 | MBytes |
| Total CASCADE memory HILT area | 78 | $mm^2$ |
| HILT die area | 143 | $mm^2$ |
| CASCADE Array HILT % of area | 54% | |
| Time to transfer HILT memory over data fabric | 16.42 | µs |

## 16 CASCADE Array HILT support in a TRIMERA

The HILT die contains HILT data arrays to feed the CASCADE arrays with activations, collect calculated sums from the output, and provide the CREST comparison logic. The weights are stored directly in the CASCADE array in the SLD.

Table 11 shows the support logic, HILT arrays, and FIFOs feeding the CASCADE arrays with activations and weights and collecting output sums. The activation HILTs feed into the centers of the broadcast latch trees of the CASCADE rows, and are positioned in the centers of CASCADE arrays to minimize 12 GHz wire lengths.

The output sums HILTs are connected to the final CASCADE array and are large enough to need to be distributed across the chip. The clock frequency of the output sum hilts can readily be reduced with negligible effect on system performance by increasing write parallelism from 128 to 256 bits.

## 17 ZettaLith Data Fabric

Table 12 shows the number of lanes and bandwidth of ZettaLith TRIMERA data fabric links.

| Table 12: TRIMERA data fabric link bandwidth | | |
|---|---|---|
| **Vertical fabric links** | **Value** | **Units** |
| UCIe 2.0 bandwidth per lane | 32 | GT/s/lane |
| Microbump pitch | 20 | μm |
| Microbumps per lane | 4 | μbumps |
| Rows of microbumps | 60 | μbumps |
| Width of rows of UCIe 2.0 bumps | 1.2 | mm |
| Horizontal chip width | 13 | mm |
| Microbumps per vertical UCIe 2.0 link | 39,000 | μbumps |
| Wire density | 3 | wires/μm |
| Length of wires (all same length) | 1.4 | mm |
| Lanes per vertical link | 9,750 | lanes |
| Bandwidth per vertical link | 312,000 | GT/s |
| UCIe 2.0 bandwidth per vertical link | 39 | TB/s |
| **Horizontal fabric links** | | |
| Vertical chip width allocated to UCIe 2.0 | 2.2 | mm |
| Columns of microbumps | 100 | μbumps |
| Total microbumps per horizontal UCIe 2.0 link | 11,000 | μbumps |
| Lanes per horizontal link | 2,750 | lanes |
| Bandwidth per horizontal link | 88,000 | GT/s |
| UCIe 2.0 bandwidth per horizontal link | 11 | TB/s |

The ZettaLith data fabric is a 2D asymmetric mesh with 39 TB/s chip-to-chip bandwidth in the vertical direction, and 11 TB/s chip-to-chip bandwidth in the horizontal direction, As ZettaLith is not a general purpose machine, there is no attempt to generalize the data fabric to an any-to-any configuration that maximizes flexibility. Instead, the data fabric is configured for the maximum usefulness for transformer inference within the constraints of the WSSCB, the TRIMERA chips, and UCIe 2.0 connections.

The vertical connections between TRIMERA chips is chosen to be the higher bandwidth connection because the horizontal connections are interrupted by the HBM4 links, and these horizontal data fabric connections need to be routed around the TRIMERA-HBM4 links in the WSSCB. The vertical connections are not interrupted by the HBM interface. For simplicity, they are identical parallel 1.4 mm USR wires.

UCIE 2.0 has a data transfer rate of 32 GT/s/lane. To achieve the 39 TB/s chip-to-chip bandwidth, 9,750 lanes are required. As each lane requires 4 wires, there are 39,000 wires between vertically adjacent TRIMERA stacks. As the TRIMERA stacks are 13 mm wide, the wire density of the vertical fabric links is 3 wires per μm. The number of RDL layers required in the WSSCB depends on the WSSCB wiring pitch. For example, if the pitch is 1 μm, then a minimum of 3 RDL layers are required. WSSCB processing is based on TSMC CoWoS-S, where this wiring pitch is readily achievable. The 4 μm pitch commonly associated with CoWoS is for CoWoS-R.

The wiring between vertically adjacent TRIMERA chips is extremely simple: 39,000 parallel wires each 1.4 mm long between matching pairs of μbumps in the BIDs of adjacent TRIMERAs. Only a few lanes of wires need to be routed and simulated, then those few wires can be replicated along the top and bottom edges of the BID footprints in the WSSCB.

The highest bandwidth requirement is the transfer of activations and output sums between adjacent TRIMERAs when calculating arrays larger than 24,576 activations in × 8,192 activations out. In this case, vertically adjacent TRIMERA stacks should be used for calculating adjacent sections of the large matrix, so the data transfers can be done simultaneously at 39 TB/s per TRIMERA stack pair.

The UCIe 2.0 interfaces are in the BID, nominally implemented using the TSMC N7 node. UCIe 2.0 Intellectual Property (IP) blocks are available for the TSMC N7 node, eliminating the need for custom interface design.

### 17.1 Inter-ZettaLith connections

ZettaLith could provide 32 channels of 800 gigabit Ethernet (GbE) connection to the outside world, with a total bandwidth of 25.6 Tb/s (3.2 TB/s). This is provided by converting mesh links at the left and right edges of the WSSCB array from the UCIe 2.0 to 800 GbE.

None of this Ethernet bandwidth is used in the transformer calculations described here: these are optional connections if transformer systems of more than 5 trillion parameters are to be inferenced. A ZettaLith can operate at the specifications described here in stand-alone configuration with no GbE connections. In comparison, GPUs may provide substantial GbE bandwidth, but the majority of this is used internally by the GPU cluster to transfer partial sums, inference the transformers, so it is not available for external connectivity.

For applications where more inter-ZettaLith data bandwidth than can be provided by 800 GbE is required (e.g., for transformer training or HPC applications), optical communications can be used. ZettaLith is a good candidate for the TeraPHY™ 8 Tb/s optical I/O chiplets and SuperNova™ multi-wavelength laser modules recently announced by Ayar labs. These optical modules connect by UCIe, so the ZettaLith data fabric is already suited for the TeraPHY system. However, 78 of the 1 TB/s TeraPHY chiplets would be required to extend each of the 39 TB/s vertical data fabric links from intra-ZettaLith to inter-ZettaLith while maintaining the full bandwidth. This would require 1,560 TeraPHY optical chiplets per ZettaLith. This illustrates how fast the TRIMERA chip-to-chip data fabric on the WSSCB is.

If it is certain that ZettaLiths will not be connected together at high bandwidth, all these Ethernet connections can be eliminated from the ZettaLith design to save manufacturing cost, design time, and complexity. Any external connectivity can then be provided by the PCIe 6.0 interfaces. A first generation ZettaLith may omit the 800 GbE interfaces to reduce TTM.

## 18 Hybrid Bonds

The SLD-HILT interface around a million hybrid bonds, as shown in Table 13. The hybrid bond pitch of 8.6 µm is above TSMC's projected minimum of 3.0 µm for the A16 node.

To achieve a very even power and ground distribution over the entire SLD chip, 787,968 of the hybrid bonds are power and ground. This minimizes the differences between PEs resulting from their position on the die, reducing the IR droop and ground bounce margins required and simplifying simulation.

Although backside power distribution will be available for the A16 and A16 nodes, it is not used for the SLD chip as the backside of the die has DRIE silicon heat-sink fines etched into it.

| Table 13: Hybrid bonds | | |
|---|---|---|
| **General** | **Value** | **Notes** |
| CASCADE Array rows | 64 | Rows |
| CASCADE Array columns | 8,208 | Columns (including spares) |
| PEs in a CASCADE Array | 525,312 | PEs |
| Activation and weight bits | 4 | FP4 |
| Partial sum bits | 8 | FP8 |
| **Hybrid bonds per CASCADE array** | | |
| Weights write data bus | 256 | Weight data bus to PEs |
| Weight write enables | 2,052 | Individual enables as decoder is in HILT |
| Activations in | 256 | Broadcast activations input |
| CREST multiplexers write data bus | 32 | Control of the CREST multiplexers |
| CREST multiplexers address decoder | 11 | CREST address decoder and write enable |
| Weight, activation, sum clocks | 6 | High frequency clock distribution |
| Ground | 1,026 | Ground return paths |
| Power | 1,026 | Power delivery to CASCADE arrays |
| **Total hybrid bonds between the SLD and HILT chips** | | |
| Total bonds for a CASCADE array | 4,665 | Hybrid bonds for a single CASCADE array |
| CASCADE Arrays | 384 | Arrays |
| Total bonds for all CASCADE arrays | 1,791,360 | Hybrid bonds for all arrays |
| Column partial sums/biases input | 65,664 | Bias or partial sum inputs for final sum |
| Column sums output | 65,664 | Sum outputs from last CASCADE array |
| Total hybrid bonds per SLD-HILT | 1,922,688 | Hybrid bonds between the SLD and HILT die |
| TRIMERA bond die areas | 143,000,000 | µm$^2$ each |
| Required hybrid bond pitch | 8.6 | µm |
| TSMC minimum hybrid bond pitch for A16 | 3.0 | µm |
| Status | OK | Hybrid bond pitch is manufacturable |

## 19 CPUs

WSSCB CPU stacks handle high-speed computations not suitable for the main TRIMERA array, as commonly required in CPU-GPU systems for AI inference and high performance computing. This flexibility extends to future implementations, where arrays of chips on a single WSSCB design can become increasingly varied and application-specific as new compatible SLDs are developed.

### 19.1 CPU stacks

A first generation ZettaLith only requires enough CPU processing to facilitate FP4 Transformer inference. The initial amount of CPU power only needs to be "adequate", not "as much as possible". The CASCADE array compute performance is so much higher than any feasible CPU performance that the CPUs are only useful for control and operations which cannot be parallelized.

An optimized CPU-SRAM-BID stack can be provided later. As the JETSTREAM cooling will already be in place, CPU stacks can be designed for high power consumption and dissipation, equal to TRIMERA. In this way, substantially greater performance can be achieved for the CPU than could be achieved with conventional power supply and cooling systems.

There are two types of CPUs that are used with ZettaLith:

- CPUs integrated into the ZettaLith system WSSCB. These have data fabric connections as fast as 39 TB/s to the TRIMERA stacks for each of the CPU, for a total of 624 TB/s for the 16 CPU stacks.
- External CPUs connected by PCIe 6.0 (2 TB/s total bandwidth) or 800 GbE (3.2 TB/s total bandwidth).

The ZettaLith CPUs can use standard ARM cores, RISC-V cores, or an arrangement largely derived from designs that the ZettaLith OEM is already familiar with. The CPUs can be designed as a single CPU chip or a multi-chip stack. If it is to be a multi-chip stack, the TRIMERA BID can be used. This would likely reduce overall system design time, as much of the BID contents is directly applicable to a CPU stack, and compatibility with the extremely high bandwidth data fabric would be maintained. As the BID also contains the HBM4 connection, this would not need to be redesigned. However, the TRIMERA HILT die is unsuitable for a CPU, and a separate SRAM die providing the CPU cores with cache memory should be designed. Alternatively, cache SRAM can be incorporated on the CPU die, and the CPU die. In this case, the CPU die would need TSVs and would be back-to-back hybrid bonded to the BID die.

### 19.2 CPU HBM

In the recommended ZettaLith configuration, the TRIMERA stacks use minimum height HBM4 stacks, but the CPU stacks use maximum height HBM4 stacks. There are many applications for the larger memory of the CPU stacks:

- KV caches;
- Reasoning model contexts;
- Parameters of transformers that are not in current use, but may be needed faster than they can be loaded from SSD;
- Video and images being generated by ZettaLith;
- Large user documents and query histories – for example, code bases, PDFs, image and video inputs, etc.; and
- Space for running relatively large user-requested programs (such as simulations) locally.

The larger memory of the CPU stacks can also be used for holding data or programs that are used frequently by ZettaLith. These may be in the form of Model Context Protocol (MCP) servers. Examples include:

- Wikipedia (a frequently updated local copy, without the large edit histories, which can be accessed externally when required);
- Company databases, for corporation-specific ZettaLith systems, or for shared systems if data security is adequate;
- 3D graphics systems such as Blender;
- Symbolic math systems such as Mathematica; and
- Engineering simulation systems such as ANSYS.

MCP is rapidly gaining acceptance, and there are already thousands of MCP servers. Frequently accessed MCP server instances may be hosted directly on ZettaLith for very high bandwidth and low latency access.

### 19.3 PCIe 6.0 links

The 16 CPU chips in ZettaLith provide 16 PCIe 6.0 links from the CPUs to SSD storage, external servers, and the Internet. Each PCIe 6.0 link has 16 lanes of 8 GB/s for a total bandwidth of 2 TB/s (16 Tb/s).

During typical transformer inference, this bandwidth is unused. High bandwidth is required to load parameters when rapidly switching to transformers which are not loaded into HBM, and to load large user contexts which are not stored on ZettaLith. Since ZettaLith has enough HBM for 20 trillion parameters (5 trillion in low cost system), it can hold multiple different trillion parameter LLMs in memory simultaneously, thereby not normally requiring any PCIe 6.0 bandwidth to switch between transformers.

## 20 Fault Tolerance

The practical implementation of ZettaLith capabilities requires careful consideration of module yield management through redundant routing, power distribution uniformity across the wafer, thermal gradient minimization, signal integrity maintenance across UCIe interfaces, known good die-attach to the WSSCB, and system-level fault tolerance and graceful degradation. These considerations are addressed through the redundant routing structures, distributed power delivery architecture, silicon spring thermal isolation, JETSTREAM cooling, and fault tolerance mechanisms.

ZettaLith incorporates multi-level fault tolerance.

- At the passive WSSCB level, all wires are individually fault tolerant, requiring highly coincidental defects on separate wiring layers to cause a single fault.
- The UCIE 2.0 and HBM4 interfaces have wire-level fault tolerance.
- Defective stacks, or stacks without power as their PSU is defective, are automatically mapped out of the compute array during power-up self-test.
- Within an SLD, spare CASCADE array columns are included to dynamically replace any faulty columns via the CREST mechanism.
- At the system level, if an entire TRIMERA stack is found to be defective, the design employs a fail-in-place strategy that maps out the faulty module without disrupting overall operation.

There are two levels in which a TRIMERA stack can be mapped out of the array – at bootup, the TRIMERA stack will automatically be mapped out if the BID doesn't boot. This uses the horizontal bypass and vertical bypass lanes of Figure 6c. At any stage during operation, a TRIMERA stack can also be mapped out of the array under software control.

### 20.1 CASCADE array fault tolerance

Fault tolerance is achieved by providing spare CASCADE array columns. Each of the 384 CASCADE arrays on an SLD chip has 16 spare columns in addition to the 8,192 active columns, giving a total of 8,208 columns per CASCADE array. Redundancy is required to achieve a practical yield in advanced CMOS processes. A total of 3,151,872 CASCADE columns can be individually replaced by 6.144 spare columns in an SLD chip allowing random defect densities of up to 2,014 defects per $cm^2$ with no effect on performance. These defects can be clustered at up to 40 defects per mm$^2$, also without affecting performance.

Beyond this, the SLD can have graceful degradation, but the software required to support SLDs with performance differences is complex and is not recommended for a first generation ZettaLith.

In the unlikely situation that correcting for 2,014 defects per cm$^2$ is not enough, it is far simpler to increase the number of spare columns than to accommodate SLDs with so many defects that performance is affected. The chosen value of 16 spare columns per CASCADE array is simply an arbitrary number far more than is required for a production process. It is chosen to allow SHAPE production on pre-production processes with very high defect densities.

Spare CASCADE columns are also used to provide CREST run-time fault tolerance. CREST fault tolerance is implemented by the inter-array CASCADE and CREST logic and does not affect the individual PE design.

## 20.2 CREST run-time fault detection and repair

The ZettaLith TRIMERA fault tolerance system enhances reliability through cyclic redundant spare testing (CREST), which includes dynamic CASCADE array redundancy and runtime error detection. This limits transformer inference errors to a low probability of a single token error, without any drop in processing throughput. If the possibility of a single token error is unacceptable, ZettaLith can re-run the inference to correct the error.

CREST implements both initial testing to correct manufacturing defects using spare columns and uses the same spare columns to detect and correct operational faults during run-time inference.

To detect faults during run-time, a subset (which may be 1) of unallocated spare CASCADE columns in a CASCADE array is dynamically assigned to cyclically test operational columns during inference. These testing columns receive identical input activations and weight parameters as their paired operational columns via identically loaded weights and synchronized control paths, ensuring identical computational trajectories. Outputs from both the active column and the testing column are compared in software. It is only one FP8 number than needs to be checked against another, for each of the columns to be tested, so this operation is not compute intensive. A configurable error threshold (e.g., ≥3 consecutive mismatches) filters transient errors such as those caused by cosmic rays. Confirmed faults trigger error logging and temporary replacement of the potentially faulty column with a spare column.

Figure 10 shows the way that CREST detects and replaces defective columns of CASCADE arrays. Only a small section of a TRIMERA array is shown with 5 rows of CASCADE arrays (CRow) and 18 CASCADE columns. A CASCADE array in the TRIMERA system has 384 CRows each with 8,208 CASCADE columns.

Each CASCADE column has 64 PEs – the number of rows in a CASCADE array. Faulty PEs are not detected and replaced individually – only the entire CASCADE column that the faulty PE is in. Each box in Figure 10 is a CASCADE column of 64 PEs, not a single PE. The numbers in the boxes are not the physical column numbers, but the column of matrix multiply data that is being processed in each physical CASCADE column. A “column” here refers to the entire column of 384 CASCADE columns – i.e., 18,432 PEs, not 64 PEs.

## 20.3 Detailed CREST operation

Figure 10a shows CREST testing column 4. The comparison system compares the output of column 4 with the output of column 18 (in a TRIMERA SLD, the column used for testing would be a known good column - typically column 8,208, but even this column may be defective and repaired using CREST, so a different column would be used for testing). The weights of column 4 have been copied into column 18 before the CREST comparison is made. If there are no defects in column 4, the results for column 18 should exactly match those of column 4, as the activations are broadcast to all columns, and the weights are the same. In this case, the comparison system reports that column 4 is OK.

Figure 10b shows CREST testing column 5. The comparison system compares the output of column 5 with the output of column 18. The weights of column 5 have been copied into column 18. Column 5 containing a defective PE in CASCADE column in CRow(3). This makes the entire CASCADE column of 64 PEs defective. The defective CASCADE column is shown in black.

The defective CASCADE column causes incorrect results to propagate to downstream CASCADE columns. In Figure 10, CASCADE columns which are not physically defective but are propagating incorrect results from upstream defects are shown in pink with an X in them. In Figure 10b, the comparison system reports that column 5 is Faulty.

A defect has been detected. The next step is to isolate that defect.

Figure 10c shows CREST testing if the defect in column 5 is in CRow(1). This is done by setting the CREST multiplexers (shown in Figure 8) to connect the output of column 5, CRow(1) to the cascading sums of column 6, starting at CRow(2). The comparison is now made between column 6 and column 18. If the comparison system now reports OK, then the fault was not in CRow(1) of column 5.

Figure 10d shows CREST testing if the defect in column 5 is in CRow(2). The comparison system reports OK.

Figure 10e shows CREST testing if the defect in column 5 is in CRow(3). This is detected because the reprogramming of the CREST multiplexers has caused the output of defective column 5, CRow(3) to be connected to the cascading sums of column 6, starting at CRow(4). This caused the output of column 6 to be corrupted and not match the test data in column 18. The comparison system reports Faulty.

Figure 10f shows CREST repairing the defect in column 5, CRow 3 using a spare column segment. The data for column 5 is now processed in column 5 of CRow(1) and CRow(2), diverts to column 6 to avoid the defective CASCADE column 5 of CRow(3) , and reverts to column 5 for subsequent CRows. In this way, the defect at CASCADE column 5 of CRow(3) has been repaired using one of the spare CASCADE columns.

Note that in the simplified explanation of Figure 10, the bulk of the weights are shuffled along for each CREST test. This would be highly inefficient in a real system. In the complete system, the weight data for untested CASCADE columns can be kept at the high end of the column numbers (starting at column 8207 and counting backwards), and only moved to lower columns as tests proceed.

### 20.4 Extent of correctable defects

Figure 10g shows 16 CRows and 18 CASCADE columns. CREST is shown testing faulty column 13 after having repaired 9 faults in the first 16 CRows of an array. The TRIMERA SLD has 6,144 spare CASCADE columns, and disperse and random configurations of defects are well corrected. With 16 spare CASCADE columns, the system can detect and correct 15 clustered defective CASCADE columns in any one CRow.

A BID controller manages testing assignments, prioritizing arrays not yet tested or flagged for retesting. Testing cycles are quantized to align with layer transitions in transformer models, enabling seamless reconfiguration during natural computational breaks. Normal CREST cyclic testing is controlled by software and/or DMA hardware in the BID and has no impact on ZettaLith performance.

### 20.5 Error containment and recovery

Faulty CASCADE columns identified by CREST are temporarily bypassed at layer boundaries, with dataflow rerouted to standby spares via BID-managed data routing. This ensures continuity of the CASCADE array's operation while isolating errors to individual tokens. Corrupted tokens can be recomputed using validated columns, flagged for heuristic validation (e.g., semantic coherence checks), or simply ignored.

Temporarily replaced CASCADE columns undergo stress testing during idle periods to diagnose root causes (e.g., by altering operational voltages, or by ramping up and down clock speeds to detect marginal timing). This testing is controlled by software running on the BID controller. Permanently faulty CASCADE columns are decommissioned, reducing the pool of unallocated spares.

### 20.6 Power on Self-Test (POST) and Built-In Self-Test (BIST) using CREST

While CREST is used for on-going real-time testing, the CREST system is also used for POST and BIST. In these situations, the testing is required to be more comprehensive than run-time error detection. Every type of fault that can effect a calculation should be detected.

To comprehensively test arrays using CREST, sets of test vectors are loaded into the activations and weights HILTs. These test every combination of activation, weight, and partial sum in every PE of every array. All the 31,407 million PEs in a ZettaLith can be comprehensively tested in 2,132 milliseconds using this approach. Tests can be repeated at voltage margins by reprogramming the power supplies, and at temperature margins by including patterns of data designed to increase or decrease the power consumption of the PEs.

### 20.7 Graceful degradation

ZettaLith experiences no performance degradation until the spare columns in an TRIMERA are consumed. At that point,

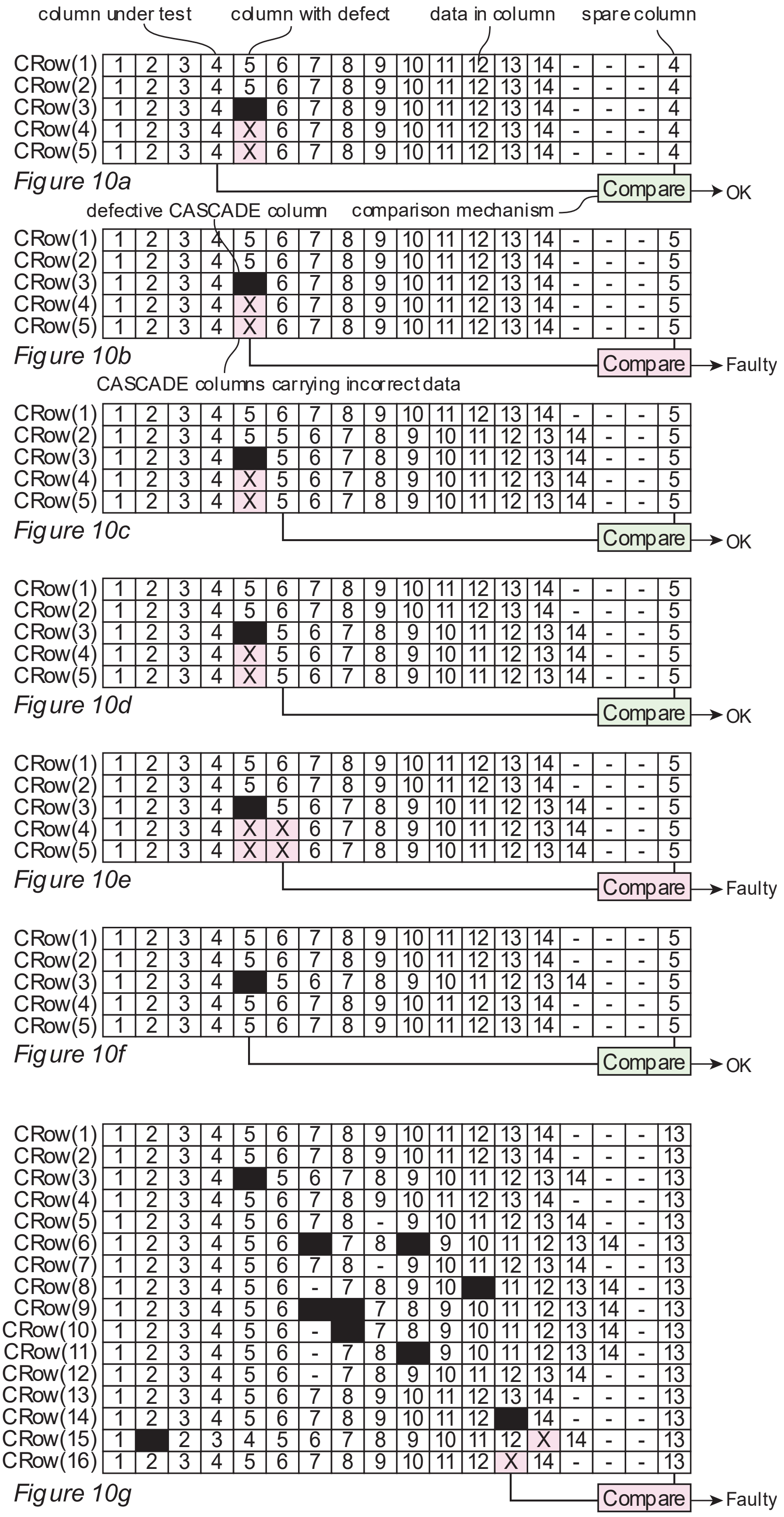

the TRIMERA can still operate, but not at full capacity, given the software drivers are written to accommodate imperfect arrays. For each column exceeding the allowance of spare columns in an TRIMERA, ZettaLith loses 0.0001% of its performance.

Once the TRIMERA loses enough arrays that it is deemed more of a computational problem than an asset, the TRIMERA is mapped out of active use. At this stage ZettaLith will lose 0.64% of its performance. ZettaLith can continue operating with graceful degradation until so many TRIMERAs fail that the ZettaLith is no longer cost effective to keep operating and is decommissioned.

As is normal for 3D microbonded chiplets and stacks, it is not expected that a faulty TRIMERA stack can be replaced without significantly risking the reliability of the overall ZettaLith, so a fail in place strategy is used.

# 21 CREST Fault Tolerance Compared to Alternative Approaches

The CREST (Cyclic Redundant Spare Testing) fault tolerance mechanism represents a departure from conventional fault tolerance methods, specifically optimized for the CASCADE architecture's regular column-oriented structure. This section compares CREST with established fault tolerance approaches to contextualize its advantages and trade-offs.

## 21.1 Error Detection and Correction (EDAC)

Conventional EDAC implementations add parity or ECC bits to detect and correct bit flips during storage or transmission, typically incurring 12.5-25% overhead for ECC memory. While EDAC operates continuously at the bit/word level, CREST operates at the column granularity (64 PEs) with spare columns consuming no power during normal operation. For transformer inference applications, the inherent tolerance to occasional bit errors eliminates the need for traditional EDAC in the CASCADE arrays, though the CPU stacks retain full EDAC capability where required.

## 21.2 Triple Modular Redundancy (TMR)

TMR systems execute three copies of each computation simultaneously and vote on results, providing immediate fault masking at the cost of 200% area and power overhead. In contrast, CREST requires only ~0.2% area overhead (16 spare columns per 8,192 active columns) with no power penalty. While TMR provides instantaneous fault masking, CREST's approach of testing and replacing faulty columns at layer boundaries should prove sufficient for transformer inference workloads where brief delays in fault correction are acceptable.

If required, a faulty inference token can be re-run.

## 21.3 Checkpoint/Restart Mechanisms

Traditional checkpoint/restart systems periodically save computational state to enable rollback upon error detection, impacting performance even during fault-free operation. CREST eliminates the need for checkpointing by continuously validating columns during inference and seamlessly transitioning to spare columns at natural computational boundaries. This approach avoids the performance penalties and complexity of state management inherent in checkpoint-based systems.

## 21.4 Built-in Self-Test (BIST)

Conventional BIST performs testing during power-on or scheduled maintenance windows, leaving systems vulnerable to runtime failures. CREST extends the BIST concept by performing continuous testing during normal operation using actual inference workloads as test vectors. This dual-purpose approach would enable detection of aging-related degradation and transient faults while maintaining full system availability.

## 21.5 Memory Array Spare Rows/Columns

Traditional memory arrays allocate spare rows and columns at manufacturing time to replace defective elements through simple address remapping. CREST extends this concept with dynamic allocation of spare columns that can address both manufacturing defects and runtime failures.

## 21.6 CREST Implementation Advantages

The CREST mechanism exploits several characteristics specific to the CASCADE architecture:

- **Column Independence:** The CASCADE architecture's column-oriented computation enables independent testing and replacement without affecting adjacent columns.
- **Natural Reconfiguration Points**: Transformer layer boundaries provide seamless opportunities for column reconfiguration without computation interruption.
- **Scalable Coverage:** With 6,144 spare columns distributed across a ZettaLith system, CREST can tolerate random defect densities up to 2,014 defects per cm² or clustered defects up to 40 defects per mm².
- **Unified Mechanism:** The same spare columns and comparison logic serve both manufacturing yield enhancement and runtime reliability.

## 21.7 Design Trade-offs

CREST's specialization for CASCADE architectures implies certain limitations:

- Column-level granularity precludes correction of individual PE failures – but 64 PE columns are still very small.
- Fault detection latency depends on the testing cycle time.
- The approach requires regular architectures with little state.

These trade-offs prove advantageous for transformer inference, where the regular computational pattern, tolerance for transient errors, and large scale (31.4 billion PEs) make traditional fine-grained approaches impractical. CREST thus represents a co-design of fault tolerance with the computational architecture, achieving robust operation with minimal overhead by exploiting the specific characteristics of transformer inference workloads.

## 22 ZettaLith Power Supply Units (PSU)

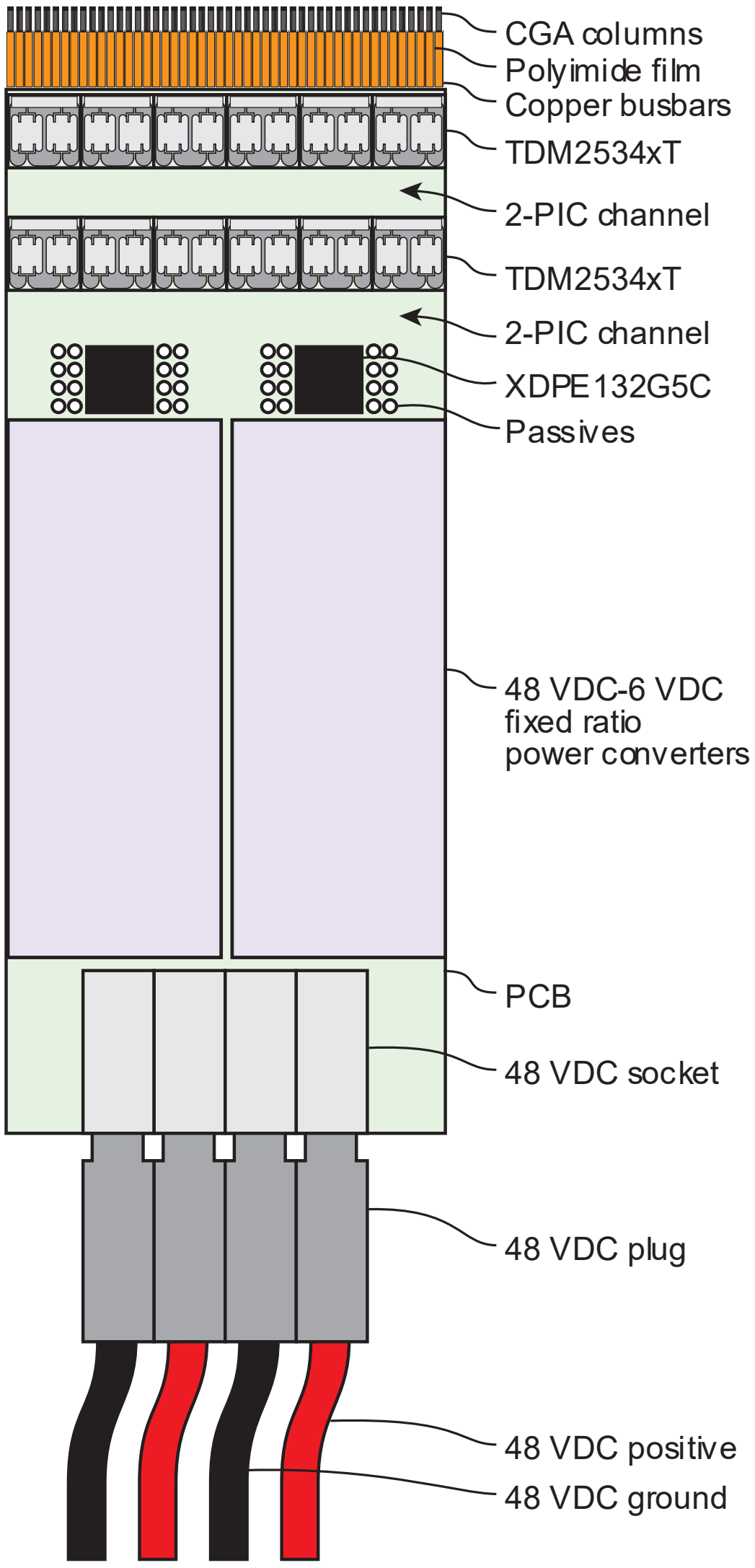

*Figure 11a:*
*Top view of PSU PCB*

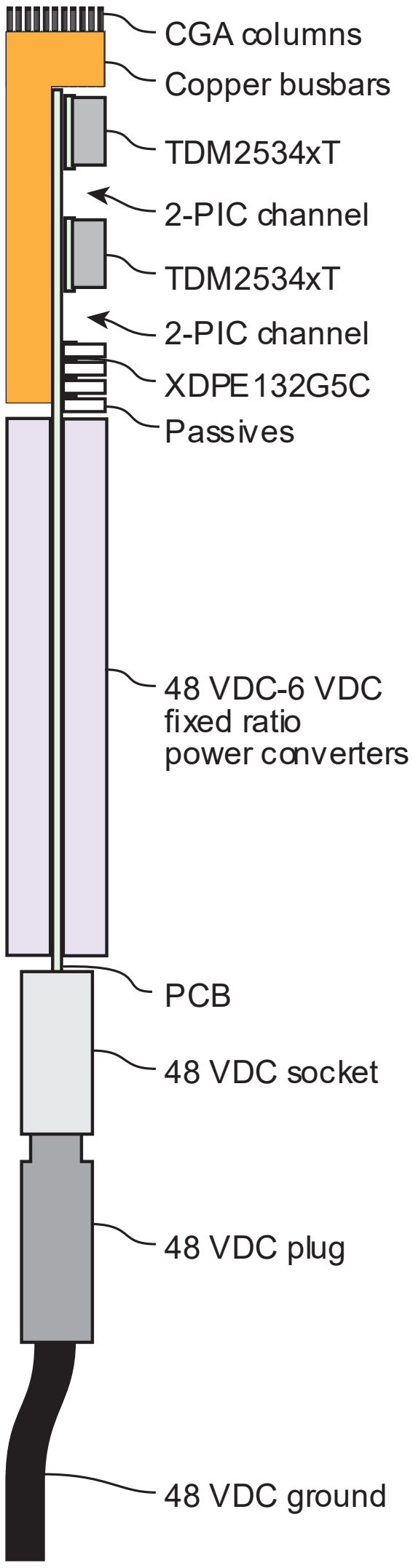

*Figure 11b:*
*Side view of PSU PCB*

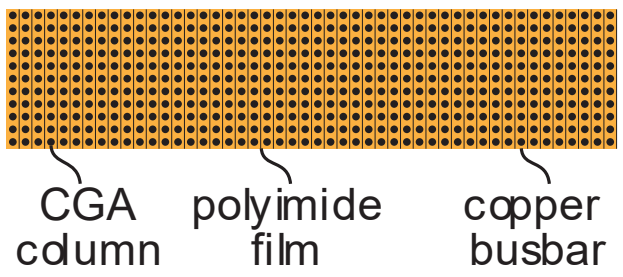

*Figure 11c:*
*CGA end view of PSU PCB*

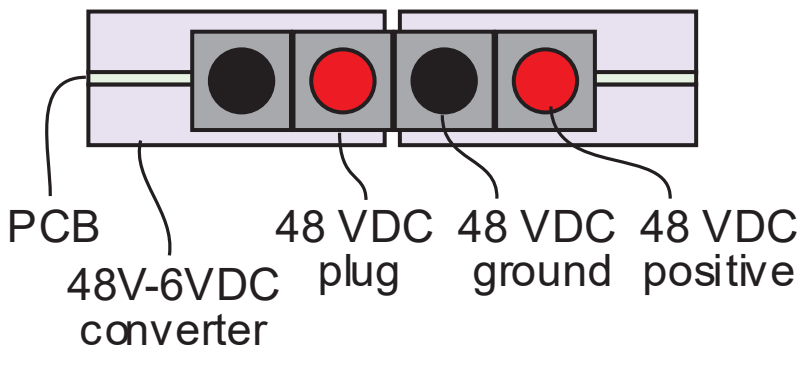

*Figure 11d:*
*Power connector end of PSU PCB*

Figure 11a shows a top view of a ZettaLith PSU PCB. The copper wire CGA columns connect the PSU PCB to the WSSCB. The copper busbars are separated by 50 µm thick polyimide film insulation. Each PSU printed circuit board contains 12× TDM2534xT power modules, 2× XDPE132G5C multiphase controllers, passive components, and 2× 48 VDC to 6 VDC fixed ratio converters. Power is connected by 48 VDC power socket and 48 VDC power plug, with the power cables being a 48 VDC positive wire and a 48 VDC ground wire. Channels allow pumped 2-PIC coolant to flow between rows of power modules.

Figure 11b shows a side view of a ZettaLith PSU PCB showing the same components as the top view.

Figure 11c shows an end view of a ZettaLith PSU PCB from the WSSCB end. The copper wire CGA columns connect the PSU PCB to the WSSCB. The power and ground busbars are separated by the polyimide film insulation.

Figure 11d shows an end view of a ZettaLith PSU PCB. From this view, the 48 VDC to 6 VDC fixed ratio converters are visible, as is the PSU printed circuit board. The 48 VDC power plug shows the 48 VDC positive wires and the 48 VDC ground wires.

**Table 14: ZettaLith power supply units (PSU)**

| Aspect | Value | Unit |
|---|---|---|
| ZettaLith TRIMERAs supplied | 2 | TRIMERAs |
| Number of PSUs | 86 | PSUs |
| Power per PSU | 980 | Watts |
| Current per PSU | 1,323 | Amps |
| Interface width | 48 | mm |
| Interface height | 11 | mm |
| Interface area | 528 | $mm^2$ |
| CGA spacing | 1 | mm |
| CGA columns | 528 | columns |
| CGA Ground columns | 264 | columns |
| CGA Power columns | 260 | columns |
| CGA Signal columns | 4 | columns |
| XDPE132G5C Multiphase controllers | 2 | chips |
| Multiphase controller phases | 16 | phases |
| Min. TDM2534xT power modules | 9 | modules |
| Actual TDM2534xT power modules | 12 | modules |
| Power modules in a ZettaLith | 1,032 | modules |
| Max Distance of TDM2534xT to SLD | 24 | mm |
| Current of power modules | 160 | Amps |
| Rows of power modules | 2 | rows |
| Length of power modules | 8 | mm |
| Length of power section of PCB | 16 | mm |
| Input voltage | 48 | Volts |
| Input current | 20 | Amps |
| Intermediate voltage | 6 | Volts |
| Intermediate current | 163 | Amps |
| HSC-IBC 8:1 converter power | 750 | Watts |
| HSC-IBC 8:1 converter module | 1 | module |
| PSU efficiency | 88% | |
| Input power of PSU | 1,114 | Watts |
| TRIMERA decoupling capacitance | 158 | µF |
| SLD decoupling capacitance | 57 | pF |

### 22.1 Busbars and lack of high current connectors

The power and ground busbars, and the various power busbars, are insulated from each other by 50 µm thick polyimide film. The ground busbars are 0.945 mm thick copper sheets accurately cut (wire EDM is recommended) into "L" shapes as shown in Figure 11b. Copper sheet which is accurately rolled to 0.945 mm thick is used so that when stacked with 50 µm polyimide film, the thickness equals the 1 mm spacing of the CGA columns, allowing 5 µm for adhesive.

The inside long edge of the L shape is chamfered to around 0.5 mm before it is reflow-soldered to the PSU PCBs so that there is no short circuit between the power and ground busbars. The power busbars are like the ground busbars, except the power busbars may be divided into multiple sub-busbars, each separated by 0.95 mm wide polyimide strips.

A custom busbar and connection system is required, as there are no commercially available solutions able to handle the high current and low-stress connections to a silicon wafer that are required. The PSU CGA pillars are soldered directly to the WSSB.

There is no plug and socket used, so the PSUs are not field replaceable. The reason that they are soldered to the WSSCB is that a standard connector able to handle the required current is far bulkier than the space available, and a connector designed to fit the space available would be a major source of failure.

### 22.2 Characteristics of the ZettaLith PSUs

Table 14 shows the basic characteristics of the precision power supply units (PSU) powering the ZettaLith and directly attached to the WSSCB. There are 86 PSUs each supplying 2 TRIMERAs. Each PSU supplies 980 Watts, in various power domains. Most of the power is for the 31,407 million active FP4 PEs, at 0.7 Volts. 1.1 Volts is used for much of the I/O such as UCIe 2.0 and the HBM4 interface, as well as for the HBM4 stacks themselves.

This example PSU uses the Infineon XDPE132G5C multiphase controller, and the Infineon TLVR TDM2534xT power modules for extremely fast transient response. There is a total of 1,032 TDM2534xT power modules, in the 86 PSUs connected to the WSSCB. Each of the 1,032 regulator modules are less than 24 mm from the active silicon that it powers. That distance is mostly through solid copper busbars.

The PSU is controlled by using the Power Management Bus (PMBus).

### 22.3 Power Losses

Table 15 shows Voltage drop and parasitic power losses of the SLD power supply power connections to the CMOS load on the SLD, and back again to the PSU PCB ground. This is the flow of positive holes – the electrons flow the other way.

Most of the voltage drop and power dissipation is in the PSU power and ground rails, as these are much longer than any other part of the interface.

| Table 15: Parasitic power losses of a TRIMERA stack between PSU power and ground | | | | | | | | | | |
|---|---|---|---|---|---|---|---|---|---|---|
| **Structure** | **Quantity** | **Current** | **Resistivity** | **Length** | **Area** | **Resistance** | **Voltage** | **Power** | **Total** | **Current Density** |
| **Units** | | **mA** | **nΩm** | **μm** | **μm²** | **mΩ** | **mV** | **μW** | **W** | **A/cm2** |
| PSU rails solder | 60 | 11,026 | 13.2 | 100 | 800,000 | 0.002 | 0.018 | 201 | 0.012 | 1,378 |
| PSU rails | 60 | 11,026 | 17.7 | 16,000 | 1,504,000 | 0.19 | 2.076 | 22,891 | 1.373 | 733 |
| CGA wires | 28,210 | 23.45 | 17.7 | 3,000 | 5,027 | 10.56 | 0.248 | 5.8 | 0.164 | 467 |
| CGA solder | 130 | 5,089 | 13.2 | 20 | 321,699 | 0.001 | 0.004 | 21 | 0.003 | 1,582 |
| SCB TSVs | 130 | 5,089 | 17.7 | 710 | 321,699 | 0.04 | 0.199 | 1,012 | 0.132 | 1,582 |
| SCB RDL | 13,000 | 50.89 | 17.7 | 40 | 225 | 3.15 | 0.160 | 8 | 0.106 | 22,617 |
| μbump solder | 264,000 | 2.51 | 13.2 | 3 | 113 | 0.35 | 0.001 | 0.002 | 0.001 | 2,216 |
| μbump CU pillar | 264,000 | 2.51 | 17.7 | 10 | 79 | 2.25 | 0.006 | 0.01 | 0.004 | 3,191 |
| BID metal stack | 264,000 | 2.51 | 17.7 | 11 | 16 | 12.48 | 0.031 | 0.1 | 0.021 | 15,662 |
| BID TSVs | 88,000 | 7.52 | 17.7 | 100 | 20 | 90.15 | 0.678 | 5.1 | 0.448 | 38,287 |
| HILT TSVs | 88,000 | 7.52 | 17.7 | 100 | 20 | 90.15 | 0.678 | 5.1 | 0.448 | 38,287 |
| HILT m. stack | 393,984 | 1.68 | 17.7 | 11 | 1 | 199.66 | 0.335 | 0.56 | 0.222 | 167,911 |
| SLD RDL | 393,984 | 1.68 | 17.7 | 100 | 40 | 44.25 | 0.074 | 0.125 | 0.049 | 4,198 |
| SLD metal stack | 393,984 | 1.68 | 17.7 | 11 | 1 | 199.66 | 0.335 | 0.56 | 0.222 | 167,911 |
| ▲ Power connection chain | | | | | | | | | | |
| █ Active load of CASCADE arrays in TRIMERA SLD | | | | | | | | | | |
| ▼ Ground connection chain (reverse of power connection chain, but wider) | | | | | | | | | | |
| SLD metal stack | 393,984 | 1.68 | 17.7 | 11 | 1 | 199.66 | 0.335 | 0.563 | 0.222 | 167,911 |
| SLD RDL | 393,984 | 1.68 | 17.7 | 100 | 40 | 44.25 | 0.074 | 0.1248 | 0.049 | 4,198 |
| HILT m. stack | 393,984 | 1.68 | 17.7 | 11 | 1 | 199.66 | 0.335 | 0.563 | 0.222 | 167,911 |
| HILT TSVs | 176,000 | 3.76 | 17.7 | 100 | 20 | 90.15 | 0.339 | 1.3 | 0.224 | 19,143 |
| BID TSVs | 176,000 | 3.76 | 17.7 | 100 | 20 | 90.15 | 0.339 | 1.3 | 0.224 | 19,143 |
| BID metal stack | 528,000 | 1.25 | 17.7 | 11 | 16 | 12.48 | 0.016 | 0.02 | 0.010 | 7,831 |
| μbump CU pillar | 528,000 | 1.25 | 17.7 | 10 | 79 | 2.25 | 0.003 | 0.00 | 0.002 | 1,595 |
| μbump solder | 528,000 | 1.25 | 13.2 | 3 | 113 | 0.35 | 0.000 | 0.001 | 0.000 | 1,108 |
| SCB RDL | 13,200 | 50.12 | 17.7 | 40 | 225 | 3.15 | 0.158 | 7.9 | 0.104 | 22,274 |
| SCB TSVs | 132 | 5,012 | 17.7 | 710 | 321,699 | 0.04 | 0.196 | 981 | 0.130 | 1,558 |
| CGA solder | 132 | 5,012 | 13.2 | 20 | 321,699 | 0.00 | 0.004 | 21 | 0.003 | 1,558 |
| CGA wires | 28,644 | 23.10 | 17.7 | 3,000 | 5,027 | 10.56 | 0.244 | 5.6 | 0.161 | 459 |
| PSU rails | 12 | 55,129 | 17.7 | 16,000 | 4,512,000 | 0.06 | 3.460 | 190,756 | 2.289 | 1,222 |
| PSU rails solder | 12 | 55,129 | 13.2 | 100 | 8,000,000 | 0.000 | 0.009 | 501 | 0.006 | 689 |
| **TRIMERA Total** | | | | | | | **10** | **mV** | **6.9** | **Watts** |
| **ZettaLith Total** | | | | | | | **10** | **mV** | **1.2** | **kW** |

The columns of this table are:

- **Structure**: this is the type of structure that the current flows through at this point in the connection chain.
- **Quantity**: this is the number of those structures that the current flows through in parallel for each SLD.
- **Current**: this is the current through each of those structures in mA.
- **Resistivity**: this is the resistivity of the material, in nΩ·m
- **Length**: this is the length of the current path through the structure, in μm.
- **Area**: this is the cross sectional area of the current path through the structure, in μm².
- **Resistance**: this is the resistance of the structure, in mΩ.
- **Voltage**: this is the voltage drop across the structure, in mV.
- **Power**: this is the parasitic power loss of the structure, in μW.
- **Total**: this is the total parasitic power loss of all the structures of this type in a single SLD, in Watts.
- **Current Density**: This is the current density in the structure, in A/cm². It is relevant for checking current density for potential electromigration problems.

The structures through which the current flows on the path from the PSU positive voltage to ground are:

- **PSU rails solder**: this is the soldered interface between the PCB and the solid copper rails carrying power to the WSSCB.
- **PSU rails**: these are the solid copper rails carrying power to the WSSCB.
- **CGA wires**: these each of the 217 copper wires forming a wire bundle that forms the CGA columns.
- **CGA solder**: this is the solder interface between the CGA columns and plating on top of the TSVs in the WSSCB.
- **SCB TSVs**: these are the TSVs in the WSSCB. The WSSCB is nearly full thickness silicon, and the TSVs are thick copper columns through the silicon matching the 1 mm pitch of the CGA columns.
- **SCB RDL**: these are the metallization columns through the RDL of the WSSCB.
- **μbump solder**: this is the thin solder layer joining the copper pillars of the microbumps to the landing pads on the

front surface of the WSSCB.

- **µbump CU pillar**: These are the copper pillars of the microbumps. They are formed on the undersurface of the BID, with one Cu pillar per WSSCB TSV.
- **BID metal stack**: this is the conventional metal stack of the mainstream CMOS BID wafer. Many metal columns are formed in the metal stack for each TSV, allowing routing between the columns.
- **BID TSVs**: these are the short and thin standard TSVs of the Base Interface die. As this is an active CMOS chip, TSVs consume area otherwise used for logic, so the total area % of TSVs is constrained.
- **HILT TSVs**: these are the short and thin standard TSVs of the HILT die. The BID and HILT wafers are back-to-back hybrid bonded, so a compliant redistribution layer (RDL) will be needed over the TSVs to prevent the thermal expansion of the entire copper TSV columns from interrupting the hybrid bonding process. Due to the RDLs, the TSVs of the HILT and BID wafers do not need to match (and may be required to anti-match, depending on the compliance of the RDLs).
- **HILT metal stack**: this is the normal metallization stack for power from HILT TSVs to the top level metallization of the HILT wafer which is hybrid bonded to the SLD wafer.
- **SLD RDL**: this is the redistribution layer of the SLD. TSMC a16 process has a thick RDL as the standard top layers. These RDLs also include decoupling capacitance.
- **SLD metal stack**: this is the normal metallization stack for power from the redistribution layer to the CMOS of the SLD. To maintain exact hard macro configuration for small groups of PEs, there are separate identical a power and ground stacks leading from the power and ground planes of the metallization down to those small groups of PEs.
- The power then reaches the **CMOS transistors** of the CASCADE arrays in the TRIMERA SLD, the active load where the power is to be delivered. The power dissipated by the CASCADE arrays is not a parasitic power loss, so it is not included in the total.

Power then returns to the ground of the PSU via the ground connection chain, which is essentially the reverse of the power connection chain. The number of ground connections is often greater than the number of power connections to reduce ground bounce.

Most of the voltage drop and power dissipation is in the PSU power and ground rails, as these are much longer than any other part of the interface.

A parasitic power loss of 1.0 kW in power distribution may seem excessive, but this is only 1.2% of the total ZettaLith power of 84 kW.

### 22.4 Electromigration

The Current Density column of Table 15 shows the current density through each structure in A/cm$^2$ . All the structures are made of copper except those identified as solder. The maximum current density for copper before electromigration is generally considered to be a problem is $10^6$ to $10^7$ A/cm$^2$. All the copper structures have current densities of less than $10^5$ A/cm$^2$, so are more than an order of magnitude below the threshold for the onset of electromigration.

Solder has an electromigration threshold of only around $10^4$ A/cm$^2$. The various solder connections are well within this limit.

Electromigration for the entire SLD die is easy to calculate due to its SHAPE architecture. All CASCADE array columns are identical.

## 23 ZettaLith High Current Density and Technique for PSU PCB Attachment

A fundamental challenge for ZettaLith implementation is delivering 114,000 Amps of precisely regulated fast response power to the computational elements.

### 23.1 CGA columns

Conventional CGA columns made of solder represent a critical failure point that would render the entire system non-functional, as they could catastrophically fail (melt) under ZettaLith's high current densities. This power delivery bottleneck represented a potential "showstopper" that could have invalidated the entire ZettaLith architecture.

The solution is a CGA column design being 217 fine copper wires in a hex-close pack configuration. Each 80 µm diameter wire contributes to a robust 640 µm copper column that simultaneously could provide:

- low resistance and low voltage drop of 0.25 mV;
- total power loss of all CGA columns of only 0.33 W;
- highs current-carrying capacity without electromigration failure;
- thermal-mechanical compliance to accommodate differential expansion;
- elimination of elastoplastic deformation common in solder columns; and
- sufficient structural integrity for reliable system assembly.

To manufacture these columns, continuous copper wire bundles are induction welded at intervals of approximately 4 mm. These welded sections are then cut through their centers and staked into the busbars. Small holes are drilled in the edge of the busbars where the CGA columns are to go. These holes are plastically enlarged by forcing hardened steel spikes into them, displacing the copper sideways. The CGA column is placed into the expanded copper hole, and the displaced copper is compressed back into place, trapping the CGA columns and forming a conductive path.

Table 16: CGA column structure

| Aspect | Value | Units |
|---|---|---|
| Diameter of CGA column | 640 | µm |
| Copper wire diameter | 80 | µm |
| Hex close pack configuration | | |
| Number of complete rings | 8 | rings |
| Number of copper wires | 217 | wires |

The CGA columns are precision-trimmed in a dedicated fixture to ensure accurate length and coplanarity. A high-temperature elastomer applied between the welded sections wicks between the 217 individual wires of a CGA column, preventing solder from later infiltrating the bundle during reflow to the WSSCB - thus preserving the critical wire flexibility required for reliable long-term operation.

### 23.2 PSU PCB Attach process

Each PCB undergoes final inspection and final electrical verification testing of voltage regulation and control systems. Verified PCBs are loaded into a precision-aligned mounting jig that maintains their positions without constraining the CGA columns. The jig assembly is dipped approximately 1 mm into a low-temperature tin-lead solder bath (Sn63/Pb37, melting point 183°C), applying a controlled amount of solder to all CGA column tips simultaneously. Alternatively, they may be printed with solder paste.

After WSSCB plasma cleaning, the complete PCB array is aligned to the WSSCB, forming all CGA connections simultaneously through low-temperature reflow that protects the attached chips and underfill materials. The 34°C melting point difference between SAC305 solder used for the PSU PCB assembly and SCB microbumps and the 183°C tin-lead solders would enable reliable attachment with adequate temperature margin. The total amount of lead used is extremely small compared to the entire system, so a RoHS exemption should be readily available.

This assembly sequence reduces populated WSSCB handling, would enable PCB inspection and repair before WSSCB attachment, limits high-temperature processes, controls solder volume, forms all CGA connections simultaneously, and reduces risk to the high value WSSCB assembly.

The multi-PCB architecture could provide distributed power delivery near the point of load, independent voltage regulation zones for WSSCB regions, PCB-level maintenance, redundant power paths through parallel CGA connections, and thermal management through the PCB attachment structure.

## 24 Cooling Requirements and 2-PIC JETSTREAM system

As is typical with modern electronic systems, power supply and the resultant necessary heat dissipation are limiting factors on system performance and size. That is certainly the case here. The ZettaLith system has a very high power density, and the waste heat must be efficiently removed.

ZettaLith's dense integration of computational elements creates significant thermal management challenges, with each SLD consuming approximately 458 Watts, resulting in a total power dissipation of around 84 kW in an extremely compact volume. The SLD TDP of 458 Watts is not particularly excessive, as some GPUs and advanced CPUs are already around 1,200 Watts. It is the high power density of 321 W/cm$^2$ that presents the problem.

Traditional cooling solutions such as forced air, direct liquid cooling, or two-phase immersion cooling are inadequate for managing the high thermal density of 321 W/cm$^2$ at the TRIMERA stack interfaces. This thermal challenge represents a major limitation in scaling transformer inference capabilities, as conventional cooling approaches cannot maintain acceptable junction temperatures at these power densities. The ability to operate at such high power densities is important for maximizing the ZettaLith performance. To maintain the advantage of all computation being in a single all-silicon domain, the entire 84 kW power required for ZettaLith computation is concentrated in a volume of only around 200 mm × 260 mm × 2 mm.

### 24.1 JETSTREAM Cooling

**Table 17: JETSTREAM cooling system**

| Aspect | Value | Units |
|---|---|---|
| 2-PIC coolant (Opteon 2P50) pressure | 100 | kPa |
| 2-PIC coolant density (ρ) | 1,456 | kg/m$^3$ |
| 2-PIC coolant specific heat capacity (cp) | 1,090 | J/kg·K |
| 2-PIC coolant thermal conductivity (κ) | 0.07 | W/(m·K) |
| 2-PIC coolant viscosity (μ) | 0.00062 | Pa·s |
| 2-PIC coolant surface tension (γ) | 0.011 | N/m |
| Heat to be removed (Q) | 84,305 | Watts |
| Incoming 2-PIC coolant temperature | 30 | °C |
| Outgoing 2-PIC coolant temperature | 49 | °C |
| Temperature difference (ΔT) | 19 | °C |
| Mass flow rate ($\dot{m} = Q/(cp \cdot \Delta T)$) | 4.07 | kg/s |
| Volume flow rate ($\dot{V} = Q/(\rho \cdot cp \cdot \Delta T)$) | 0.0028 | m$^3$/s |
| Volume flow rate in litres/minute | 168 | litres/min |
| Nozzle width | 11 | mm |
| Nozzle height | 0.5 | mm |
| Nozzle area | 5.5 | mm$^2$ |
| Total area of all nozzles (A) | 946 | mm$^2$ |
| Nozzle 2-PIC coolant velocity | 2.96 | m/s |
| Discharge coefficient (Cd) | 0.9 | |
| Pressure difference ($\Delta P = \dot{m}^2/(2\rho \cdot Cd^2 \cdot A^2)$) | 7.85 | kPa |
| 2-PIC coolant cycle time | 10 | seconds |
| 2-PIC coolant required to circulate | 41 | kg |
| 2-PIC coolant in chamber | 93 | kg |
| Pump redundancy | 3 | pumps |
| Pump motor power (each) | 2 | kW |

JET Surface Thermal Regulation via Evaporative Array Manifold (JETSTREAM) uses two-phase immersion cooling (2-PIC) with individual tuned submerged jets of liquid coolant directed to each logic chip stack on the ZettaLith WSSCB.

### 24.2 JETSTREAM manifold

To achieve the required mass flow rate evenly to each SLD or CPU, ZettaLith employs a separate 2-PIC coolant jet interfacing with the silicon heatsink fins etched into the back side of each TRIMERA stack. This would enable effective heat removal at the required power densities while maintaining acceptable junction temperatures across the entire WSSCB and its attached chip stacks.

To address potential local temperature non-uniformities across the WSSCB, the system includes a 3D-printed JETSTREAM manifold made of titanium powder fused via laser melting. This manifold is specifically designed to incorporate individually optimized nozzles to jet 2-PIC coolant evenly to each TRIMERA stack.

By jetting a carefully metered flow of 2-PIC coolant to each chip location, the JETSTREAM manifold ensures effectively identical coolant velocities and pressure drops to each TRIMERA stack, irrespective of their position on the WSSCB. As a result, heat removal remains consistent from die to die, avoiding the common problem of some chips receiving less coolant flow, or chips located at trailing edges of coolant flows receiving coolant already heated by chips closer to the coolant inlet, or of some chips being in thermal hot spots.

The uniform distribution of 2-PIC coolant by jets tuned by individual static 3D printed baffles bolsters the ability to operate each SLD at the high power densities described in this disclosure, without compromising reliability or performance due to uneven cooling.

## 24.3 Heat transfer

Table 17 shows various aspects of the ZettaLith JETSTREAM cooling system.

The back-side of the SOTA wafer is patterned with an array of deep channels defining heat-sink fins in silicon. The fins are etched to within approximately 25 µm of the CMOS layer to minimize temperature difference through the silicon.

The 2-PIC coolant is individually jetted directly into the heat-sink silicon fin arrays etched into each of the 172 SLDs. This could provide an optimal and consistent temperature and mass flow for every SLD. In comparison, most current systems flow coolant over a larger area, where chips nearer the coolant inlet receive "fresh" coolant, while chips closer to the exit receive coolant already heated by prior chips. This results in hot-spots in the design, which ZettaLith eliminates. The HBM4 stacks generate comparatively little heat and are cooled by minor 2-PIC coolant flow patterns of each nozzle.

A precision 3D-printed JETSTREAM manifold manages the flow of 2-PIC coolant to and from all 172 WSSCB locations for TRIMERA stacks and CPUs. The JETSTREAM manifold is manufactured using additive manufacturing of metal (e.g. laser melting of titanium powder) that has a very high precision and rigidity, and minimum interaction with 2-PIC coolant.

The complex internal geometry of the JETSTREAM manifold incorporates flow distribution channels and 3D printed baffles. These are designed and optimized using computational multiphysics simulation in ANSYS or other suitable engineering simulation software to ensure uniform 2-PIC coolant delivery jetted to each TRIMERA stack.

This optimization process integrates thermal, mechanical, and fluidic simulations to achieve optimal flow distribution across all chip locations, with individually optimized baffle and/or nozzle structures for each SLD position on the WSSCB to ensure the appropriate 2-PIC coolant flow. The CPU logic stacks will consume a different amount of power than the CASCADE arrays, and this difference can be accommodated in the JETSTREAM manifold design.

**Table 18: 2-PIC PCHE heat exchanger**

| Aspect | Value | Units |
|---|---|---|
| ZettaLith heat to be removed | 84,305 | Watts |
| PSU heat to be removed | 11,496 | Watts |
| Total heat to be removed (Q) | 95,801 | Watts |
| Condensation heat transfer coefficient (h) | 50,000 | W/(m²·K) |
| Opteon 2P50 boiling point | 49 | °C |
| Average condenser temperature | 30 | °C |
| Opteon temperature difference (ΔT) | 19 | °C |
| 2-PIC heat exchange area (A=Q/(h·ΔT)) | 0.1 | m² |
| Water inlet temperature | 25 | °C |
| Water outlet temperature | 35 | °C |
| Water temperature difference (ΔT) | 10 | °C |
| Water heat transfer coefficient (U) | 2,000 | W/(m²·K) |
| Water heat exchange area (A=Q/(U·ΔT)) | 4.8 | m² |
| Maximum of water and Opteon PCHE area | 4.8 | m² |
| Channel surface area density | 3,000 | m²/m³ |
| PCHE volume | 0.00160 | m³ |
| Cylindrical PCHE diameter | 380 | mm |
| Cylindrical PCHE minimum height | 14 | mm |

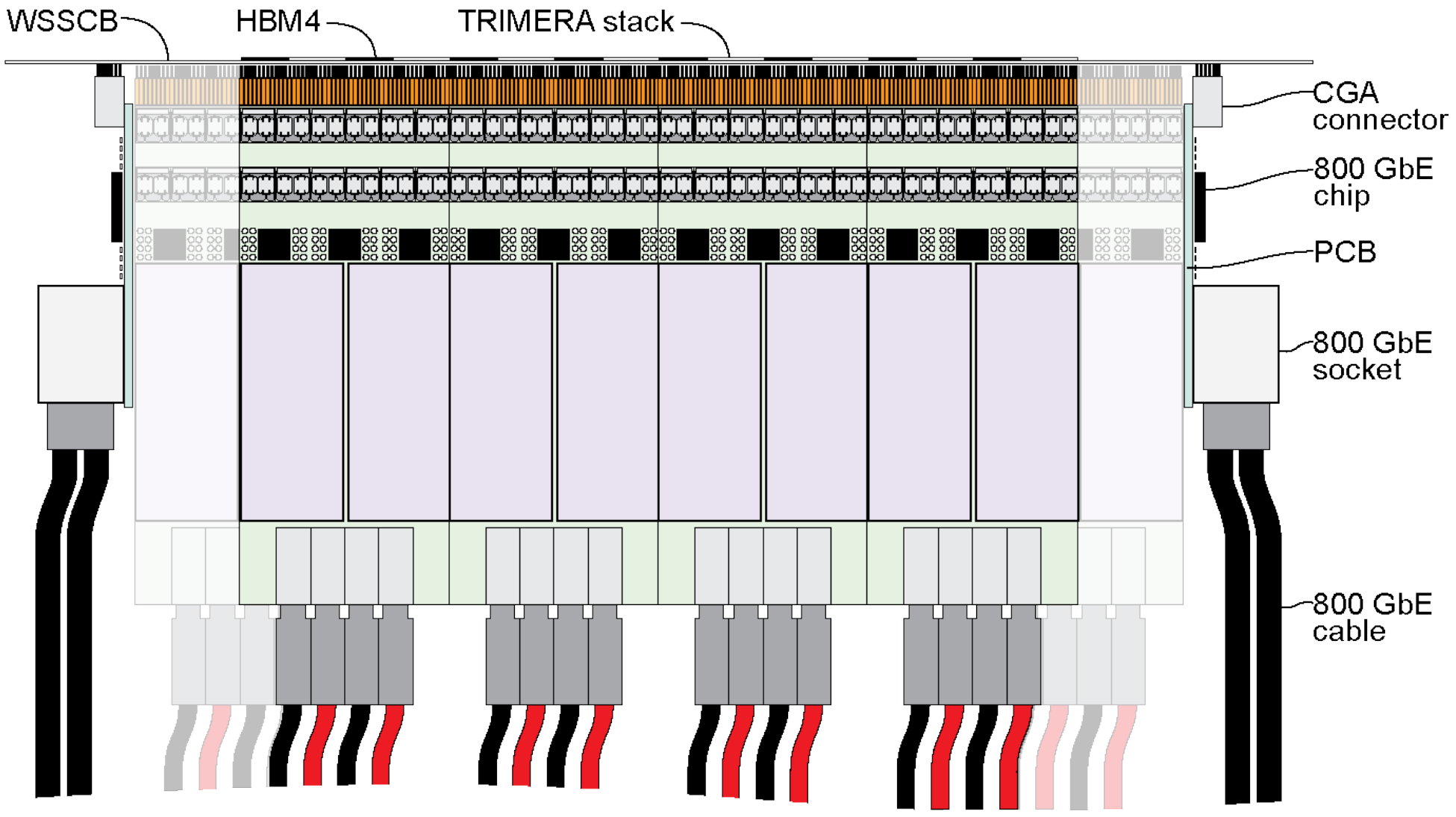

*Figure 12a: ZettaLith with power supplies: PSU front view*

The JETSTREAM cooling system has redundant pumps circulating 2-PIC coolant through the PCHE and JETSTREAM manifold. The system includes three high-reliability pumps, each able to pump the entire required 2-PIC coolant flow. Thus, any pump can fail without causing a system failure. The faulty pump can then be replaced during regular system maintenance.

If the valves and sealing design can be made sufficiently reliable, then the pumps can be made hot-swappable. However, the current design uses high reliability pumps that are replaced in maintenance cycles, to avoid potential problems with hot-swapability.

### 24.4 ZettaLith PSU stack front view

Figure 12a shows a front view of a ZettaLith power supply array showing a row of PSU PCBs connected to a WSSCB wafer, with attached HBM4 stacks and TRIMERA stacks. Parts of a second row of PSU PCBs are visible behind the first row as the array is not square, to better fit the circular 300 mm wafer used in WSSCB fabrication. There are a total of 86 PSU PCBs attached to the WSSCB.

Figure 12a also shows a side view of 800 GbE PCBs. These PCBs are connected by CGA connectors to the WSSCB, through which UCIe 2.0 connections connect 800 GbE controllers to the BID dies on the WSSCB. These UCIe 2.0 connections are programmed for reduced speed compared to the UCIe 2.0 connections on the WSSCB, as they go through WSSCB vias, CGA columns, and onto PCBs. The 800 GbE connections are for expansion and are not involved in normal transformer inference. The 800 GbE PCB is connected by 800 GbE sockets to 800 GbE cables leading to connectors through the 2-PIC tank walls, and thence to a TOR switch (not shown).

In Figure 12a, the PCIe 6.0 PCBs are not shown, as these would obscure the view of the PSU PCBs.

### 24.5 ZettaLith PSU stack side view

Figure 12b shows a side view of a ZettaLith power supply array showing a row of side views of PSU PCBs connected to a WSSCB wafer, with attached TRIMERA stacks. The HBM4 stacks are obscured in this view.

Figure 12b also shows a side view of PCIe 6.0 PCBs. These PCBs are connected by CGA connectors to the WSSCB, through which UCIe 2.0 connections connect PCIe 6.0 controllers to the CPU dies on the WSSCB. As with the 800 GbE connections, these UCIe 2.0 connections are programmed for reduced speed compared to the UCIe 2.0 data fabric on the WSSCB. The PCB is connected by PCIe 6.0 sockets to PCIe 6.0 cables leading to connectors through the pressure vessel walls, and thence to SSDs and other PCIe 6.0 equipment as required (not shown).

In Figure 12b, the 800 GbE PCBs are not shown, as these would obscure the side views of the PSU PCBs.

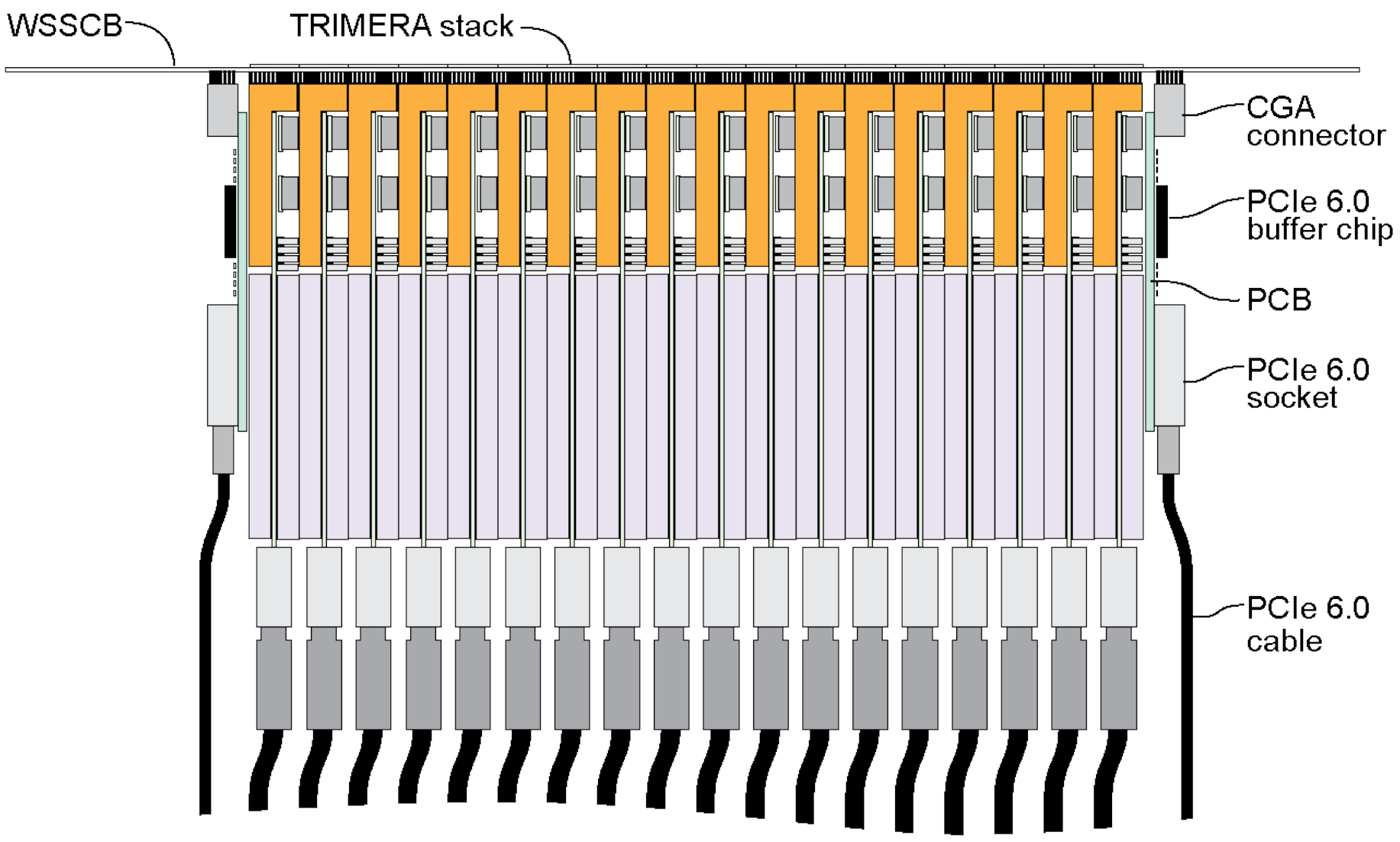

*Figure 12b: ZettaLith with power supplies: PSU side view*

### 24.6 PSU stack end view

Figure 13 shows an end view of a ZettaLith PSU PCB array, including an end view of the PSU PCBs. The end view of 800 GbE PCBs with 800 GbE cables is shown. Also shown is the end view of PCIe 6.0 PCBs and PCIe 6.0 connectors. The WSSCB wafer appears in the background.

### 24.7 2-PIC Fluids

The selection of an appropriate dielectric coolant is critical for 2-PIC efficacy and safety. Historically, the 2-PIC heavily relied on engineered fluids from 3M™, namely the Novec™ and Fluorinert™ product lines. These fluorinated compounds (including fluorocarbons, hydrofluoroethers, and fluoroketones) offered advantageous properties such as:

1. Excellent dielectric strength (electrical insulation).
2. Tailored boiling points suitable for passive heat transfer from typical semiconductor operating temperatures (e.g., ~50-60°C).
3. Good material compatibility with data center hardware.
4. Non-flammability.

However, many of these legacy fluids fall under the regulatory definition of perfluoroalkyl and polyfluoroalkyl substances (PFAS). Due to concerns regarding their environmental persistence and potential health risks, regulatory bodies (e.g., US EPA, ECHA) are increasingly restricting PFAS use. Compounding this, 3M announced its intention to exit all PFAS manufacturing by the end of 2025, including the Novec™ and Fluorinert™ ranges, prompting an urgent industry transition towards sustainable alternatives.

### 24.8 Chemours Opteon™ 2P50

The 2-PIC coolant market is currently adapting to this shift. While no single fluid has achieved universal dominance as a replacement, several alternatives are emerging. Chemours, for example, is actively developing and positioning next-generation coolants. Chemours Opteon™ 2P50 is a proprietary hydrofluoroolefin (HFO)-based fluid specifically engineered for 2-PIC applications. It features a very low Global Warming

Potential (GWP ≈ 10) and zero Ozone Depletion Potential (ODP). Following successful trials (e.g., with NTT Data) and regulatory approvals, it is anticipated to become commercially available in the 2025 timeframe.

The availability of this and other emerging low-GWP, PFAS-free (or less regulated PFAS) dielectric coolants reinforces the viability of 2-PIC as a practical cooling solution for a power-scaled ZettaLith implementation. This document assumes the use of Opteon 2P50 but can be readily changed to alternative fluids. Such a change may require redesign of the JETSTREAM manifold.

## 24.9 Mechanical configuration

Figure 14 illustrates the physical configuration of the JETSTREAM version of ZettaLith.

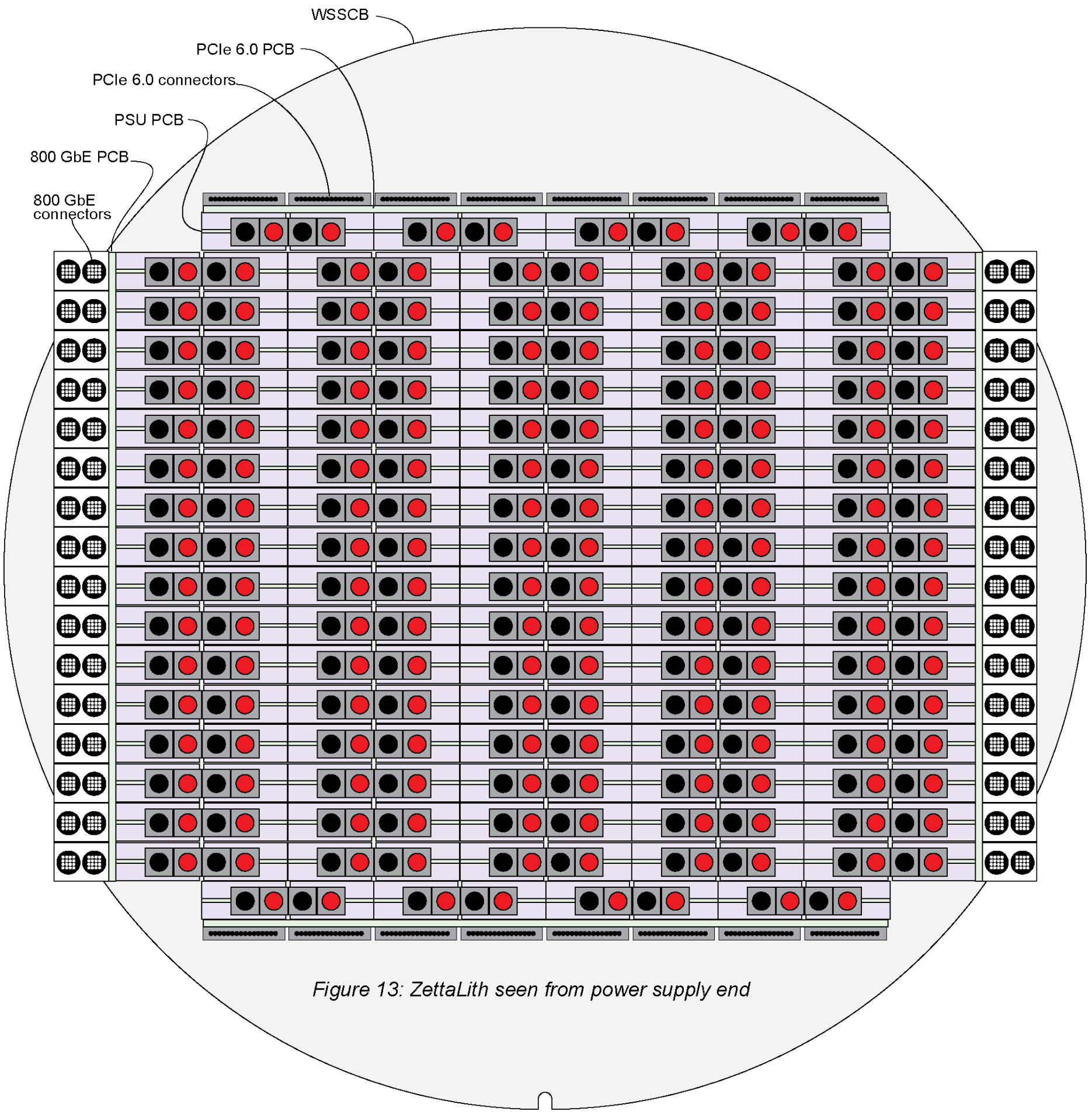

*Figure 13: ZettaLith seen from power supply end*

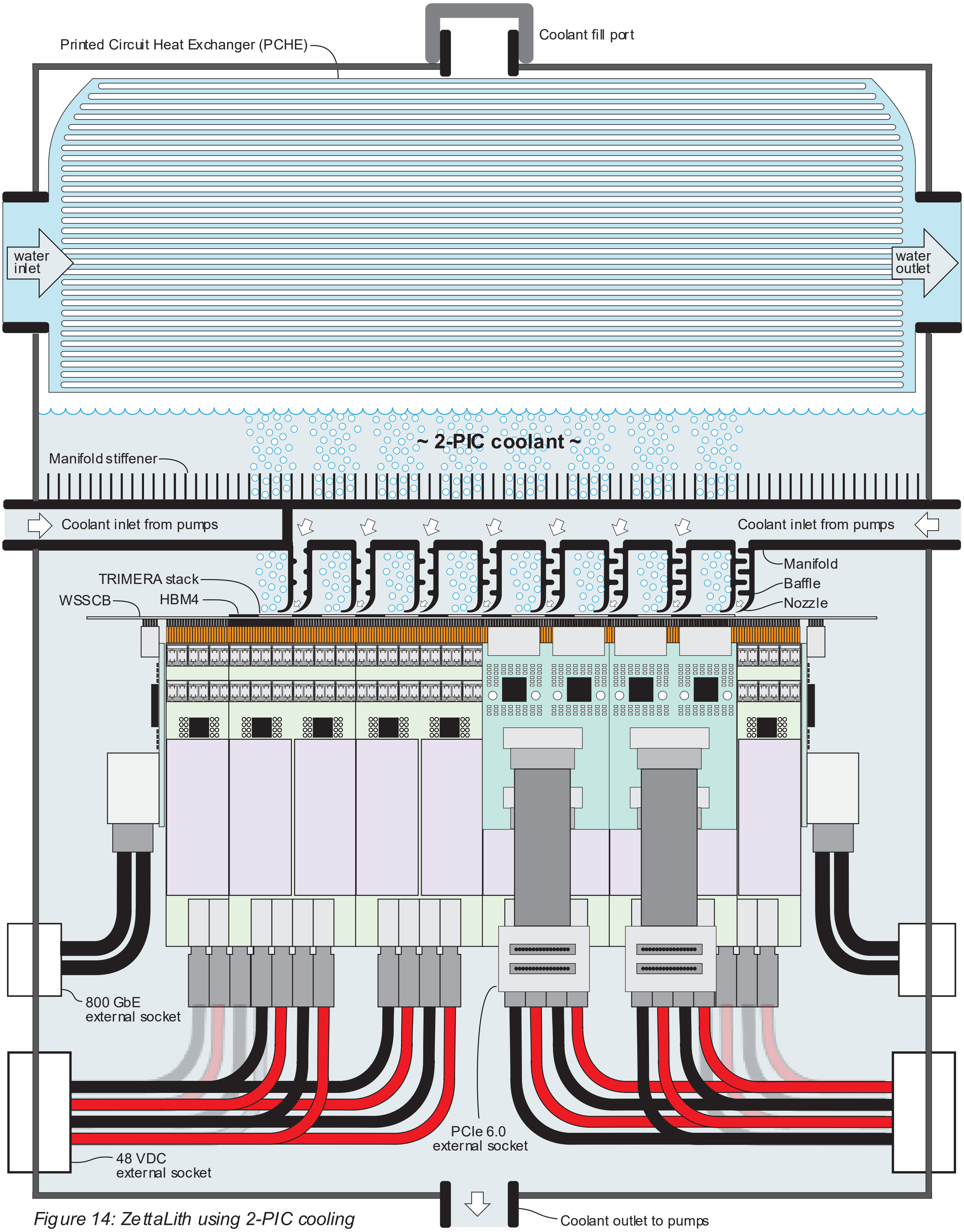

*Figure 14: ZettaLith using 2-PIC cooling*

# 25 Potential Future Logic Technologies for Enhanced ZettaLith Performance

The ZettaLith architecture, when implemented using projected near-term CMOS technology like TSMC's A16 node, is anticipated to be power-limited rather than compute-limited. Consequently, its inference performance could potentially be substantially higher if not constrained by thermal design power (TDP) limits. The advent of more power-efficient logic technologies beyond conventional silicon CMOS presents a significant opportunity. Such advancements could allow future ZettaLith iterations to either:

- **Reduce overall power** consumption while maintaining the baseline performance level.
- **Increase inference performance** significantly within the same power envelope.
- **Achieve a combination** of improved performance and reduced power.

Crucially, the ZettaLith architecture presented in this document *does not rely on any post-CMOS technologies.* The performance estimates are based on technologies like advanced CMOS (e.g., A16) and mature silicon lithography and SOTA 3D packaging technologies.

However, ZettaLith's modularity and system partitioning are deliberately designed to accommodate future post-CMOS advancements:

- **SHAPE**: Could provide the framework for integrating diverse chiplet types into the WSSCB.
- **WSSCB**: Acts as a high-bandwidth integration platform, capable of connecting numerous advanced logic die (SLDs).

ZettaLith separates the types of structures that are required in a complete system:

- **Transformer parameters** reside in standard HBM stacks.
- **Local memory** (replacing SRAM) is in the HILT layer.
- **I/O, mixed-signal, and power** delivery functions are in the BID layer.

This separation allows the core logic (TRIMERA SLDs) to be potentially replaced with post-CMOS equivalents without requiring those new technologies to immediately replicate complex memory or I/O functions.

## 25.1 Manufacturing compatibility

Assuming future post-CMOS logic continues to use silicon wafers as a mechanical substrate and supports hybrid bonding, integration into the ZettaLith stack should be feasible. Once PEs can be effectively implemented in a new technology, corresponding TRIMERA SLDs can be manufactured and integrated. Hybrid bonding is a relatively low temperature process, with initial contact at room temperature and annealing at 200°C to 250°C, resulting in potential compatibility with new materials such as bismuth, which has a melting point of 271.5°C. Care must be taken with the dual damascene metallization of the SLD, as dielectric layers can be deposited between 200°C and 450°C depending on deposition method. A low temperature PECVD processes should be used.

## 25.2 Emerging post-CMOS technologies

Several emerging device technologies are under active research and development. These hold the potential to drastically reduce the power requirements and associated heat dissipation challenges inherent in scaling compute density. Key candidates include:

## 25.3 Bismuth-based gate-all-around (GAA) FETs

Recent advances in 2D semiconductor materials and GAA architectures offer a promising path beyond silicon. (Tang et al., 2025) reported wafer-scale integration of bismuth FETs using epitaxially grown $Bi_2O_2Se$ as the 2D channel material and $Bi_2SeO_5$ as a high-κ native oxide dielectric. These FETS had a with a 30 nm gate length, operated at 0.5 V, with a low intrinsic delay of 1.9 ps and an energy-delay product of $1.84 \times 10^{-27}$ Js $\mu m^{-1}$.

## 25.4 Spiking neural networks (SNNs)

SNNs represent a different computational paradigm inspired by biological neurons. They process information using discrete, asynchronous events (spikes) rather than continuous values, primarily relying on temporal coding and integrate-and-fire dynamics. Implementing SNNs on ZettaLith's SLDs could become a compelling route to drastically reduce system power consumption, however, the entire TRIMERA stack would need to be reinvented.

## 25.5 Carbon nanotube field-effect transistors (CNTFETs)

CNTFETs are a relatively mature post-CMOS candidate. While basic commercial processes might emerge in the mid-to-late 2020s, high-yield manufacturing suitable for complex SLDs will likely take longer. Early work includes (Tans et al., 1998). (Liu et al., 2024) provides a recent manufacturing overview.

## 25.6 Reversible logic

Reversible computing aims to overcome the fundamental Landauer limit on energy dissipation by designing logic operations that are logically and thermodynamically reversible, theoretically allowing energy used in a computation step to be recovered. While offering profound long-term potential for energy reduction, reversible logic faces immense practical hurdles. Current approaches like Vaire Computing's resonator-based design are incompatible with ZettaLith's density requirements due to large physical footprints (e.g., 34 µm diameter resonator). Should breakthroughs make reversible logic dense and practical, ZettaLith's modular SLD approach could potentially integrate it, but this is considered a long-term prospect.

### 25.7 Transition metal dichalcogenides (TMDs)

TMDs and other 2D materials beyond graphene represent another active research frontier. Commercial viability for complex logic is uncertain, likely post-2030. Seminal work includes (Radisavljevic et al., 2011), with a recent perspective in (Li et al., 2024).

### 25.8 Tunneling field-effect transistors (TFETs)

TFETs operate based on quantum mechanical tunneling, allowing them to potentially achieve subthreshold slopes steeper than the 60 mV/decade Boltzmann limit of MOSFETs. TFETs could offer substantial power savings if ON-current and integration challenges are resolved. Foundational work includes (Zhang et al., 2006), with recent III-V analysis in (Chen et al., 2024).

### 25.9 Superconducting electronics

Logic families like Rapid Single Flux Quantum (RSFQ) operate using quantized magnetic flux pulses. Requires cryogenic cooling (typically ~4K), adding significant system complexity, cost, and infrastructure overhead. Furthermore, cryogenic cooling is fundamentally incompatible with 2-PIC cooling used in JETSTREAM, and the entire packaging, power supply, and cooling technologies of ZettaLith would need to be re-imagined. Seminal work: (Likharev and Semenov, 1991); recent review: (Shibata et al., 2023).

### 25.10 Graphene transistors

While graphene exhibits exceptional electron mobility, its lack of a natural bandgap makes it inherently difficult to use for conventional digital logic (achieving a high ON/OFF current ratio is problematic). Graphene is currently more promising for analog/RF applications or as an advanced interconnect material (replacing copper) rather than as a direct replacement for silicon in ZettaLith's logic PEs. Foundational work: (Schwierz, 2010); recent perspective: (Mak et al., 2024).

### 25.11 Accommodating High Defect Densities in Early Post-CMOS Technologies

A major challenge for any nascent semiconductor technology is achieving high manufacturing yields. Early post-CMOS processes will likely exhibit significantly higher defect densities than even bleeding-edge CMOS.

- **ZettaLith's Resilience**: The SHAPE architecture and CREST fault tolerance mechanism are inherently suited to managing this. While the baseline design assumes bleeding-edge CMOS-level defect rates (allocating 16 spare columns per 8,192-column CASCADE array), this allocation can be readily increased.
- **Example Scenario**: If, for example, 10% of a post-CMOS SLD's area were dedicated to spare columns, each CASCADE array could have approximately 900 spare columns. This would provide tolerance of potentially thousands of defects per square centimeter. Specifically, up to 120,705 uncorrelated defects/cm² could theoretically be tolerated across the 172,608 spare CASCADE columns on an SLD chip, though correlated defects would reduce this substantially in practice.
- **Implication**: This adaptability significantly lowers the barrier for adopting promising but initially high-defect-rate post-CMOS technologies within the ZettaLith framework, potentially accelerating their deployment for high-performance computing.

Table 20: ZettaLith comparison to SOTA GPU rack

| Aspect | GPU rack | ZettaLith | Units | Ratio |
|---|---|---|---|---|
| Number of accelerators | 72 | 156 | chips | 2.17 |
| Number of CPUs | 36 | 16 | chips | 0.44 |
| Inference (FP4 sparse) | 1,440 | 1,507,534 | PFLOPS | 1047 |
| Inference (FP4 dense) | 720 | 753,767 | PFLOPS | 1047 |
| Active PE cycles per second | 360 | 376,883 | PHz | 1047 |
| Accelerator HBM stacks | 576 | 156 | HBMs | 0.27 |
| Accelerator HBM memory | 13,824 | 9,984 | GBytes | 0.72 |
| Total DRAM chips or stacks | 1,152 | 172 | DRAM | 0.15 |
| Accelerator memory bandwidth | 576 | 256 | TB/s | 0.44 |
| Max in-rack transformer inference | 28 | 20 | TP | 0.72 |
| Weights bandwidth from HBM | 1,152 | 512 | TW/s | 0.44 |
| Interchip data fabric | Proprietary | UCIe 2.0 | type | NA |
| Bandwidth of interchip data fabric | 259 | 7,800 | TB/s | 30 |
| Ethernet network | 400 GbE | 800 GbE | Gb/s | 2.00 |
| Ethernet connections | 72 | 32 | GbEs | 0.44 |
| Ethernet bandwidth | 28,800 | 25,600 | Gb/s | 0.89 |
| PCIe for SSDs etc. | PCIe 5.0 | PCIe 6.0 | version | NA |
| PCIe links | 36 | 16 | links | 0.44 |
| PCIe bandwidth | 2,304 | 2,048 | GB/s | 0.89 |
| Total active PEs | 217 | 31,407 | million | 145 |
| Power efficiency | 12 | 17,882 | TF/W | 1490 |
| PE power | 86 | 72 | kW | 0.83 |
| CPU and other power | 34 | 13 | kW | 0.38 |
| Total power consumption | 120 | 84 | kW | 0.70 |
| Cooling | Water | 2-PIC | type | NA |

## 26 Comparison to Current GPU Rack

ZettaLith is not a general purpose GPU, so comparison to GPUs is *only* meaningful if the application is *exclusively* inference of transformers in FP4 format. ZettaLith is a hardwired array of CASCADE arrays dedicated to transformer inference with FP4 weights. ZettaLith is not intended for non-inference workloads.

At the system level, the key performance metrics in Table 20 demonstrate ZettaLith's capabilities. Table 20 shows a balanced system which could provide the memory capacity, memory bandwidth, CPU capacity, CPU memory, chip-to-chip fabric bandwidth and the fabric topology required for the system to keep up with the TRIMERA arrays.

## 27 ExaLith: ZettaLith Chips on the Desktop

While the full ZettaLith architecture targets the large scale and performance demands of hyperscale data center, smaller desktop systems are also useful.

Users might include small-to-medium businesses (SMBs), research institutions, AI developers, and creative professionals who require substantial local AI inference capabilities but lack the budget and infrastructure for multi-rack GPU clusters or dedicated data center solutions.

ExaLith is conceived as a direct application of the core ZettaLith chips and technologies to provide exascale-class FP4 inference performance within the familiar form factor and power envelope of a high-end workstation or desktop PC component.

There are several feasible formats using ZettaLith silicon in desktop, workstation, or departmental environments:

- **PCIe card:** for integration into standard workstations and servers.
- **AI Workstation:** a complete, pre-integrated desktop/tower system built around one or more ExaLith accelerators.
- **Network Attached AI Accelerator (NAA)**: a standalone, network-accessible box containing a single ExaLith accelerator.
- **Multi-Accelerator Appliance**: a dedicated chassis housing multiple (e.g., 2-8) ExaLith accelerators for shared, high-throughput network access.
- **Server Blade/Module**: Integrating the ExaLith accelerator onto a standard blade form factor for denser rack deployments. This format is particularly suited for private clouds which don't require full ZettaLith performance.

The power consumption of a ZettaLith TRIMERA stack is too high to be used in a notebook computer. For this application, new silicon would be required, and the NEXAI concept is more appropriate.

### 27.1 ExaLith PCIe card

The core concept of ExaLith is to leverage the modularity and efficiency of the ZettaLith architecture, specifically utilizing the chips to be developed for ZettaLith (the SLD, HILT, BID and CPU, dies as defined previously) and most of the software stack, integrated onto a single PCIe board.

This approach crucially avoids the need for fundamentally new silicon development for the core compute elements, instead focusing on innovative integration and memory configuration at the board level.

ExaLith uses a high-performance PCIe printed circuit board (PCB) which serves as a carrier for a compact Silicon Circuit Board (SCB) module. This SCB module, fabricated using ZettaLith's WSSCB process but on a smaller scale, integrates the core compute and memory elements. A typical configuration places the following components onto this SCB module:

**Table 21: ExaLith PCIe card characteristics**

| Aspect | Value | Units |
|---|---|---|
| TRIMERA stack on SCB | 1 | TRIMERA stack |
| CPU stack on SCB | 1 | CPU stack |
| Operational clock frequency | 8 | GHz |
| Total active PEs in ExaLith | 201 | million PEs |
| Performance of 1 PE (1 MAC = 2 Ops) | 16 | GFLOPS |
| ExaLith performance (sparse) | 6.4 | exaFLOPS |
| ExaLith performance (dense) | 3.22 | exaFLOPS |
| FP4 parameters in memory (HBF) | 1 | TP |
| Minimum latency for 1 TP LLM | 0.5 | seconds |
| TRIMERA-CPU data link (UCIe on SCB) | 39 | TB/s |
| HBM4 memory | 16 | GB |
| HBM4 bandwidth | 1.64 | TB/s |
| HBF memory | 512 | GB |
| HBF bandwidth | 1 | TB/s |
| PCIe 6.0 bandwidth | 128 | GB/sec |
| TRIMERA SLD power density | 214 | W/cm$^2$ |
| ExaLith CASCADE array power | 306 | W |
| Power limited CPU stack power | 130 | W |
| HBM power | 30 | W |
| HBF power | 30 | W |
| ExaLith total compute power | 496 | W |
| Multiphase buck converter efficiency | 92% | |
| Total PCIe card power | 539 | W |

- A CPU stack, paired with high-bandwidth memory (HBM4) to run a subset of the transformer inference code developed for ZettaLith, and to store KV caches, intermediate activations, and frequently accessed data, mirroring a portion of the full ZettaLith configuration.
- a TRIMERA stack is coupled with emerging High-Bandwidth Flash (HBF) memory technology (such as that announced by SanDisk). This HBF stack serves as a large, cost-effective, and non-volatile repository primarily for storing the vast parameter sets of trillion-parameter-scale transformer models.

This HBM+HBF combination allows ExaLith to inference transformer models up to 1 trillion FP4 parameters locally achieving a target inference performance of around 3.22 exaFLOPS (dense FP4, approximately 6.4 exaFLOPS sparse) - performance comparable to multiple racks of current-generation AI accelerators - within a single PCIe card footprint.

Performance projections, memory configurations, power breakdowns, and cost estimates for an ExaLith PCIe card are provided in Table 21.

### 27.2 High bandwidth within ExaLith

The SCB module facilitates a high-bandwidth connection, nominally 39 TB/s using UCIe 2.0 over dense RDL wiring, directly between the Base Interface Dies (BIDs) of the CPU stack and the TRIMERA stack, enabling rapid data exchange. This bandwidth is far higher than the combined HBM and HBF bandwidths, effectively making them directly part of the TRIMERA stack high speed memory environment. The TRIMERA stack can also communicate with CPU cache

SRAM at this speed.

## 27.3 Power consumption

Achieving this level of performance within a PCIe card necessitates careful thermal and power management. Calculated ExaLith total board power is 539 W, under the 600 W limit for PCIe cards. Cooling is envisioned using advanced air-cooling solutions incorporating phase-change heat pipe technology and high-efficiency fans, like those employed in flagship consumer and workstation GPU cards. While demanding, this remains within the established capabilities of desktop/workstation thermal design, avoiding the 2-PIC JETSTREAM cooling requirements of the full ZettaLith system.

12V power delivery utilizes the 16-pin 12VHPWR connector from a ATX 3.0 compliant PSU. The 12V input is regulated to the TRIMERA, CPU, HBM, and HBF requirements by an on-board multiphase controller (such as the Infineon XDPE192C4C programable digital multi-phase controller) with 12 interleaved phases driving power stages such as the Infineon TDA21590, Monolithic Power MP86956, or Renesas RAA220105.

Multiphase buck converters are selected for their high efficiency and cost-effectiveness at PCIe power levels, compared to the TLVRs chosen for ZettaLith's high current regulation needs.

## 27.4 Use cases

By offering performance equivalent to multi-million-dollar GPU clusters in a single PCIe card, ExaLith fundamentally changes the economics for numerous users. Potential use cases include:

- **SMBs**: Running complex, proprietary AI models for data analysis, customer service, or internal process automation without relying on expensive cloud services or compromising data privacy.
- **Researchers & Academia**: Enabling experimentation and development with state-of-the-art large models on local hardware, accelerating research cycles. While ExaLith can't be used for transformer training, it is well suited to developing agentic and reasoning models using pre-trained LLMs.
- **AI Developers**: Providing a powerful local platform for developing, testing, and debugging AI applications destined for various deployment targets.
- **Creative Professionals**: Facilitating high-fidelity, low-latency local generation of AI-driven content (images, video, audio, 3D assets) enabling faster creative iteration.
- **Specialized Workstations**: Powering domain-specific AI tasks in fields like financial modeling, medical image analysis, scientific simulation, and engineering design requiring secure, high-performance local AI processing. Such a workstation would require HPC computing separate from ExaLith. This could be provided by a GPU card.
- **Smart games**: Adding a GPU card (or integrating mid-level GPU capabilities in the ExaLith) could enable a new class of games which focus on intelligent interactions with virtual people or virtual societies.

## 27.5 ExaLith PCIe card block diagram

Figure 15 is a high-level block diagram of an ExaLith PCIe card integration. A Silicon Circuit Board (SCB) essentially using two modules of ZettaLith WSSCB contains four chip stacks:

1. A TRIMERA stack using a BID, a HILT die and an SLD with FP4 CASCADE PEs.
2. A HBF stack connected to the TRIMERA stack BID by HBF channels (which are almost identical to HBM channels).
3. A CPU stack using a BID (identical to the TRIMERA BID), an optional SRAM cache die, and a CPU die. If an SRAM cache die is not used, a smaller amount of SRAM cache would be implemented directly on the CPU die, and the CPU die takes the place of the cache SRAM die.
4. A HBM stack connected to the CPU stack BID by HBM channels.

The TRIMERA BID and CPU BID are connected by the vertical UCIe connections between two BIDs. This could provide a 39 TB/s BID-BID data link, as it uses the same ultra-high bandwidth UCIe 2.0 data fabric connection used in ZettaLith. 39 TB/s is far higher than the sum of the HBM and HBF bandwidths, and this would enable the TRIMERA stack to utilize the CPU cache SRAM at very high bandwidth.

The ExaLith PCB contains a UCIe to PCIe conversion chiplet, which is used to connect the ExaLith computational engine on the SCB to the PCIe edge connector. The UCIe to PCIe conversion chiplet is preferably the same as used for ZettaLith. Although not currently commercially available as off-the-shelf components, UCIe-PCIe chiplets are likely to be available in time for use by ZettaLith and ExaLith. If not, an appropriate PCIe 6.0 interface can be integrated in the BID design, at the expense of some horizontal UCIe 2.0 fabric bandwidth.

Power supply is standard for a 600W PCIe card. 12 V DC Power is provided from the system PSU via a 12VHPWR connector. A multiphase controller drives ~12 power stages in multiple phases.

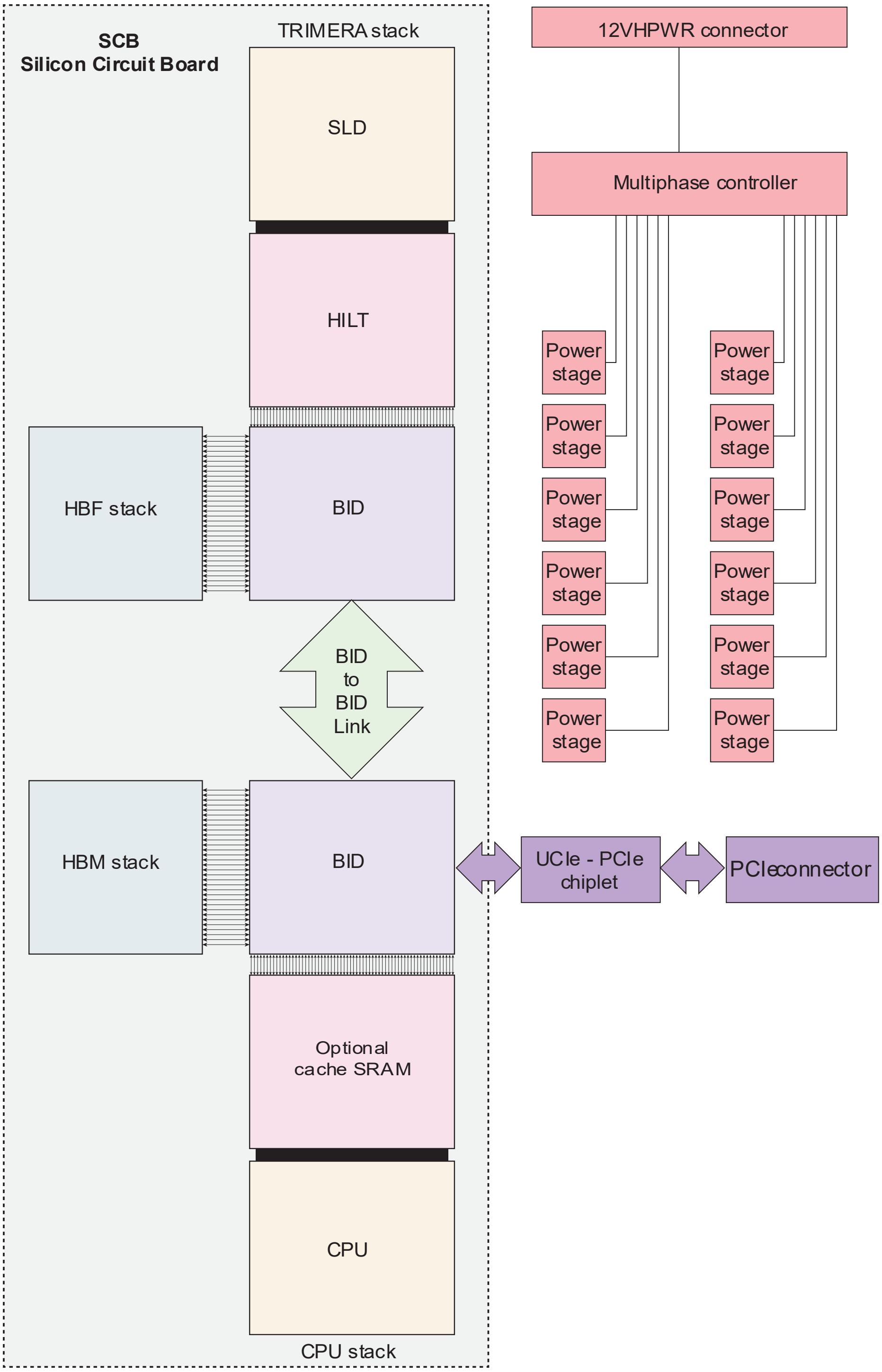

*Figure 16: Block diagram of ExaLith PCIe card*

## 28 NEXAI: ZettaLith at the Edge

The exponential growth of generative AI has created enormous demand for high-performance inference engines in edge devices - autonomous cars, humanoid robots, medical systems, smart PCs, factory automation, and augmented reality platforms. While data-center solutions like ZettaLith leverage 156 HBM stacks and 12 GHz CASCADE compute arrays to deliver zettaFLOPS-scale performance, edge devices face strict power, thermal, and cost constraints.

This proposed NEXAI (Neural Engine eXecution Accelerator for Inference) IP block adapts some of ZettaLith's core approaches - CASCADE, SHAPE, HILT, and CREST - together with SanDisk's HBF into a compact, edge-optimized IP block that would enable AGI-scale transformer inference for next-generation edge AI applications.

Intended to start with next generation SoCs using TSMC's N2 CMOS process, NEXAI integrates CASCADE arrays with a total of 524,288 active PEs clocked at 12 GHz achieving 12,583 dense TFLOPS (FP4) at just 1.43 W.

**Table 22: NEXAI IP blocks in Edge SoCs**

| Aspect | Value | Units |
|---|---|---|
| Performance (dense) | 12,583 | TFLOPS |
| Target CMOS process | TSMC N2 | node |
| Logic density | 313 | MTr/mm$^2$ |
| Weights and activations format: FP4 | 4 | bits |
| Primary NEXAI clock | 1.5 | GHz |
| Processing Element (PE) area | 0.77 | µm$^2$ |
| HILT unit cell area | 0.012 | µm$^2$ |
| HILT area overhead (including latch tree) | 22% | |
| CASCADE local clock speed | 12 | GHz |
| Rest of NEXAI IP block clock speed | 1.5 | GHz |
| Batch size x input token length in HILT | 4,096 | B x L |
| Active CASCADE array columns | 512 | columns |
| Spare CASCADE columns for CREST | 8 | columns |
| Columns per CASCADE array | 520 | columns |
| Rows per CASCADE array | 64 | rows |
| CASCADE arrays in NEXAI IP block | 16 | arrays |
| Total CASCADE rows NEXAI IP block | 1,024 | rows |
| PEs in NEXAI IP block | 532,480 | PEs |
| Active PEs in NEXAI IP block | 524,288 | PEs |
| Weight bits in CASCADE PEs | 2,097,152 | bits |
| Activations HILT bits | 16,777,216 | bits |
| Output sums HILT bits | 16,777,216 | bits |
| Number of SanDisk HBF NAND Flash stacks | 1 | stack |
| Capacity of HBF stacks | 512.0 | GBytes |
| Likely bandwidth of HBF stacks | 1.2 | TB/s |
| CASCADE array chip area | 0.41 | mm$^2$ |
| Activations HILT chip area | 0.25 | mm$^2$ |
| Output sums HILT chip area | 0.25 | mm$^2$ |
| Total chip area for NEXAI IP block | 0.91 | mm$^2$ |
| Total NEXAI IP block memory | 4.46 | MBytes |
| CASCADE system power consumption | 1.43 | Watts |
| Transformer inferenced | **DeepSeek V3/R1** | |
| Typical weights activated per MoE inference | 37 | billion |
| Input token sequence | 2,048 | tokens |
| CASCADE limited transformer inference time | 9.0 | ms |
| HBF limited transformer inference time | 17.0 | ms |
| Max inference rate, limited by HBF | 59 | tokens/sec |

### 28.1 High Bandwidth Flash

NEXAI uses a SanDisk High Bandwidth Flash (HBF) stack providing 512 GB of parameter storage at around 1.2 TB/s. This would enable real-time inference of 1 trillion FP4 weights worth of a mix of LLMs, multimodal transformers, and reasoning AIs such as DeepSeek R1 at a HBF-bandwidth limited rate of 59 tokens/second - performance rivaling rack-scale GPUs in a mobile edge device.

### 28.2 Source of advantage

NEXAI's capability lies in its FP4-optimized pipelines and ZettaLith-derived HILT memory. Unlike SRAM-based edge AI accelerators, HILT's latch-tree topology is designed to achieve very high bandwidth at very low power, with a footprint smaller than SRAM (n.b.: HILT is *not* a general purpose SRAM replacement).

SanDisk's recently announced HBF combines the low cost/TByte and non-volatility of Flash with HBM scale bandwidth. Combining CASCADE's efficient large arrays of fast tiny PEs, NEXAI fits alongside CPUs/GPUs and I/O in edge SoCs, making it suited for latency-critical applications like robotic motion planning, real-time AI generated video and VR, and self-driving cars.

### 28.3 Efficient silicon

NEXAI can deliver data-center level AI inference in a form factor that can take up less than 1 mm$^2$ of SoC area. With the ability to inference 2,048-token prompts of an AI with DeepSeek intelligence in 9 ms, people will be able to have intelligent conversations with their personal humanoid robots – without their private information ever traversing the internet. Sophisticated transformer models for real-time speech recognition and synthesis can be run concurrently, so people can converse naturally with the device at full speed and without cloud connectivity. Vision models and movement can also be run simultaneously, where appropriate.

### 28.4 Avoiding hotspots

CASCADE columns have a very high power density, if they are run at 12 GHz for high performance. If the NEXAI IP were to be provided as a single hard macro around 1 mm$^2$, power would be highly concentrated, and a hot-spot would be created. However, the CASCADE architecture allows it to be efficiently divided into 18 blocks, which can be spread over the SoC die to minimize hotspots, using the silicon substrate as a heat spreader. These 18 blocks can be provided as a set of 16 identical hard macros for the CASCADE arrays, and different hard macro for the output HILTs and a control CPU, with minimal wiring required between them. This minimizes localized hotspots while maintaining efficient high frequency operation.

# 29 Design Validation

## 29.1 Validation Requirements and Open Questions

The following critical aspects require experimental validation before the ZettaLith architecture can be considered viable:

- PE Timing Closure: SPICE simulation using TSMC A16 PDK to verify 12 GHz operation
- Thermal Feasibility: CFD analysis of 321 W/cm² heat dissipation with JETSTREAM cooling
- WSSCB Mechanical Integrity: FEA of silicon spring stress relief under thermal cycling
- CASCADE Array Functionality: RTL simulation of reduced-scale array
- CREST Effectiveness: Statistical analysis of fault coverage under realistic defect models

## 29.2 Design Knobs

ZettaLith's performance targets result from "pushing the envelope" along multiple axes, with the expectation that each axis may be relaxed during detailed design and simulation to recover margin elsewhere. The major "knobs" include:

- **PE Clock Frequency**
  • Nominal: 12 GHz ⟷ Reduced: 8 GHz (or intermediate)
  • Margin gained: timing-closure headroom, ~2/3 dynamic power
- **Chiplet size**: 11 mm x 13 mm is somewhat arbitrary. 11 mm matches the HBM/HBF size, and 13 mm is chosen to fit 6 per reticle.
  • Nominal: 143 mm$^2$ ⟷ Reduced or increased
  • Margin gained: chiplet stack yield, ratio of HBM to compute, system partitioning
- **WSSCB BID-BID data bandwidth**
  • Nominal: 39 TB/s ⟷ Reduced: 20 TB/s or increased: 78 TB/s
  • Margin gained: Area of BID consumed by UCIe 2.0, data fabric power consumption, and WSSCB RDL layers
- **CASCADE Array Shape**
  • Nominal: 64 rows × 8,192 columns ⟷ e.g. 128 rows × 4,096 columns
  • Margin gained: simpler floor planning, lower inter-array adder latency, different numbers and sizes of HILTs
- **Spare-Column Allocation** (CREST) spares/8,208 total ⟷ up to 160 spares/8,352 total
  • Margin gained: much higher tolerance to clustered defects while still at <2 % area overhead. Unlikely to be required
- **Power-Grid Density** (CGA & Busbars)
  • Nominal: 217-wire CGA columns (8 rings) ⟷ e.g. 61-wire (4 rings)
  • Margin gained: simpler wire bundling, with a reduction in CGA flexibility and increase in mechanical and thermal stress
- **HILT Hierarchy Overhead**
  • Nominal: 16 % HILT cell overhead ⟷ 25 % (for relaxed clock-tree slew)
  • Margin gained: lower latch drive, simplified RDL routing, at the expense of slightly larger chip area used for HILT

## 29.3 Iterative Design-Simulation Flow

Each trade-off can be dialed-in using an iterative PPA-closure methodology:

1. **Define PPA Targets & Corners**
   Establish worst-case timing, dynamic/static power, IR-drop (±10 % budgets).
2. **Macro-Level Floorplan**
   Fix SLD tiling pitch, WSSCB module layout and CGA grid dimensions.
3. **PE-Level SPICE**
   Simulate single-PE macro including local RDL parasitics at slow/typical/fast PVT corners.
4. **Full-Chip Extraction**
   Place & route CASCADE arrays, extract netlist for timing, IR-drop and EM analysis.
5. **Thermal & CFD Validation**
   Model JETSTREAM manifold flow and WSSCB silicon-spring thermal isolation in ANSYS (or equivalent).
6. **Margin Recovery**
   If any constraint is violated, back off the corresponding knob (lower clock, add spares, loosen bond pitch, etc.) and repeat until all PPA metrics simultaneously close.

This systematic knob-turning could ensure that the final ZettaLith implementation meets its zettaFLOPS, power-efficiency, thermal-reliability and yield targets with well-quantified margins.

## 30 ZettaLith Risk Mitigation

ZettaLith includes multiple risk mitigation strategies to address technical, manufacturing, and operational challenges, summarized in Table 23.

**Table 23: ZettaLith risk mitigation**

| Risk | Mitigation | Effect of mitigation |
|---|---|---|
| Overall risk too high | Start with ExaLith | Delays datacenter version |
| Not interested in datacenter version | Use ExaLith or NEXAI | Scaled performance and cost |
| 5 trillion weights not enough | Use maximum memory | 59% more expensive |
| 20 trillion weights not enough | Use combination HBM, HBF | Increased software complexity |
| ZettaLith compute not enough | Connect units via 800 GbE | Complex, decreased efficiency |
| Processing speed > HBM speed | Use large batch sizes | Already incorporated in ZettaLith |
| TRIMERA HILT insufficient | Extreme bandwidth fabric | Already incorporated in ZettaLith |
| Partial sum transfer bottleneck | Eliminate using CASCADE | Already incorporated in ZettaLith |
| Timing closure can't be achieved | Reduce clock speed | Proportionally fewer PFLOPS |
| Don't want to reduce clock speed | Use dataflow architecture | Complex simulation |
| TRIMERA clock distribution | Clock domain 0.367 mm^2 | Already incorporated in ZettaLith |
| Transistor cushion exceeded | Increase transistor cushion | 0.19% / transistor fewer PFLOPS |
| EUVL not available | Use 7 nm for SLD | Approx 78% fewer PFLOPS |
| Thermal expansion of WSSCB | MEMS silicon springs | Already incorporated in ZettaLith |
| Warpage of WSSCB | MEMS silicon springs | Already incorporated in ZettaLith |
| Crack propagation in WSSCB | MEMS silicon springs | Already incorporated in ZettaLith |
| Fragile handling of WSSCB | MEMS silicon springs | Already incorporated in ZettaLith |
| Unclean environment for WSSCB | MEMS silicon springs | Already incorporated in ZettaLith |
| Hardware faults in CASCADE arrays | CREST fault tolerance | Already incorporated in ZettaLith |
| Marginal CASCADE arrays | CREST fault tolerance | Already incorporated in ZettaLith |
| Errors in transformer inference | CREST fault tolerance | Already incorporated in ZettaLith |
| Arrays fail, but are correctable | Ignore | None |
| More than correctable arrays fail | Graceful degradation | 0.0001% fewer PFLOPS |
| TRIMERA fails | Fail in place | 0.78% per TRIMERA fewer PFLOPS |
| CPU SLD fails | Fail in place | <6.25% per CPU fewer PFLOPS |
| Power supply unit fails | Fail in place | 1.2% per PSU fewer PFLOPS |
| Power density too high | Reduce clock speed | Proportionally fewer PFLOPS |
| Electromigration in SLD | SHAPE design | Already incorporated in ZettaLith |
| Electromigration in BID, HILT | Careful design, simulation | Normal design process |
| Activation broadcast SSN | Decouple CASCADE columns | Already incorporated in ZettaLith |
| HBM4 unobtainable | Use HBM3E | 39% fewer weights |
| TSMC A16, A14 unavailable | Use TSMC N2 | 30% fewer PFLOPS |
| Poor TRIMERA or HBM attach yield | Set acceptance threshold | 0.78% per TRIMERA fewer PFLOPS |
| Hybrid bond pitch too fine | Consolidate power and ground | More simulation required |
| Poor hybrid bond yield | CREST fault tolerance | Already incorporated in ZettaLith |
| Processors not replaceable | Fail in place | Not repairable |
| Power supplies not replaceable | Fail in place | Not field repairable |
| Not field repairable | Swap, return to factory | Downtime unless local spare |
| Potential low yield of TRIMERAs | SHAPE, CREST, small die | Already incorporated in ZettaLith |
| SOTA CMOS design complexity | SHAPE design | Already incorporated in ZettaLith |
| Complex software stack | Develop focused subset | None (ZettaLith is not a GPU) |

## 31 Glossary of New Terms

This glossary explains terms that are new to the ZettaLith technology.

- **All-silicon domain**: a contiguous region of silicon-fabricated circuitry in which the active processing elements and their high–bandwidth interconnects are integrated on silicon, thereby excluding conventional board–level and rack-level interconnection mechanisms such as printed circuit boards, backplanes, ethernet cables and optic fibers. WSSCB enables large all-silicon domains.
- **BID**: Base Interface Die. A semiconductor die incorporating high-speed I/O, controller, test circuits, and various logic circuits designed to support a TRIMERA stack or CPU stack and provide standardized interfaces between internal and external connections including HBM and HBF memory stacks and the data fabric using UCIe 2.0 links.
- **BN ZettaLith**: a ZettaLith with an array of TRIMERA stacks that contain CASCADE arrays of BitNet 1.58 PEs instead of FP4 PEs.
- **CASCADE**: Column-Array Systolic Computation with Accumulation During Execution. A column-oriented matrix multiply architecture that eliminates data skewing and inter-chip partial sum transfers by performing independent vertical computation down each of many parallel columns.
- **CREST**: Cyclic Redundant Spare Testing. A fault-tolerance system integrated into the ZettaLith architecture that ensures operational reliability by continuously monitoring and repairing CASCADE arrays during transformer inference.
- **ExaLith**: An exa-scale transformer inferencing system for desktop and workstation applications. ExaLith uses small numbers of ZettaLith chips in a low-power formats such as PCIe cards. ExaLith is also software compatible with ZettaLith.
- **FA spiral**: A type of silicon spring design that combines the properties of Fermat and Archimedean spirals to create a structure that can elastically release stress in X, Y, and Z directions simultaneously while maintaining a compact footprint.
- **Folded beam**: A type of silicon spring design that is a compromise between mechanical properties and signal routing.
- **HILT**: Hierarchical Integrated Latch Tree. A sequential-access memory structure composed of pipelined latch arrays, multiplexed via transmission gates in a hierarchical tree topology. It replaces traditional SRAM in ultra-high-bandwidth applications like transformer inference, though it is not a general SRAM alternative.
- **JETSTREAM:** JET Surface Thermal Regulation via Evaporative Array Manifold. A two phase immersion cooling (2-PIC) system that incorporates a precise manifold to direct an array of liquid 2-PIC coolant jets to microchannel heatsink fins etched in the back surface of silicon chips.
- **NEXAI**: Neural Engine eXecution Accelerator for Inference. A high-performance, edge-optimized semiconductor IP block designed for transformer inference and AI workloads in power- and thermally constrained environments, derived from ZettaLith technologies.
- **SCB**: Silicon Circuit Board. A silicon substrate that serves as a circuit board, replacing traditional PCBs. An SCB is somewhat like a silicon interposer but is more robust and doesn't carry the high speed signals.
- **SHAPE**: Simple Hybrid Array of Processing Elements. A novel processing architecture that uses a SOTA Logic Die (SLD) - a high-density array of ultra-simple processing elements that can be full custom designed in a new process before the availability of standard cell libraries, SRAM, analog, mixed signal, or IP blocks. All circuits requiring these are on other die, hybrid bonded to the SLD die.
- **Silicon springs**: Micromechanical structures in silicon that provide thermal and mechanical stress relief. This stress relief can isolate sources of thermal and mechanical stress by orders of magnitude, effectively limiting propagated stress to small regions of approximately 1 $cm^2$.
- **SLD**: SOTA Logic Die, part of a TRIMERA stack. A state-of-the-art semiconductor die manufactured in the most advanced available process node at the time (e.g., TSMC A16 or A14), containing digital logic circuits optimized for high performance and low power consumption.
- **TRIMERA**: TRIchip Module for Exascale Reasoning Applications. A high-performance 3D integrated circuit architecture consisting of three vertically stacked silicon dies that are hybrid bonded together to create a dedicated transformer inference accelerator.
- **ZettaLith**: A zetta-scale transformer inferencing system that combines a passive WSSCB with CASCADE arrays in SHAPE format in TRIMERA stacks, CREST fault tolerance, and advanced cooling.
- **V-beam:** A type of silicon spring design optimized for signal routing.
- **WSSCB**: Wafer-Scale Silicon Circuit Board. A passive silicon substrate analogous to a printed circuit board (PCB) but fabricated using semiconductor processes. As with a standard FR4 PCB, WSSCB contains no transistors - only interconnects. A WSSCB supports attachment of chiplets and chip stacks with standard microbumps. The WSSCB takes the place of PCB, package substrate and silicon interposer.

.

## 32 Conclusion

ZettaLith represents an architectural exploration of extreme specialization for AI inference that could theoretically achieve 1.507 zettaFLOPS, pending extensive validation of its component technologies. Compared to state-of-the-art GPU clusters, ZettaLith could potentially achieve 1,047× higher throughput, 1,490× better energy efficiency, and 2,325× more cost-effective inference.

To my knowledge, this is the first architecture to host up to 20 trillion parameters in a single silicon-only domain.

### 32.1 Transforming AI economics

By specializing exclusively in FP4 transformer inference - and leveraging new software innovations such as multi-head latent attention and advanced sparsity exploitation - ZettaLith could help reduce end-to-end inference costs by up to five orders of magnitude relative to 2024 baselines. At that scale, large-model reasoning becomes effectively "free" at the point of use, opening new classes of AI applications: on-device multi-agent reasoning, real-time high-fidelity VR/AR, local humanoid robot intelligence, and ubiquitous semantic search with sub-second latencies.

### 32.2 Addressing data-center energy

Global data-center power draw is on track to exceed 7% of world electricity by 2030 - equivalent to the entire electricity consumption of India's 1.4 billion people - driven largely by AI workloads. ZettaLith's 1,490 × gain in PFLOPS/Watt would enable scaling to zettascale-class inference within existing power envelopes, sharply reducing the carbon footprint per query and easing pressure on grid and cooling infrastructure.

### 32.3 From racks to edge

The same design principles scale elegantly down - from ExaLith desktop cards (~6 exaFLOPS sparse @ 500 W) to NEXAI mobile IP blocks in 1 mm² of SoC area delivering $>10^{13}$ ops/sec. Thus, ZettaLith can power everything from hyperscale data centers to power-constrained edge devices and smartphones, all sharing a unified inference engine.

### 32.4 Next steps

This work is a pre-implementation design study. The immediate engineering milestones include:

- SPICE-level PE design & timing-closure at the TSMC A16 (or A14) node
- Full-chip RTL synthesis, physical layout and IR-drop analysis of CASCADE+CREST
- Thermo-mechanical simulation of WSSCB stress relief and JETSTREAM 2-phase cooling manifold
- Prototype fabrication of a CASCADE column in an MPW shuttle for silicon validation

Early use of AI-accelerated RTL generation and layout tools may cut development time and cost. Moving from paper to silicon will require a disciplined PPA-driven flow and multi-disciplinary teams, but the potential 1,000 × leap in AI inference efficiency makes ZettaLith a compelling next step toward sustainable, widely available AGI-scale computing.

### 32.5 Why no simulation yet?

The absence of simulation data reflects a deliberate methodological choice to first establish architectural feasibility across all subsystems before detailed validation. The effort required for simulation can prematurely "crystallize" the overall design into a configuration that fails for reasons far outside the scope of the simulation. The author has deliberately maintained ZettaLith in a flexible pre-simulation state until there is confidence that fundamental problems in achieving ZettaLith's performance, power, and cost goals that can be identified at this stage are addressed. This approach has proven effective in the author's previous work on complex multidisciplinary systems.